\documentclass[twocolumn]{aastex701}

\usepackage{graphicx,amsmath,amssymb}
\usepackage[export]{adjustbox}
\usepackage{bm}
\usepackage{mathrsfs}
\usepackage{multirow}
\usepackage{xcolor}
\usepackage{ragged2e}
\usepackage{tabularx}
\usepackage{booktabs}

\def\mnras{Mon. Not. R. Astron. Soc. }

\def\apj{Astrophys. J.}
\def\apjl{Astrophys. J. Lett.}
\def\apjs{Astrophys. J., Suppl. Ser.}

\def\aap{Astron. Astrophys.}
\def\aapr{Astron. Astrophys. Rev.}

\shorttitle{Erythrohenosis}
\shortauthors{Amaro Seoane}

\begin{document}

\title{Erythrohenosis - The crimson chronicles of two giants}

\author[orcid=0000-0003-3993-3249,gname='Pau',sname='Amaro Seoane']{Pau Amaro Seoane}
\affiliation{Universitat Politècnica de València, C/Vera s/n, València, 46022, Spain}
\affiliation{Max Planck Institute for Extraterrestrial Physics, Giessenbachstra{\ss}e 1, Garching, 85748, Germany}
\email[show]{amaro@upv.es}

\begin{abstract}
We investigate ``erythrohenosis''—the collision and merger of two red giants—establishing an end-to-end model for this fundamental evolutionary channel in dense stellar environments. Combining three-dimensional SPH simulations of a binary with analytical modeling, we characterize the event from initial encounter to terminal explosion. We demonstrate that grazing encounters induce tidal capture and rapid orbital decay, accompanied by large-amplitude, nonlinear stellar oscillations. The subsequent inspiral spins up the common envelope into a stable, non-spherical equilibrium, powering a luminous precursor with quasi-periodic bursts. The terminal explosion, modeled with angular momentum conservation, produces an intrinsically flattened remnant that preserves a geometric memory, or \textit{morphomnesia}, of its binary origin. The associated gravitational wave signal features a rapid, drag-dominated frequency evolution, identifiable by a unique time-varying apparent chirp mass. These results define a distinctive multi-stage observational fingerprint—linking transient optical precursors, asymmetric nebulae, and anomalous gravitational wave chirps—to guide identification in current and future multi-messenger surveys.
\end{abstract}


\tableofcontents

\section{Introduction}
\label{sec.intro}

Environments such as globular clusters and galactic nuclei feature exceptionally high stellar densities, reaching from one million to one hundred million stars per cubic parsec. In these crowded regions, stellar relative velocities can be on the order of $\sim \text{a~few~}10\,\text{km/s}$ for globular clusters and climb to $\sim 100-1000\,\text{km/s}$ within galactic nuclei \citep{NeumayerEtAl2020,Spitzer87,BinneyTremaine08}. Under such extreme conditions, which are unparalleled elsewhere in a host galaxy, the effects of stellar encounters become significant. The term ``collisional'' here refers broadly to mutual gravitational deflections that result in an exchange of energy and angular momentum, but it also encompasses direct physical impacts between stars. The crucial role that stellar collisions might play in explaining specific astronomical observations and in shaping the overall evolution of dense stellar systems has been the focus of numerous numerical investigations \citep{SS66,DDC87a,Sanders70b,BenzHills1987,DDC87b,DaviesEtAl1991,BenzHills1992,MCD91,LRS93,LRS95,LRS96,1999MNRAS.308..257B,DaviesEtAl1998,BaileyEtAl1999,LombardiEtAl2002,2002ASPC..263....1S,AdamsEtAl2004,TracEtAl2007,DaleEtAl2009,2020ApJ...901...44W,Mastrobuono-BattistiEtAl2021,2021A&A...649A.160V}. {Collisions involving giant stars are particularly significant as they often initiate a common envelope evolution (CEE) phase, a critical but poorly understood stage in binary evolution responsible for the formation of compact binaries \citep[see e.g.][]{Paczynski1976, Ivanova2013,Iben1993}. While CEE is typically studied starting from an unstable mass transfer scenario \citep[e.g.,][]{PassyEtAl2012, OhlmannEtAl2016}, direct collisions offer an alternative pathway to initiate this phase. Furthermore, stellar mergers are now observationally confirmed as the progenitors of Luminous Red Novae (LRNe), a class of transients characterized by intermediate luminosities and red colors, such as V1309 Sco \citep{TylendaEtAl2011} and V838 Mon \citep{WoodwardEtAl2021}. Understanding the dynamics and observational signatures of red giant mergers is therefore crucial for interpreting these transients and constraining CEE physics.}

To quantify the frequency of these events, the study of blue stragglers provides an excellent proxy for inferring the amount of stellar collisions in globular clusters, as these stars are widely believed to be the product of such encounters \citep[][]{Maeder1987,Bailyn1995,Leonard1989}. For instance, \cite{Leonard1989} estimated a collision rate of $10^{-8}\,\text{yr}^{-1}$ within a single globular cluster, assuming a small primordial binary population. Extrapolating this to a typical galaxy like the Milky Way, with approximately 100 globular clusters, yields a rate of $10^{-6}\,\text{yr}^{-1}$ per galaxy. This figure could be an underestimate, as collisions involving binaries are more important \cite{LeonardFahlman1991}. A few years later, a detailed theoretical and numerical study by \cite{SigurdssonPhinney95} suggested a broader range for main-sequence stellar collisions, from $10^{-6}$ to $10^{-4}$ per year per galaxy, also assuming 100 globular clusters.

It is important to recognize that the average number of globular clusters in early-type galaxies correlates with the mass of their central massive black hole \citep{BurkertTremaine2010}. As shown in their Fig.~(1), this number can increase by several orders of magnitude; for example, NGC 4594 contains approximately $2\times 10^3$ globular clusters, which would substantially increase the overall collision rate. On a larger scale, within a reference distance of 100 Mpc, we find massive structures such as the Virgo Cluster and the Coma Cluster (Abell 1656), each with over $10^3$ identified galaxies, and superclusters like the Laniakea Supercluster with about $10^5$ galaxies \citep{TullyEtAl2014} and the CfA2 Great Wall at a distance of only $\sim 92\,\text{Mpc}$ \citep{GellerHuchra1989}. The potential collision rates are therefore significant; if we took an average of 1000 clusters per galaxy and the higher rate of $10^{-4}$ from \cite{SigurdssonPhinney95}, the number of collisions would be a thousand times larger than the baseline estimate. Although these authors did not address collisions involving red giants, the much larger cross-section of these stars and the lower relative velocities in globular clusters imply that their collision rates must be much higher.

This work builds upon the initial analytical article of \cite{AmaroSeoane2023}, which focused on stellar collisions (main sequence and red giants) in galactic nuclei. We now address the collision of two red giants in a globular cluster. Because of the slow process of the collision, we cannot present an analytical work and resort to numerical simulations, which we study and present in detail, as well as their implications. We suggest the term \textit{erythrohenosis}: ``red-unification'' or ``the becoming-one of the red'' as the crimson cannibalism of red giants. This is based on \textit{erythrós}, red, which evokes the red giant's spectral class plus \textit{henosis}, unification, from hen - ``one''.

The remainder of this paper is structured as follows. In Section~\ref{sec.headongrazing}, we analyze the statistics of stellar encounters, determining the probability of head-on versus grazing collisions in different environments. Section~\ref{sec.creation} details the creation of our realistic red giant models using MESA and their mapping into three-dimensional SPH initial conditions. The setup and initial dynamics of the collision are presented in Section~\ref{sec.bracing_collision}. We analyze the evolution of the degenerate cores and the surrounding gas shells in Section~\ref{sec.coreinnershells}, including an analytical model for the electromagnetic bursts and their duty cycle. Section~\ref{sec.wavelets} employs a continuous wavelet transform to analyze the non-stationary energy dissipation during the inspiral. In Section~\ref{sec.Airy}, we use the Airy function formalism to model the core-envelope interface and quantify the accretion rates. We present initial semi-analytical models for the subsequent evolution of the cores and the explosion morphology in Sections~\ref{sec.UlteriorNaive} and \ref{sec.homothetic}. A refined, quantitative analysis of the common envelope evolution, the final inspiral, and the explosion is detailed in Section~\ref{sec.quantitative_analysis}. {Section~\ref{sec.ImprovedModel} introduces an improved model for the final inspiral, including predictions for the gravitational wave signature and a detection algorithm (the TVACM diagnostic).} Section~\ref{sec.induced_dynamics} analyzes the tidally induced nonlinear oscillations resulting from a grazing encounter. Finally, we summarize our findings and discuss the observational fingerprints of erythrohenosis in Section~\ref{sec.summary_conclusions}.

\section{Head-on or grazing?}
\label{sec.headongrazing}

{Before modeling a specific collision, we first establish the statistical likelihood of different encounter geometries in the relevant astrophysical environments.} The probability distribution of the minimum approach distance, $d_\text{min}$, during a stellar encounter requires an analysis grounded in statistical mechanics and the dynamics of the two-body problem. We expand upon the analysis presented in \cite{AmaroSeoane2023}.

We consider a stellar system characterized by a distribution of velocities. We seek the probability density function (PDF) $f(d_\text{min}) \equiv dP/d(d_\text{min})$. The probability that $d_\text{min}$ lies within the interval $[d_1, d_2]$ is given by the definite integral

\begin{equation}
\label{eq.prob_definition_full}
P(d_1 \le d_\text{min} \le d_2) = \int_{d_1}^{d_2} f(d_\text{min}) d(d_\text{min}).
\end{equation}

\noindent
The PDF is proportional to the differential collision rate $d\Gamma/d(d_\text{min})$. The total collision rate $\Gamma$ is determined by the integral of the product of the number density $n$, the relative velocity $V_\text{rel}$, and the cross-section $\sigma$, averaged over the distribution of relative velocities, $\Gamma = n \langle V_\text{rel} \sigma \rangle$. The differential rate with respect to $d_\text{min}$ is therefore

\begin{equation}
\label{eq.prob_density_rate}
f(d_\text{min}) \propto \frac{d\Gamma}{d(d_\text{min})} = n \left\langle V_\text{rel} \frac{d\sigma}{d(d_\text{min})} \right\rangle.
\end{equation}

\noindent
To determine $d\sigma/d(d_\text{min})$, we analyze the dynamics of an encounter between two masses $m_1$ and $m_2$ (total mass $M=m_1+m_2$). Conservation of energy and angular momentum in the center-of-mass frame relates the impact parameter $b$ to $d_\text{min}$ and $V_\text{rel}$,

\begin{equation}
\label{eq.b_dmin_expanded}
b^2 = d_\text{min}^2 \left(1 + \frac{2GM}{d_\text{min} V_\text{rel}^2}\right) = d_\text{min}^2 \left(1 + \Theta(d_\text{min})\right),
\end{equation}

\noindent
where $\Theta(r) = V_\text{esc}^2(r) / V_\text{rel}^2$ is the Safronov number, and $V_\text{esc}^2(r) = 2GM/r$ is the square of the escape velocity.

The differential cross-section is $d\sigma = 2\pi b db$. We perform a change of variables by differentiating Eq.~(\ref{eq.b_dmin_expanded}) with respect to $d_\text{min}$, holding $V_\text{rel}$ constant,

\begin{equation}
\frac{d(b^2)}{d(d_\text{min})} = 2b \frac{db}{d(d_\text{min})} = \frac{d}{d(d_\text{min})} \left( d_\text{min}^2 + \frac{2GM}{V_\text{rel}^2} d_\text{min} \right).
\end{equation}

\noindent
The differentiation yields

\begin{equation}
2b \frac{db}{d(d_\text{min})} = 2d_\text{min} + \frac{2GM}{V_\text{rel}^2}.
\end{equation}

\noindent
The differential cross-section $d\sigma/d(d_\text{min})$ is consequently

\begin{equation}
\label{eq.diff_cross_section}
\frac{d\sigma}{d(d_\text{min})} = \pi \left(2d_\text{min} + \frac{2GM}{V_\text{rel}^2}\right).
\end{equation}

\noindent
We substitute this expression into Eq.~(\ref{eq.prob_density_rate}) to obtain the general form of the probability density function,

\begin{align}
\label{eq.f_dmin_general}
f(d_\text{min}) &\propto \left\langle V_\text{rel} \, \pi \left(2d_\text{min} + \frac{2GM}{V_\text{rel}^2}\right) \right\rangle \nonumber \\
&= 2\pi d_\text{min} \langle V_\text{rel} \rangle + 2\pi GM \langle V_\text{rel}^{-1} \rangle.
\end{align}

\noindent
This expression unifies the statistics of close encounters. The PDF is composed of two terms: a linear term representing the contribution from the geometric cross-section, and a constant term arising from gravitational focusing. The relative importance of these terms depends on the moments of the relative velocity distribution, $\langle V_\text{rel} \rangle$ and $\langle V_\text{rel}^{-1} \rangle$.

We define a critical distance $d_\text{crit}$ where the two terms in Eq.~(\ref{eq.f_dmin_general}) are equal,

\begin{equation}
\label{eq.d_crit}
d_\text{crit} = GM \frac{\langle V_\text{rel}^{-1} \rangle}{\langle V_\text{rel} \rangle}.
\end{equation}

\noindent
The PDF can be written as $f(d_\text{min}) \propto d_\text{min} + d_\text{crit}$. We analyze the asymptotic behavior in two distinct dynamical regimes.

Regime I: Weak Gravitational Focusing (e.g., Galactic Nuclei).
In environments with high velocity dispersions, the typical kinetic energy is large. For encounters where $d_\text{min} \gg d_\text{crit}$, the linear term in Eq.~(\ref{eq.f_dmin_general}) dominates. The probability density function simplifies to

\begin{equation}
\label{eq.f_dmin_weak}
f(d_\text{min}) \approx 2\pi d_\text{min} \langle V_\text{rel} \rangle \propto d_\text{min}.
\end{equation}

\noindent
The cumulative probability $P(d_\text{min} < d) = \int_0^d f(x) dx$ scales quadratically, $P(d_\text{min} < d) \propto d^2$. This scaling reflects the geometric cross-section $\pi d^2$, indicating that gravitational deflection is negligible. Distant encounters are statistically favored over close ones.

Regime II: Strong Gravitational Focusing (e.g., Globular Clusters).
In environments with low velocity dispersions, gravitational focusing dominates. For encounters where $d_\text{min} \ll d_\text{crit}$, the constant term in Eq.~(\ref{eq.f_dmin_general}) dominates. The probability density function simplifies to

\begin{equation}
\label{eq.f_dmin_strong}
f(d_\text{min}) \approx 2\pi GM \langle V_\text{rel}^{-1} \rangle = \text{constant}.
\end{equation}

\noindent
The probability density is uniform. The cumulative probability scales linearly, $P(d_\text{min} < d) \propto d$. This arises because the total cross-section in this regime scales as $\sigma(<d) \propto d/V_\text{rel}^2$, leading to a collision rate linear in $d$. The uniform distribution implies an equipartition of encounter probabilities; a head-on collision ($d_\text{min} \approx 0$) is as likely as a grazing encounter within an equivalent infinitesimal interval $d(d_\text{min})$.

The implications of these distinct distributions are significant for the nature of physical interactions, such as direct collisions or tidal disruptions. We quantify this by analyzing the statistics of collisions, defined by $d_\text{min} \le R_\text{coll}$, where $R_\text{coll}$ is the collision radius. We calculate the mean minimum approach distance $\langle d_\text{min} \rangle$ for these events.

In a galactic nucleus (weak focusing), we assume the environment is characterized by high velocities such that the scaling $f(d_\text{min}) \propto d_\text{min}$ applies. The normalized conditional probability distribution for $d_\text{min} \in [0, R_\text{coll}]$ is obtained by normalizing the PDF,

\begin{equation}
f(d_\text{min} | \text{weak}) = \frac{d_\text{min}}{\int_0^{R_\text{coll}} x dx} = \frac{2 d_\text{min}}{R_\text{coll}^2}.
\end{equation}

\noindent
The mean minimum approach distance is

\begin{align}
\label{eq.mean_dmin_weak}
\langle d_\text{min} \rangle_\text{weak} & = \int_0^{R_\text{coll}} d_\text{min} f(d_\text{min} | \text{weak}) d(d_\text{min}) = \nonumber \\
 &\frac{2}{R_\text{coll}^2} \left[\frac{d_\text{min}^3}{3}\right]_0^{R_\text{coll}} = \frac{2}{3} R_\text{coll}.
\end{align}

\noindent
This result quantifies the statistical preference for grazing collisions ($d_\text{min} \approx R_\text{coll}$) over head-on collisions in high-velocity environments.

In a globular cluster (strong focusing), we assume the relevant interactions occur where $f(d_\text{min}) = \text{constant}$. The normalized conditional probability distribution is uniform,

\begin{equation}
f(d_\text{min} | \text{strong}) = \frac{1}{\int_0^{R_\text{coll}} dx} = \frac{1}{R_\text{coll}}.
\end{equation}

\noindent
The mean minimum approach distance is

\begin{align}
\label{eq.mean_dmin_strong}
\langle d_\text{min} \rangle_\text{strong} & = \int_0^{R_\text{coll}} d_\text{min} f(d_\text{min} | \text{strong}) d(d_\text{min}) = \nonumber \\
& \frac{1}{R_\text{coll}} \left[\frac{d_\text{min}^2}{2}\right]_0^{R_\text{coll}} = \frac{1}{2} R_\text{coll}.
\end{align}

\noindent
Since $\langle d_\text{min} \rangle_\text{strong} < \langle d_\text{min} \rangle_\text{weak}$, strong gravitational focusing shifts the distribution of encounters towards closer, more deeply penetrating interactions. This enhancement of close interaction rates fundamentally alters the dynamical evolution and the production rates of phenomena dependent on small $d_\text{min}$, such as gravitational wave sources or tidal capture binaries, in different astrophysical environments.

In this low-velocity regime, where the relative speed of the stars is significantly smaller than their mutual escape velocity, the governing physics diverges fundamentally from the high-velocity encounters described in \cite{AmaroSeoane2023,RyuEtAl2024}. This is not merely a quantitative difference; the transition from a violent, shock-driven impact to a more gentle, gravitationally-mediated merger introduces an entirely new set of hydrodynamic and stellar evolutionary processes. The distinct nature of these slow collisions warrants the detailed and dedicated investigation that follows.

\section{Creation of a red giant}
\label{sec.creation}

{Having established the statistical context for stellar encounters (Section~\ref{sec.headongrazing}), we now detail the construction of the specific stellar models used in our simulations.}

\subsection{Realistic one-dimensional calculations of red giants}
\label{subsec.1D_calcs}

\begin{figure}
          {\includegraphics[width=0.55\textwidth,center]{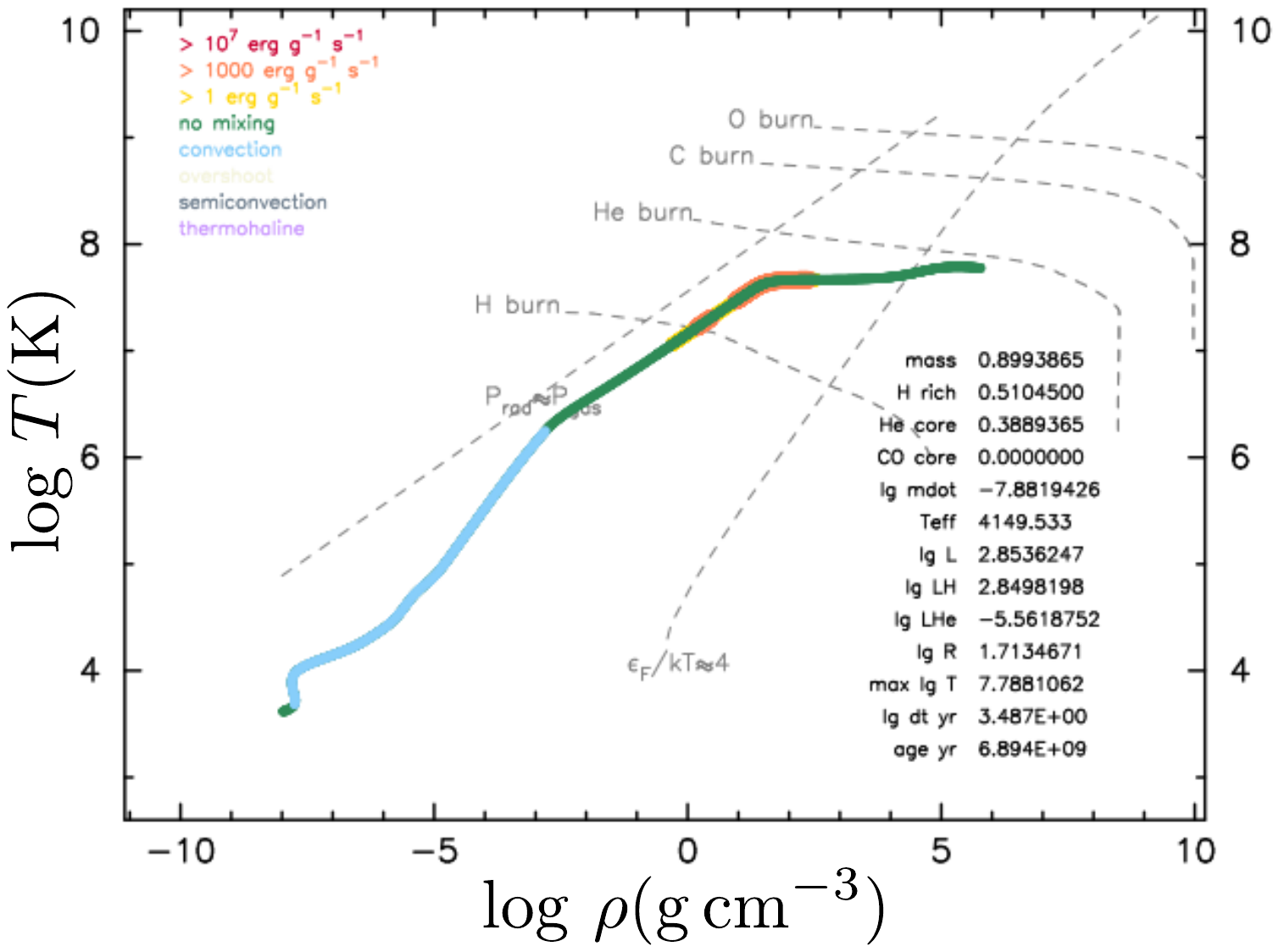}}
\caption
   {
Temperature as a function of the density for the initial conditions after $\sim 6.89\times 10^9\,\text{yrs}$. This moment captures a $1\,M_{\odot}$ star firmly in its red giant phase, with a degenerate core, active hydrogen shell burning, and an inflated envelope—all hallmarks of post-main-sequence evolution in low-metallicity environments. The integration beyond this point allows us to map the entire RGB trajectory until numerical constraints halt the simulation.
   }
\label{fig.LogT_LogRho}
\end{figure}

In Fig.~(\ref{fig.LogT_LogRho}) we display the evolution of a star with an initial mass of $1\,M_{\odot}$ and metallicity $Z=0.002$ (typical of globular clusters) that has evolved into a red giant. The figure corresponds $\sim 6.89 \times 10^9\,\text{yrs}$. The Fermi criterion $\epsilon_F/(kT) \gtrsim 4$ has already been crossed, indicating that the helium core has become electron-degenerate, a defining feature of the red giant phase. This degeneracy suppresses the core's ability to contract further despite heating, forcing the envelope to expand dramatically, as evidenced by the large radius ($\log(R/R_{\odot}) = 1.71$) and low effective temperature ($T_{\text{eff}} = 4149\,\text{K}$).  

The figure's colour-coding represents distinct physical regimes: regions with nuclear energy generation rates $>10^7\,\text{erg}\,\text{g}^{-1}\,\text{s}^{-1}$, while lower rates ($>1000$ or $>1\,\text{erg}\,\text{g}^{-1}\,\text{s}^{-1}$) correspond to less active zones. Mixing processes are also indicated: no mixing (radiative zones), convection (e.g., in the envelope), overshoot (beyond formal convective boundaries), semiconvection (partial mixing in composition gradients), and thermohaline (instability from inverse $\mu$-gradients). The hydrogen-burning shell, with its high energy generation ($\log(L_{\text{H}}/L_{\odot}) = 2.85$), dominates the luminosity, while the inert degenerate core ($M_{\text{He}} = 0.39\,M_{\odot}$) awaits helium ignition.  

We continued integrating the star's evolution until the timestep became prohibitively short, capturing multiple red giant models at different stages. This particular snapshot represents one moment in the red giant branch (RGB) phase, where the star's structure—a compact degenerate core surrounded by an extended, cool envelope—is fully established. The absence of helium burning ($\log(L_{\text{He}}/L_{\odot}) = -5.56$) confirms it has not yet reached the helium flash. The ``knee'' in the evolutionary track marks the onset of this RGB phase, triggered by core degeneracy and shell burning. Each subsequent model extracted during integration provides further insight into the star's progression toward later phases, such as the horizontal branch or asymptotic giant branch, depending on mass loss and nuclear burning evolution.  

\begin{figure*}
          {\includegraphics[width=1\textwidth,center]{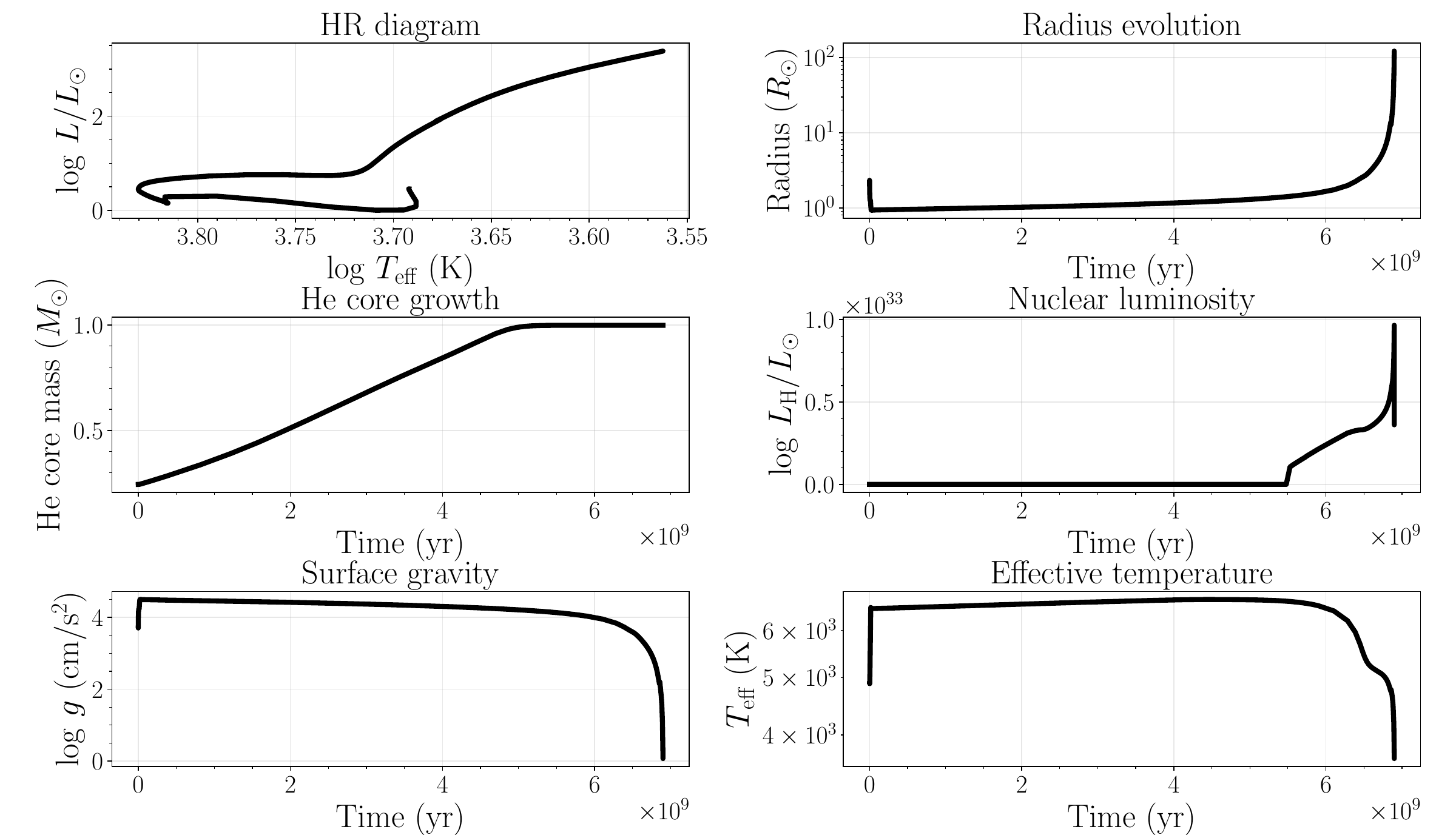}}
\caption
   {
Multi-panel visualization of a $1\,M_{\odot}$, $Z=0.002$ star's evolution into a red giant. From left to right, top to bottom: (1) HR diagram ($\log T_{\text{eff}}$ against $\log L$) showing the transition from main sequence to red giant branch; (2) Radius evolution ($R/R_{\odot}$ against time) highlighting envelope expansion; (3) He core mass growth ($M_{\text{He}}$ against time); (4) Nuclear luminosity ($L_{\text{H}}$, $L_{\text{He}}$ against time) tracking shell/core burning; (5) Surface gravity ($\log g$ against time) documenting envelope dilution; (6) Effective temperature ($T_{\text{eff}}$ against time) cooling evolution. Together, these panels capture the star's structural transformation, with the snapshot at $6.89\,$Gyr. See main text for detailed analysis. 
   }
\label{fig.RG_z0p002}
\end{figure*}

We have more details in Fig.(\ref{fig.RG_z0p002}), which provides us with a comprehensive view of the star's evolution from the main sequence to the red giant phase. This corresponds to the full integration.

The first panel shows the Hertzsprung-Russell diagram, plotting $\log T_{\text{eff}}$ against $\log L/L_{\odot}$. Here, the star begins on the main sequence, where hydrogen burning in the core maintains equilibrium. As the core hydrogen depletes, the star moves rightward and upward, entering the red giant branch. The effective temperature drops while luminosity increases, reflecting the expansion of the outer envelope and the onset of hydrogen shell burning. The current snapshot, with $\log T_{\text{eff}} \approx 3.62$ and $\log L/L_{\odot} \approx 2.85$, lies firmly in the red giant region, characterized by low surface temperatures and high luminosities due to the star's inflated radius.  

The second panel displays the evolution of the stellar radius over time, plotted as $\log R/R_{\odot}$ versus age. During the main sequence, the radius remains nearly constant, but as the star transitions to the red giant phase, the radius increases dramatically. This expansion is driven by the envelope's response to the contracting helium core and the energy generated by hydrogen shell burning. The snapshot shows $\log R/R_{\odot} \approx 1.71$, corresponding to a radius of roughly $51.5\,R_{\odot}$, a clear signature of the red giant phase.  

The third panel tracks the growth of the helium core mass, $M_{\text{He}}/M_{\odot}$, over time. Initially, the helium core is negligible, but as hydrogen burning shifts to a shell around the core, the helium mass increases steadily. The snapshot reveals a well-developed helium core with $M_{\text{He}} \approx 0.39\,M_{\odot}$, indicating advanced evolution. The core's growth is a direct consequence of hydrogen shell burning, which deposits helium ash into the core while the envelope expands.  

The fourth panel illustrates the nuclear luminosity contributions from hydrogen ($L_{\text{H}}$) and helium ($L_{\text{He}}$) burning, plotted as $\log L/L_{\odot}$ versus age. The main sequence is dominated by core hydrogen burning, but as the star evolves, shell burning takes over, leading to a sharp rise in $L_{\text{H}}$. The snapshot shows $\log L_{\text{H}}/L_{\odot} \approx 2.85$, while helium burning remains inactive ($\log L_{\text{He}}/L_{\odot} \approx -5.56$), confirming that the star is still on the red giant branch and has not yet ignited helium in the core.  

The fifth panel presents the surface gravity, $\log g$, as a function of time. During the main sequence, $\log g$ remains relatively high, but as the star becomes a red giant, the surface gravity drops significantly due to the envelope's expansion. The snapshot shows $\log g \approx 1.7$, consistent with the low surface gravity expected for a star with such an extended envelope.  

The sixth panel shows the evolution of the effective temperature, $T_{\text{eff}}$, plotted against time. The main-sequence phase exhibits stable temperatures, but as the star ascends the red giant branch, $T_{\text{eff}}$ declines. The snapshot records $T_{\text{eff}} \approx 4149\,\text{K}$, a typical value for a red giant with an extended, cool envelope.  

Together, these panels capture the star's transition from the main sequence to the red giant phase. The data reveal a degenerate helium core, an active hydrogen-burning shell, and an inflated envelope—all hallmarks of red giant structure. The integration was continued until the timestep became too short, allowing us to extract multiple models representing different stages of red giant evolution.

\subsection{Moving to three dimensions}
\label{subsec.3D_mapping}

The three-dimensional stellar model is constructed from one-dimensional MESA \citep{Eggleton1983,MESA01,MESA02,MESA03,MESA04} stellar evolution calculations to ensure consistency with realistic stellar structure. The MESA output provides a spherically symmetric profile, including the mass coordinate $m(r)$, logarithmic radius $\log_{10} r$, temperature $\log_{10} T$, density $\log_{10} \rho$, pressure $\log_{10} P$, and mass fractions of hydrogen $X$, helium $Y$, and metals $Z$. These quantities are used to initialize the three-dimensional smoothed particle hydrodynamics (SPH) representation of the star.  

The SPH model consists of $N = 10^5$ particles, each carrying mass $m_i$, position $\mathbf{r}_i$, velocity $\mathbf{v}_i$, internal energy $u_i$, and smoothing length $h_i$. The density at particle $i$ is computed via a smoothing kernel. The initial particle distribution is generated by mapping the MESA radial profile onto a three-dimensional configuration, ensuring that the cumulative mass distribution matches the one-dimensional input.  

To stabilize the model, the system is evolved under a damped dynamical relaxation phase. This procedure minimizes the bulk kinetic energy $K$, reducing spurious oscillations until the star settles into hydrostatic equilibrium. The gravitational potential energy $W$ and internal energy $U$ are monitored throughout this phase. The relaxation continues until the total energy $E_{\text{tot}} = K + W + U$ converges to a stable minimum value, and the system satisfies the virial theorem (e.g., $2U+W \approx 0$ for an ideal monatomic gas), typically within a low relative tolerance.
The final SPH model reproduces the MESA-derived density, pressure, and temperature profiles. This ensures that the star is dynamically stable before being used in subsequent collisional simulations, providing a physically consistent initial condition for studying interactions involving evolved stars.

\begin{figure}
          {\includegraphics[width=0.45\textwidth,center]{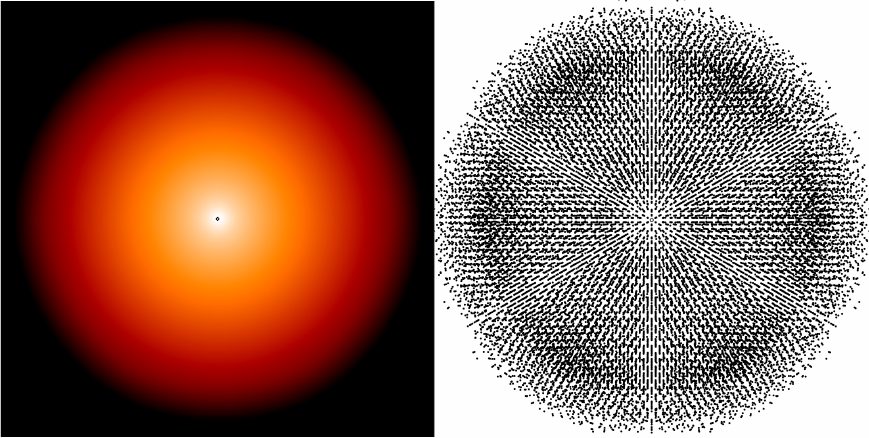}}
\caption
   {
Initial conditions of a representative red giant model. Left: Rendered visualization showing $\log \rho$, showing the density structure of the stellar envelope. Right: Raw SPH particle distribution in the $x$-$y$ plane, with the compact core visible at the centre, as on the left panel.    
   }
\label{fig.ICs_z0p002_rendered_raw}
\end{figure}

In Fig.~(\ref{fig.ICs_z0p002_rendered_raw}) we depict the initial conditions of the red giant at a selected evolutionary time in two representations; a rendered visualization and a raw particle distribution. The rendered version displays the logarithm of the gas density ($\log \rho$) as the rendering variable, enhancing the contrast in density structure across the stellar envelope. The raw view shows the unprocessed SPH particle positions in the $x$-$y$ plane, with each point corresponding to a single gas particle.  

The SPH dataset consists of $99,954$ gas particles and one compact object particle, representing the core of the red giant. The rendered plot employs a smoothing kernel to interpolate the discrete particle data into a continuous density field, taking into account the steep radial density gradient characteristic of red giant envelopes. In contrast, the raw particle plot provides a direct view of the spatial distribution, confirming uniform sampling and the absence of numerical artifacts.

\section{Bracing for collision}
\label{sec.bracing_collision}

{With the stellar models defined (Section~\ref{sec.creation}), we now set up the initial conditions for the 3D SPH simulation and present the initial dynamics of the encounter.}

The two colliding red giants have initial masses of 0.95 $M_{\odot}$ and 0.85 $M_{\odot}$ with radii of 28.48 $R_{\odot}$ and 70.23 $R_{\odot}$ respectively. Their luminosities measure $2.8 \times 10^{2}$ $L_{\odot}$ and $1.1 \times 10^{3}$ $L_{\odot}$, while their effective temperatures are 4470 K and 3980 K. The more compact 0.95 $M_{\odot}$ star shows no convective core, while the more extended 0.85 $M_{\odot}$ star exhibits a shallow convective envelope extending down to about 0.5 $M_{\odot}$. Both stars are in the hydrogen shell-burning phase, with central temperatures of $1.3 \times 10^{7}$ K and surface gravities of $\log g \sim 0.5$. Initially separated by 500 $R_{\odot}$ on a hyperbolic orbit, they reach a pericenter distance of 0.5 $R_{\odot}$, ensuring a direct collision. The larger star's extended envelope (70.23 $R_{\odot}$) and higher luminosity ($1.1 \times 10^{3}$ $L_{\odot}$) will dominate the early interaction dynamics, while their comparable masses lead to significant angular momentum transfer during the merger. The post-collision remnant's properties will be shaped by the 0.95 $M_{\odot}$ star's more compact core interacting with the 0.85 $M_{\odot}$ star's expansive envelope.

\begin{figure}
          {\includegraphics[width=0.45\textwidth,center]{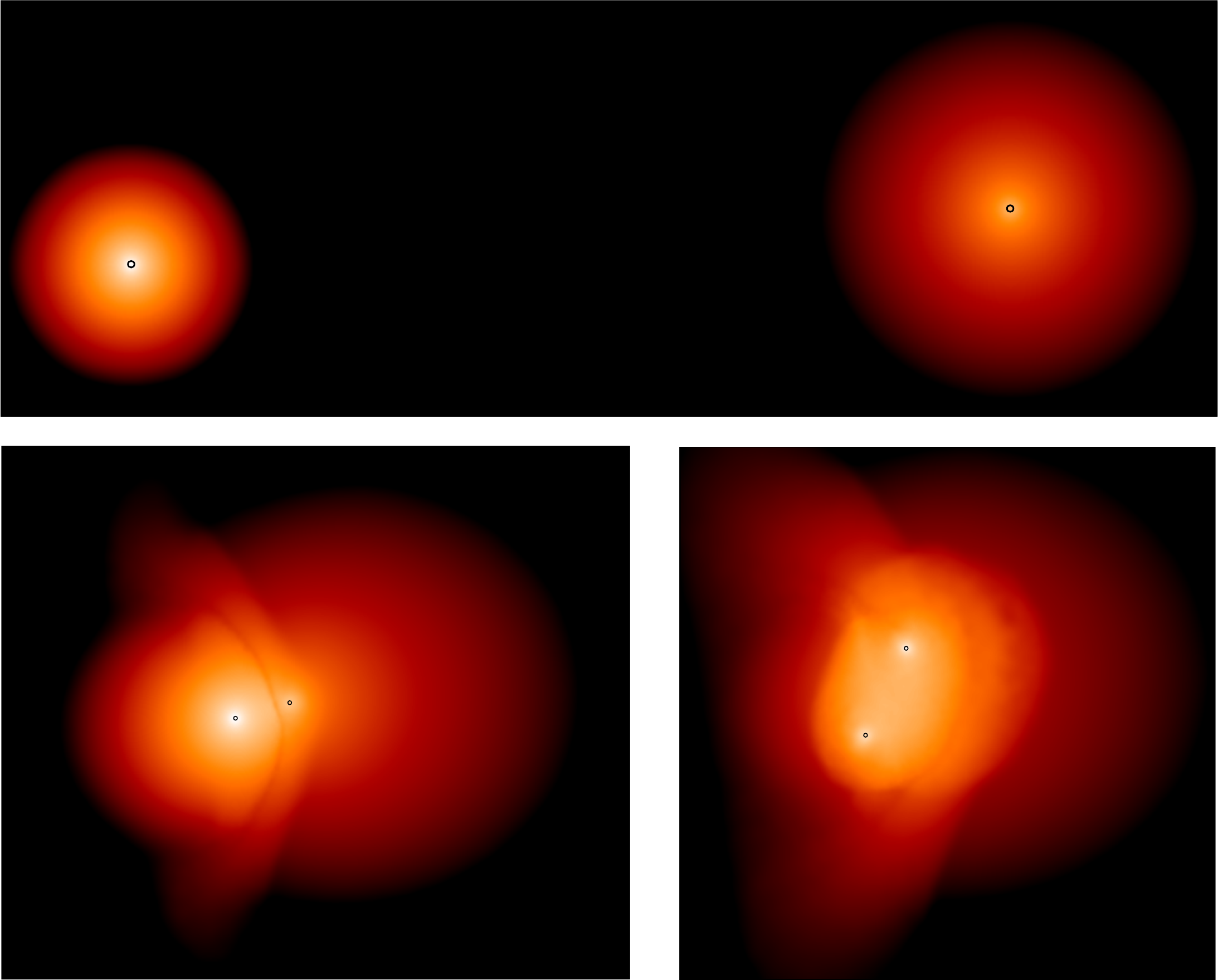}}
\caption
   {
Erythrohenosis at its best: Three key stages in the first encounter of the collision of two red giants with different initial properties. The top panel shows the initial configuration, with the larger star (right) and smaller companion (left) prior to interaction. The lower panels depict: (left) the first collision moment, and (right) a subsequent phase of this encounter. All panels display log-density renderings.
   }
\label{fig.FirstEncounter}
\end{figure}

In Fig.~(\ref{fig.FirstEncounter}) we display three evolutionary stages of a red giant collision, simulated using the methods outlined previously. The top panel shows the initial conditions, with the larger red giant and its lower-mass companion separated by their initial orbital distance. The lower panels capture the dynamical phases of the first encounter.
In the first collision phase (lower left), the more massive red giant’s extended envelope begins to engulf its companion, creating a pronounced asymmetry in the gas distribution. This interaction triggers violent hydrodynamic instabilities at the interface between the two stars, leading to the ejection of a significant fraction of their combined envelope material. By the later stage (lower right), the system has shed a substantial amount of gas—visible as diffuse, unbound ejecta—while the cores (marked by white circles) continue their inspiral within the shared envelope. The asymmetric mass transfer, driven by the initial size and density contrast between the two stars, reshapes both envelopes. The cores' subsequent orbital evolution, including potential merger or binary formation, will be examined later, where we discuss the long-term fate of the system.

\section{Core and inner shells evolution}
\label{sec.coreinnershells}

{Following the initial collision and the formation of a common envelope (Section~\ref{sec.bracing_collision}), we analyze the subsequent evolution of the degenerate cores and the surrounding gas.}

\subsection{Evolution of the degenerate cores}
\label{sec.evo_cores}

\begin{figure}
          {\includegraphics[width=0.55\textwidth,center]{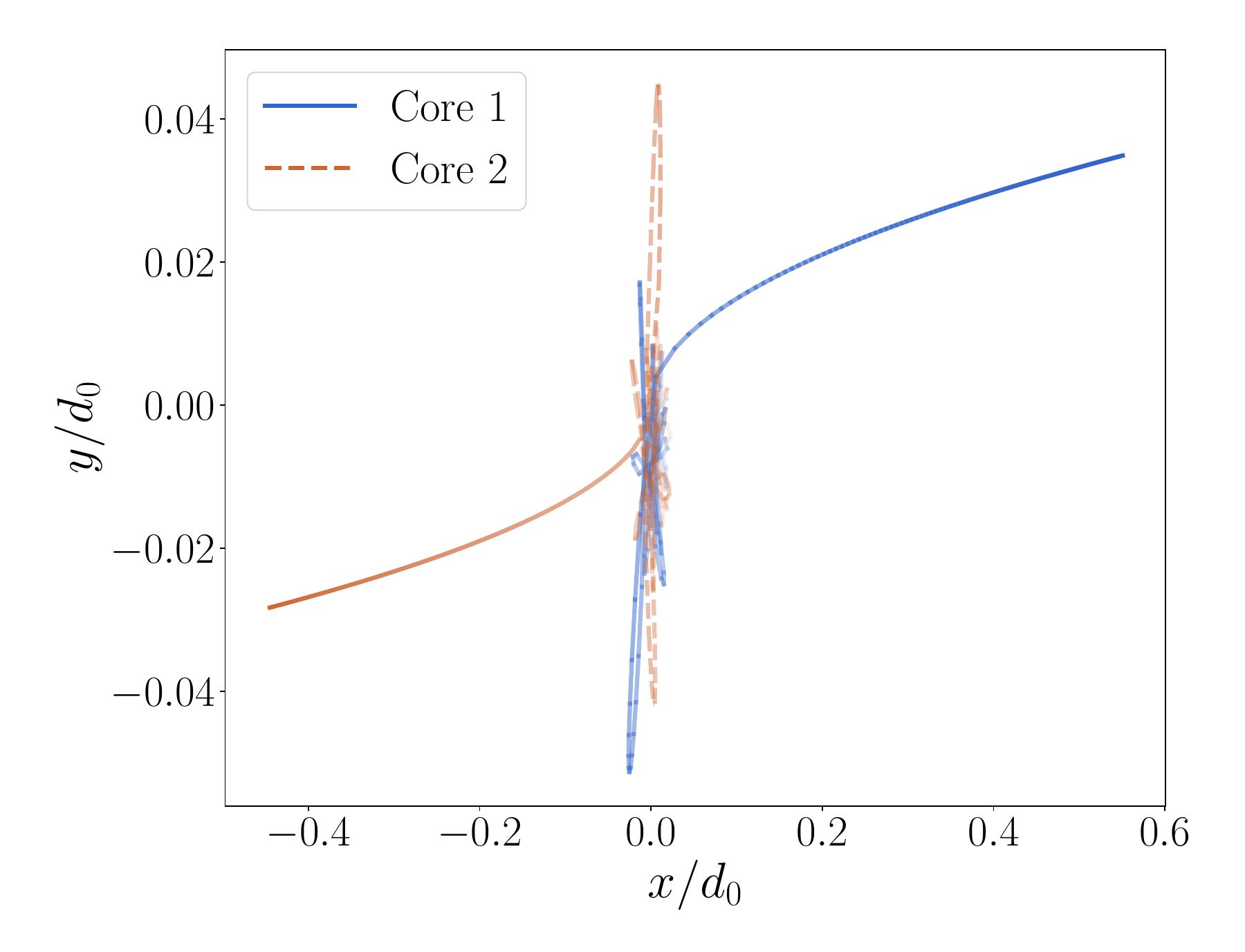}}
\caption
   {
Trajectories of the two compact cores in the $xy$-plane, extracted from the red giant progenitors. The solid blue line (dark to light gradient) shows the primary core's path, while the dashed orange line (dark to light gradient) tracks the secondary core. Both trajectories are normalized by the initial separation $d_0$ and colored by simulation time (darker shades indicate earlier times). The fading color scheme reveals how friction from the surrounding gas envelope causes the cores to spiral inward, with their closest approach occurring in the later, lighter-colored phases. The distinct line styles ($-$ for primary, $--$ for secondary) help identify interactions where the cores exchange momentum. The axes show normalized coordinates to the initial separation ($x/d_0$, $y/d_0$).
   }
\label{fig.CoreOrbitsXY}
\end{figure}

In Fig.~(\ref{fig.CoreOrbitsXY}) we display the trajectory in the $xy$-plane of the two compact cores extracted from the red giant progenitors. These cores, initially embedded in the extended gaseous envelope, have been isolated in the figure to study their dynamical interaction due to the dissipative effects of the surrounding gas.
The coordinates are normalized by the initial separation $d_0$ between the cores, where $d_0$ is calculated from their initial positions $\mathbf{r}_1(0)$ and $\mathbf{r}_2(0)$ as $d_0 = ||\mathbf{r}_2(0) - \mathbf{r}_1(0)||$. The fading color scheme reveals that close encounters between the cores, visible as trajectory crossings or near approaches, typically occur during the later, more transparent phases of the evolution.

The surrounding gas envelope plays a crucial role in the orbital evolution of the binary cores through dynamical friction. As the cores move through the dense stellar material, they experience a drag force that efficiently dissipates orbital energy. This dissipative process arises from gravitational interactions between the cores and the surrounding gas particles, which leads to the exchange of momentum and energy. 

This energy dissipation causes the cores to spiral inward over time, decreasing their separation. The efficiency of this process depends strongly on the gas density—higher densities lead to stronger drag and faster orbital decay. As a result, the cores evolve toward a tighter configuration. The extracted trajectories in the $xy$-plane illustrate this inspiral, with the fading colors highlighting the progression toward closer encounters.  

We can see better how the two cores approach in Fig.~(\ref{fig.CoreDistance}), which reveals a characteristic sequence of ``bounces'' during the inspiral, where the separation between cores reaches local minima before temporarily increasing again. This oscillatory behavior emerges from the competition between gas-driven orbital decay and the cores' finite sizes, $R_{\rm min} \approx R_1 + R_2 + \delta_{\rm gas}$, where $R_1$ and $R_2$ are the core radii, and $\delta_{\rm gas}$ represents the minimum gas layer thickness between them. Each bounce occurs when the cores approach this physical limit, with hydrodynamic pressure and tidal forces temporarily halting further contraction. 

Ideally one would continue the numerical integration until the two cores collide, to find out on what timescale this happens and what the associated gravitarional radiation would be but the integration would become very long, which would lead to a significant accumulation of numerical error in the calculation of the trajectories of the core particles. Also, we need to take into account the limitations of the numerical scheme. While at relatively large distances the integrator kernel leads to realistic distributions of gas around the cores, this is not necessarily true at short distances. Nevertheless, this is an interesting question to address. Before we present the scheme, we deem it important to first present the analysis relative to the electromagnetic phenomena.

\subsection{The role of the surrounding gas shells}
\label{subsec.gas_shells}

The simulation resolves these finite-size effects through the smoothed particle hydrodynamics formalism, where each nucleus is represented by thousands of particles. As the separation decreases below $\sim \text{few}\,(R_1+R_2)$, several physical processes become significant, such as tidal deformation, which migh raise the cores' internal energy, gas compression, which creates a pressure barrier and shock heating, a way to dissipate orbital energy.

\begin{figure*}
          {\includegraphics[width=1\textwidth,center]{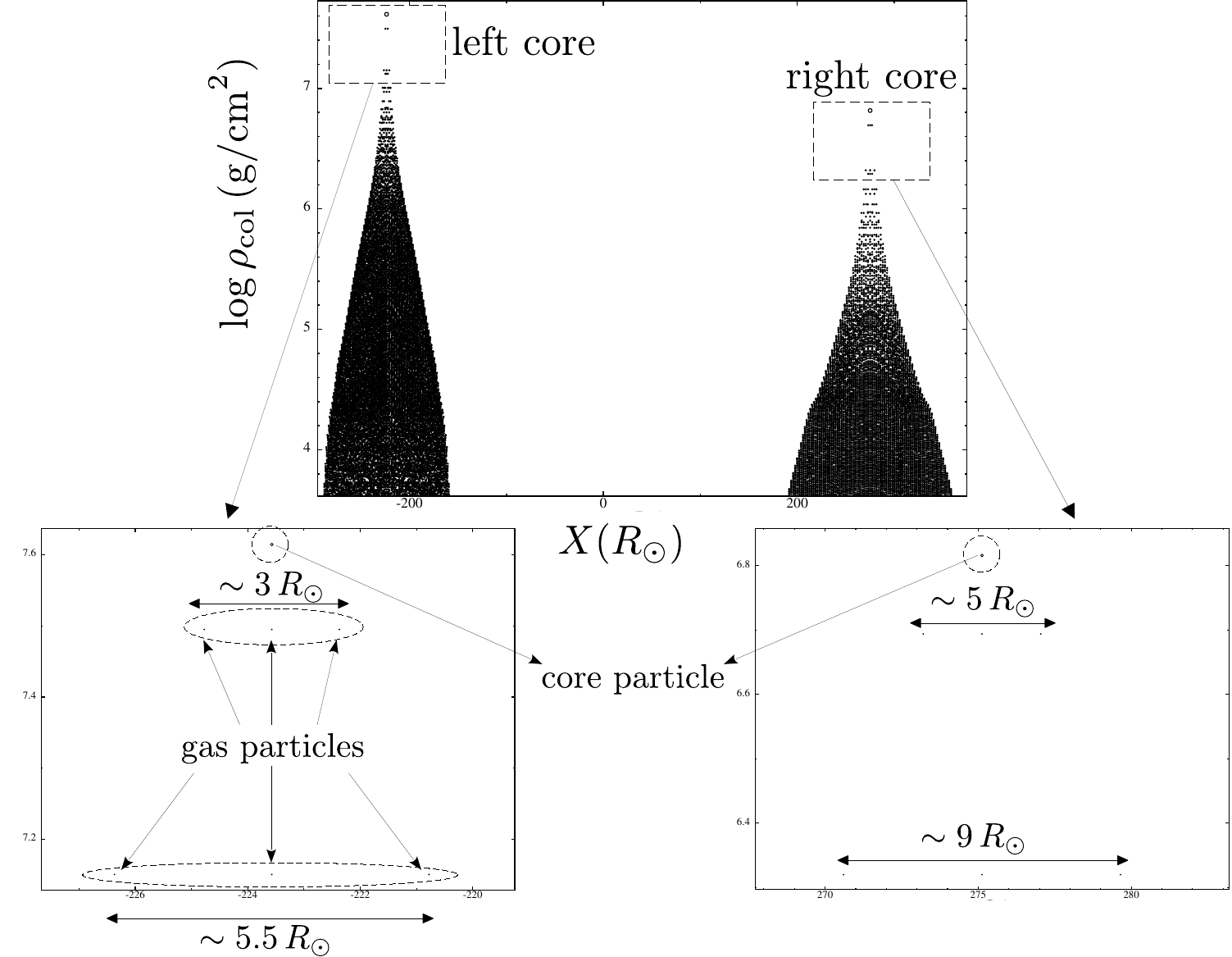}}
\caption
   {
Column density as a function of the distance in the x-xis in solar radii. The top panel shows the column density of both initial red giants. The bottom left panel provides a zoomed-in view of the left red giant's core, displaying the core particle's position along with the six densest surrounding gas particles, while also indicating the diameters (in solar radii) of the three densest gas particles as well as the next three ones. The integrator uses all of these six particles along with the core particle to define the core of a red giant. The bottom right panel presents the same measurements for the right red giant's core, maintaining identical quantities for direct comparison between both stellar cores.  
   }
\label{fig.densities_RG}
\end{figure*}

In Fig.~(\ref{fig.densities_RG}) we illustrate our numerical definition of the cores and the gas distributed around them by showing their initial column densities. We note that column density $\Sigma$ describes mass per unit area integrated along a line of sight, with units $\text{g/cm}^2$, expressed as $\Sigma = \int \rho\, dl$ where $\rho$ is the density and $dl$ is the differential path length. For a medium with uniform density, this simplifies to $\Sigma = \rho L$ where $L$ is the total path length. In the simulations, the core consists of a single particle with zero internal energy ($u=0$), a critical feature for its equations of motion (see eqs. A11 and A12 of \citealt{Gaburov2010} and eqs. 6 and 7 of \citealt{Lombardi2006}). This highest-density particle is surrounded by the six densest gas particles in the star, which define the gas shell surrounding the core. Together, these seven particles (the core particle plus its six densest neighbors) constitute the red giant's physical nucleus in our simulations.

We can see in the same figure that the left core and inner gas shell has a diameter in the x-axis of about $5.5\,R_{\odot}$, while the right core and inner gas shell has a diameter of $\sim 9\,R_{\odot}$. Since the left red giant has evolved for a longer time than the right one, it is smaller. 

With this identification we can definy the gas envelopes surrounding each degenerate core. We use a density-based criterion that identifies the above-mentioned six densest gas particles associated with each core particle. For a given core particle located at position $\vec{r}_i$, we first compute the distances to all gas particles and assign each gas particle to its nearest core. This partitioning creates two distinct gas populations, one for each core. 

For each core's gas population, we sort the particles by density $\rho$ in descending order and select the six densest particles. Let $\{ \vec{r}_{i,1}, ..., \vec{r}_{i,6} \}$ be the positions of these six particles, with corresponding densities $\rho_{i,1} \geq ... \geq \rho_{i,6}$. We then calculate their distances to the sink as $d_{i,j} = ||\vec{r}_{i,j} - \vec{r}_i||$ for $j=1,\ldots,6$.
These distances are sorted such that $d_{i,(1)} \leq ... \leq d_{i,(6)}$. The envelope radius $R_i^{\rm env}$ is defined as the distance to the third-farthest particle $R_i^{\rm env} = d_{i,(3)}$. This choice provides a measure of the envelope extent that is less sensitive to outliers than using the maximum distance. The interacting mass between two envelopes is computed when their separation $D = ||\vec{r}_1 - \vec{r}_2||$ satisfies $D \leq R_1^{\rm env} + R_2^{\rm env}$. The overlapping volume is approximated as a cylindrical region with height $\delta = R_1^{\rm env} + R_2^{\rm env} - D$ and radius $\bar{R} = (R_1^{\rm env} + R_2^{\rm env})/2$, giving $V_{\rm int} = \pi \delta^2 \bar{R}$. The mean density in the interaction region is taken as the average of the two envelope densities $\bar{\rho} = (\rho_1 + \rho_2)/2$, where $\rho_i$ is calculated from the six densest particles, $\rho_i = {\sum_{j=1}^6 m_j}/\left[{4\pi (R_i^{\rm env})^3/3}\right]$. The interacting mass is then $M_{\rm int} = \bar{\rho}\, V_{\rm int}$. The relative velocity $v_{\rm rel}$ is computed as the projection of the sink velocity difference along the line connecting them,

\begin{equation}
\label{eq.vrel}
v_{\rm rel} = \left| (\vec{v}_1 - \vec{v}_2) \cdot \frac{\vec{r}_1 - \vec{r}_2}{||\vec{r}_1 - \vec{r}_2||} \right|.
\end{equation}

We can see an example of this representation in Fig.~(\ref{fig.Spheres}), where we have used different colours for different stars and
their cores.

\begin{figure}
          {\includegraphics[width=0.55\textwidth,center]{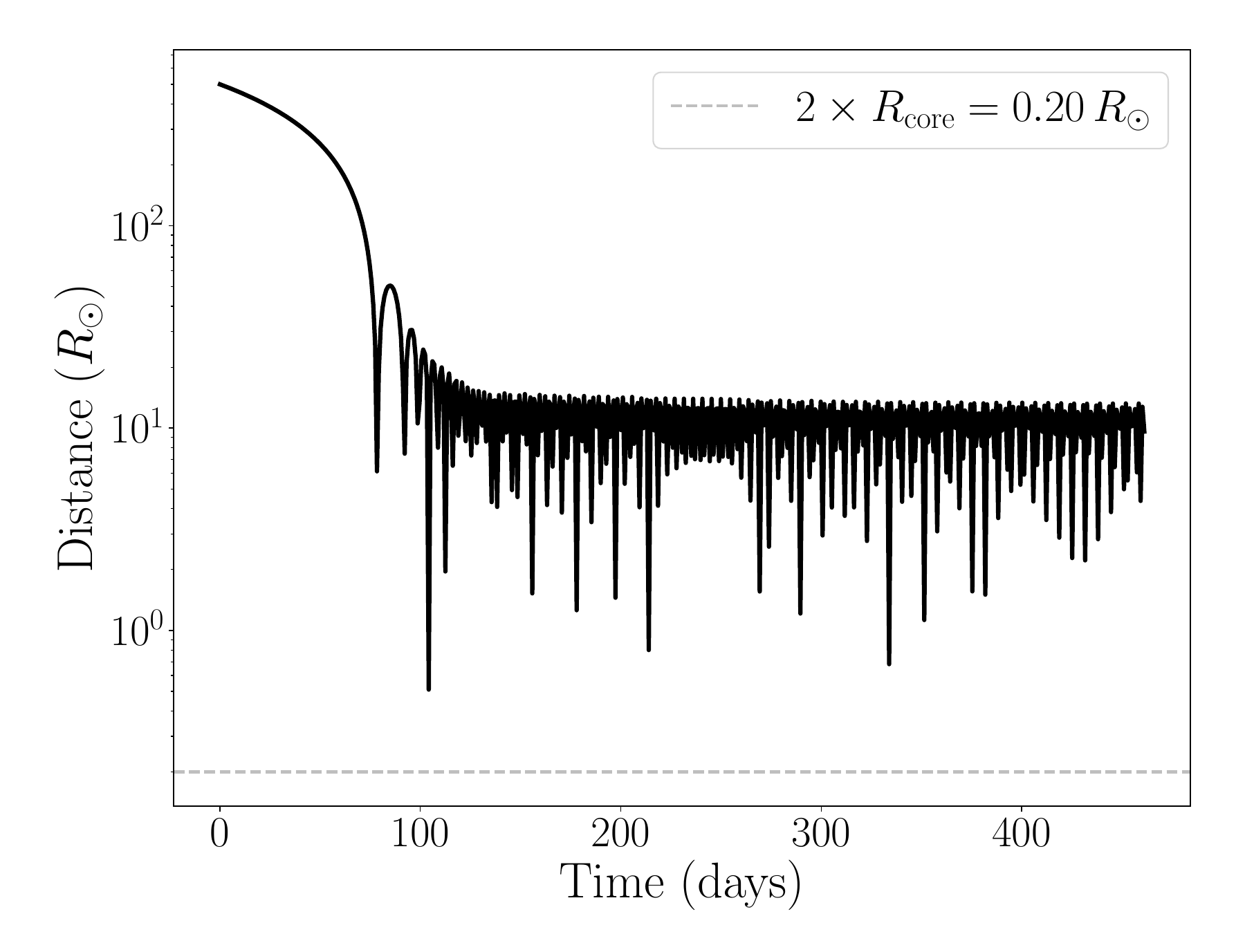}}
\caption
   {
Evolution of the separation $R$ between the two cores, normalized by their initial distance $d_0$, as a function of normalized time $T/T_{\rm tot}$ (where $T_{\rm tot}$ is the total integration time). The black curve shows the complete inspiral history, including all gas-mediated interactions. The initial plateau near $R/d_0 = 1$ represents the early phase where gas drag has minimal effect, followed by a rapid decay phase showing efficient energy dissipation through gas friction. The characteristic ``knee'' in the curve at $R/d_0 \approx 0.5$ marks the transition to the binding process as the cores enter the denser inner envelope regions. While the figure visually removes the gas component, the simulation fully accounts for its gravitational influence and dissipative effects - the normalization preserves the physical timescales of orbital decay while making the universal inspiral pattern evident. We also display the sum of the two cores' radii,
of about $\sim 0.1\,R_{\odot}$ \citep[see Fig.(29) at about $7\times 10^9\,\text{yrs}$ of][]{AmaroSeoane2023}.
   }
\label{fig.CoreDistance}
\end{figure}

\begin{figure}
          {\includegraphics[width=0.5\textwidth,center]{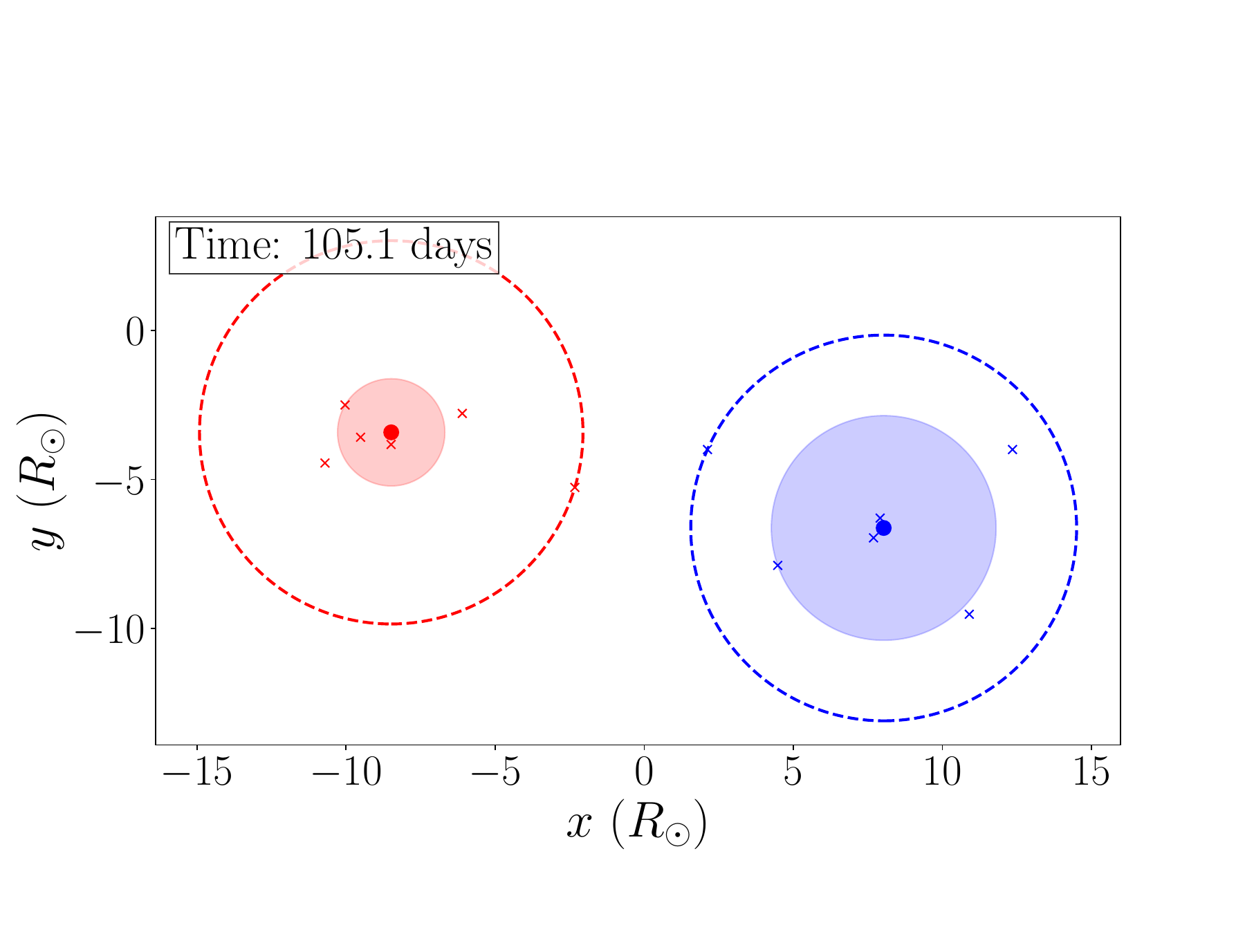}}
\caption
   {
Interacting gas envelopes surrounding two degenerate cores. We show the $x$-$y$ plane with positions in solar radii ($R_{\odot}$). Each core is represented by a colored marker (red and blue circles) surrounded by its six densest gas particles (corresponding crosses). The solid dashed circles indicate the maximum extent of the envelopes ($d_{i,(6)}$), while the shaded regions show the interaction radius ($R_i^{\rm env} = d_{i,(3)}$) of denser gas density.
   }
\label{fig.Spheres}
\end{figure}

\subsubsection{Electromagnetic bursts: A dynamical friction model}
\label{subsubsec.EM_bursts_model}

We model the interaction as one core, the perturber, moving through the extended gaseous envelope of the second star, which acts as the ambient medium. The luminosity arises from the dissipation of the perturber's orbital energy via gravitational drag, or dynamical friction. The rate of energy dissipation is equivalent to the power exerted by the drag force, a process that can be described by the formalism of Bondi-Hoyle-Lyttleton (BHL) accretion.

The fundamental quantity is the rate at which the perturber gravitationally focuses and interacts with the mass of the ambient medium. For a perturber of mass $m_1$ moving with a relative velocity $v_{\text{rel}}$ through a medium of uniform density $\rho_0$, the BHL mass interaction rate is
\begin{equation}
\label{eq.bhl_rate}
\dot{M}_{\text{BHL}} = \frac{4 \pi (G m_1)^2 \rho_0}{v_{\text{rel}}^3}.
\end{equation}

\noindent
This expression defines the flux of gas passing through the Bondi-Hoyle accretion radius, $R_A = 2Gm_1/v_{\text{rel}}^2$, which is the effective gravitational cross-section of the perturber. The power dissipated by the gravitational drag force, $F_d \approx \dot{M}_{\text{BHL}} v_{\text{rel}}$, is then given by $L = F_d v_{\text{rel}}$. Substituting Eq.~(\ref{eq.bhl_rate}) yields the burst luminosity,
\begin{equation}
\label{eq.bhl_luminosity}
L_{\text{burst}} \approx \dot{M}_{\text{BHL}} v_{\text{rel}}^2 = \frac{4 \pi (G m_1)^2 \rho_0}{v_{\text{rel}}}.
\end{equation}

\noindent
This model predicts that the luminosity is directly proportional to the density of the medium and inversely proportional to the relative velocity of the cores.

To evaluate this expression, we use parameters derived from a representative SPH snapshot during a close encounter phase. The core masses at this time are $m_1 = 0.478\,M_{\odot}$ and $m_2 = 0.418\,M_{\odot}$. We identify the more massive core as the primary perturber. The relative velocity, calculated from the particle data in the snapshot, is $v_{\text{rel}} \approx 241$ km/s. The ambient medium is the shared envelope; from the SPH data, a representative density in the interaction region is $\rho_0 \approx 10^{-8}$ g cm$^{-3}$.

Substituting these values into Eq.~(\ref{eq.bhl_luminosity}), we obtain an estimated luminosity of
\begin{equation}
L_{\text{burst}} \sim  2 \times 10^{37}~\text{erg s}^{-1}.
\label{eq.Lburst}
\end{equation}

\noindent
We must note that this analytical model, while providing a physically grounded estimate, is subject to several important limitations inherent in its simplifying assumptions. A critical assessment of these assumptions allows for an interpretation of the model's result as either a lower or an upper limit on the true luminosity.

The Bondi-Hoyle-Lyttleton formalism assumes a perturber moving through an infinite, homogeneous medium in a steady state. The common envelope environment of the inspiraling cores violates these conditions. The medium is finite and highly stratified, with a steep radial density gradient. The interaction is not steady-state but is instead highly dynamic and episodic, with energy dissipation strongly modulated by the eccentric orbit of the binary cores. Furthermore, the model neglects the effects of gas pressure, which in a red giant envelope is non-negligible and will alter the flow of gas around the perturber, reducing the effective gravitational focusing.

These limitations suggest that the model's estimate is likely a lower limit on the peak burst luminosity. The calculation uses a single, representative density $\rho_0$ characteristic of the interaction region. However, during the closest pericenter passages, the cores will plunge into regions where the local gas density is significantly higher than this average value. Since the luminosity scales directly with density, $L_{\text{burst}} \propto \rho_0$, the true peak luminosity during these brief moments of maximum compression will exceed the model's prediction.

Convrsely, the model likely provides an upper limit on the time-averaged luminosity over the course of an orbit. The cores spend the majority of their orbital period at larger separations where the ambient density is much lower than the value used in the calculation. Because the model does not account for this orbital modulation, its steady-state calculation based on a close-encounter density will overestimate the average power dissipated over the full inspiral. We will later address a more realistic way of estimating the burst luminosity.

\subsubsection{Duty cycle}
\label{subsubsec.duty_cycle}

An analytical estimate for the duty cycle can be derived by considering the geometry of the binary orbit and the physical condition for interaction. The duty cycle, $\eta$, is defined as the fraction of the orbital period during which the gas envelopes of the two cores are in direct contact, leading to luminous bursts.

The interaction is initiated when the separation between the cores, $d$, becomes less than or equal to the sum of their effective envelope radii, $R_1$ and $R_2$. We model the orbit of the cores as a Keplerian ellipse with semi-major axis $a$ and eccentricity $e$. The separation $d$ is a function of the orbit's eccentric anomaly, $E$,
\begin{equation}
d(E) = a(1 - e \cos E).
\end{equation}

\noindent
The condition for a burst to occur is therefore $d(E) \le R_1 + R_2$. This defines a critical value for the eccentric anomaly, $E_{\text{crit}}$, which marks the boundary of the interaction arc. The cores are in contact when the inequality
\begin{equation}
a(1 - e \cos E) \le R_1 + R_2
\end{equation}

\noindent
is satisfied. Rearranging this expression gives the condition on the cosine of the eccentric anomaly,
\begin{equation}
\cos E \ge \frac{1}{e} \left( 1 - \frac{R_1 + R_2}{a} \right).
\end{equation}

\noindent
The interaction is thus confined to the range of eccentric anomalies $-E_{\text{crit}} \le E \le E_{\text{crit}}$, where
\begin{equation}
\label{eq.Ecrit}
E_{\text{crit}} = \arccos \left[ \frac{1}{e} \left( 1 - \frac{R_1 + R_2}{a} \right) \right].
\end{equation}

\noindent
The duration of the burst phase, $T_{\text{burst}}$, is the time it takes for the system to traverse this arc of the orbit. This time can be calculated by integrating Kepler's equation, which relates the mean anomaly $M = (2\pi/P_{\text{orb}})t$ to the eccentric anomaly, $M = E - e \sin E$. In its differential form, this is
\begin{equation}
dt = \frac{P_{\text{orb}}}{2\pi} (1 - e \cos E) dE,
\end{equation}

\noindent
where $P_{\text{orb}}$ is the orbital period. Integrating this expression over the interaction arc from $-E_{\text{crit}}$ to $E_{\text{crit}}$ yields the total burst duration,
\begin{align}
T_{\text{burst}} &= \int_{-E_{\text{crit}}}^{E_{\text{crit}}} \frac{P_{\text{orb}}}{2\pi} (1 - e \cos E) dE \nonumber \\
&= \frac{P_{\text{orb}}}{\pi} (E_{\text{crit}} - e \sin E_{\text{crit}}).
\end{align}

\noindent
The duty cycle $\eta = T_{\text{burst}} / P_{\text{orb}}$ is therefore given by the final, purely geometric expression,
\begin{equation}
\label{eq.duty_cycle_analytical}
\eta = \frac{1}{\pi} \left( E_{\text{crit}} - e \sin E_{\text{crit}} \right),
\end{equation}

\noindent
where $E_{\text{crit}}$ is defined by Eq.~(\ref{eq.Ecrit}). This analytical formula connects the observable duty cycle of the precursor luminosity directly to the fundamental orbital parameters of the inspiraling binary.

To evaluate the value, we derive representative orbital parameters and interaction radii from the SPH simulation results. The core separation plot indicates a quasi-stable inspiral phase with a periastron of $d_{\text{min}} \approx 5\,R_{\odot}$ and an apastron of $d_{\text{max}} \approx 15\,R_{\odot}$. From the relations $d_{\text{min}} = a(1-e)$ and $d_{\text{max}} = a(1+e)$, we solve for a semi-major axis of $a=10\,R_{\odot}$ and an eccentricity of $e=0.5$. The interaction radii are taken from the diameters of the inner gas shells surrounding the cores, yielding $R_1 = 2.75\,R_{\odot}$ and $R_2 = 4.5\,R_{\odot}$, for a total interaction distance of $R_1+R_2 = 7.25\,R_{\odot}$.

With these parameters, we first calculate the argument of the arccosine in Eq.~(\ref{eq.Ecrit}),
which gives a critical eccentric anomaly of $E_{\text{crit}} = \arccos(0.55) \approx 0.988$ radians. We then substitute this value into the final expression for the duty cycle, Eq.~(\ref{eq.duty_cycle_analytical}),
\begin{align}
\eta &= \frac{1}{\pi} \left( E_{\text{crit}} - e \sin E_{\text{crit}} \right) \nonumber \\
&\approx 0.182.
\end{align}

\noindent
The analytical estimate of $\eta \approx 0.182$ indicates that the system is actively bursting for approximately 18.2\% of its orbital period. 

\section{Wavelet analysis of the electromagnetic bursts}
\label{sec.wavelets}

{To characterize the complex, non-stationary energy dissipation revealed in the previous section (Section~\ref{sec.coreinnershells}), we employ a continuous wavelet transform analysis of the luminosity signal.}

The hydrodynamic evolution of colliding stellar cores is a fundamentally non-stationary process. The resulting luminosity signal, $L(t)$, is characterized by transient, high-energy bursts from shock waves at periastron passages, superimposed on slower modulations from the orbital motion. A standard Fourier transform, $S(\omega) = \int L(t) e^{-i\omega t} dt$, would identify the characteristic frequencies $\omega$ present in the signal but would average over the entire time domain, obscuring the crucial information of \textit{when} specific events occur. To overcome this limitation and resolve the time-localization of energy dissipation, we employ a continuous wavelet transform (CWT). The CWT provides a time-scale decomposition of the signal, revealing how the energy content at different physical timescales evolves throughout the merger.

The fundamental motivation for employing a wavelet transform for this problem stems from the intrinsic non-stationary character of the luminosity signal $L(t)$. A conventional Fourier analysis, while capable of identifying the constituent frequencies of the entire signal, provides no information regarding their temporal localization. For a dynamic event such as a stellar merger, where transient phenomena are of principal scientific interest, this time-averaged representation is insufficient. The utility of the CWT, therefore, lies in its ability to deconstruct the signal in a manner that preserves both time and scale information, analogous to applying a bank of scaled, band-pass filters to the time series. This method overcomes the uncertainty principle of Fourier analysis by accepting a trade-off: rather than frequency resolution with no time resolution, it provides simultaneous, localized resolution in both domains.

The physical interpretation of this analysis is consequently direct. The resulting wavelet power spectrum, $|W(\tau,t)|^2$, can be viewed as a two-dimensional map of energy dissipation. Power concentrated at small scales ($\tau$) signifies rapid, impulsive energy release, which in this context is attributed to shock fronts formed during close encounters. Conversely, power at large scales corresponds to the slower, quasi-periodic modulation of the system, driven by the binary's orbital mechanics. This time-scale decomposition is what makes the analysis useful; it permits one to trace the flow of energy from the large-scale gravitational drivers of the orbit down to the small-scale hydrodynamic processes where it is ultimately thermalized and radiated. It allows for a direct, quantitative connection between the orbital phase and the efficiency and character of the physical dissipation mechanisms.

The CWT of the signal $L(t)$ is defined as the convolution of the signal with a scaled and translated mother wavelet function, $\psi(t)$,

\begin{equation}
W(\tau,t) = \frac{1}{\sqrt{\tau}} \int L(t') \psi^*\left(\frac{t'-t}{\tau}\right) dt'.
\label{eq.cwt}
\end{equation}

\noindent
The function $\psi(t)$ is a localized wave packet that serves as the prototype for the analysis. To be a valid wavelet, it must satisfy the admissibility condition, which requires it to have a zero mean ($\int\psi(t)dt=0$) and finite energy. The zero-mean property is crucial, as it ensures the transform is sensitive to fluctuations and transient events in the signal rather than its absolute, time-averaged value. All wavelets used in the convolution are simply scaled and translated versions of this single mother function.
Here, $\tau$ is the wavelet scale, which corresponds to a characteristic physical timescale, and $t$ is the time, retaining the temporal localization. We select the complex Morlet wavelet as our basis function,
\begin{equation}
\psi(t) = \pi^{-1/4} e^{i\omega_0 t} e^{-t^2/2}.
\label{eq.morlet}
\end{equation}

\noindent
The Morlet wavelet is essentially a complex plane wave, $e^{i\omega_0 t}$, multiplied by a Gaussian envelope, $e^{-t^2/2}$. Its structure is therefore analytically similar to the expected physical signatures within the luminosity signal: quasi-periodic oscillations (from orbital motion and pulsations) that are localized in time (due to the transient nature of shocks and periastron encounters). The Gaussian envelope provides the time-localization necessary to pinpoint when an event occurs, while the complex sinusoidal component allows the transform to measure its characteristic frequency and phase. We use the dimensionless frequency $\omega_0=6$. This choice ensures the wavelet has a zero mean (the admissibility condition) and offers an optimal trade-off between time and frequency resolution. The complex nature of this wavelet also provides phase information, which is critical for linking hydrodynamic events to the orbital phase.

The wavelet power spectrum, $|W(\tau,t)|^2$, represents the energy density of the signal in the time-scale plane. 
The physical interpretation of the wavelet coefficient $W(\tau,t)$ is that it quantifies the content of the signal $L(t)$ at a specific timescale $\tau$ and localized at a specific time $t$. The transform itself is analogous to a sophisticated template matching process. Consider the mother wavelet $\psi(t)$ as a template for a characteristic physical signature, much like a single specific beat in a complex acoustic signal. The CWT scales this template to a duration $\tau$---searching for either a rapid, high-frequency beat (small $\tau$) or a slow, low-frequency one (large $\tau$)---and translates it across the entire signal to check for a match at every moment $t$. The resulting complex coefficient $W(\tau,t)$ measures the degree of this match: its magnitude, $|W(\tau,t)|$, indicates the strength or intensity of the physical process operating on that timescale at that instant, while its phase provides critical information about the timing and synchronization of the event. Thus, by mapping these coefficients, one constructs a complete time-scale representation of the signal's underlying physical processes.

Integrating this power over the entire observational duration $T$ for each scale yields the global wavelet spectrum,
\begin{equation}
E(\tau) = \int_0^T |W(\tau,t)|^2 dt,
\label{eq.energy}
\end{equation}

\noindent
which quantifies the contribution of each timescale $\tau$ to the total variance of the signal (i.e. it represents the total integrated power at each timescale $\tau$). To compare the relative importance of these processes, we define a scale-invariant normalized energy distribution, $\mathcal{P}(\tau)$, as the probability density of energy over logarithmic scale:
\begin{align}
\mathcal{P}(\tau) &= \frac{E(\tau)}{E_{\text{total}}} \\
E_{\text{total}} &= \int E(\tau) d(\ln\tau).
\label{eq.norm}
\end{align}

\begin{figure}
          {\includegraphics[width=0.5\textwidth,center]{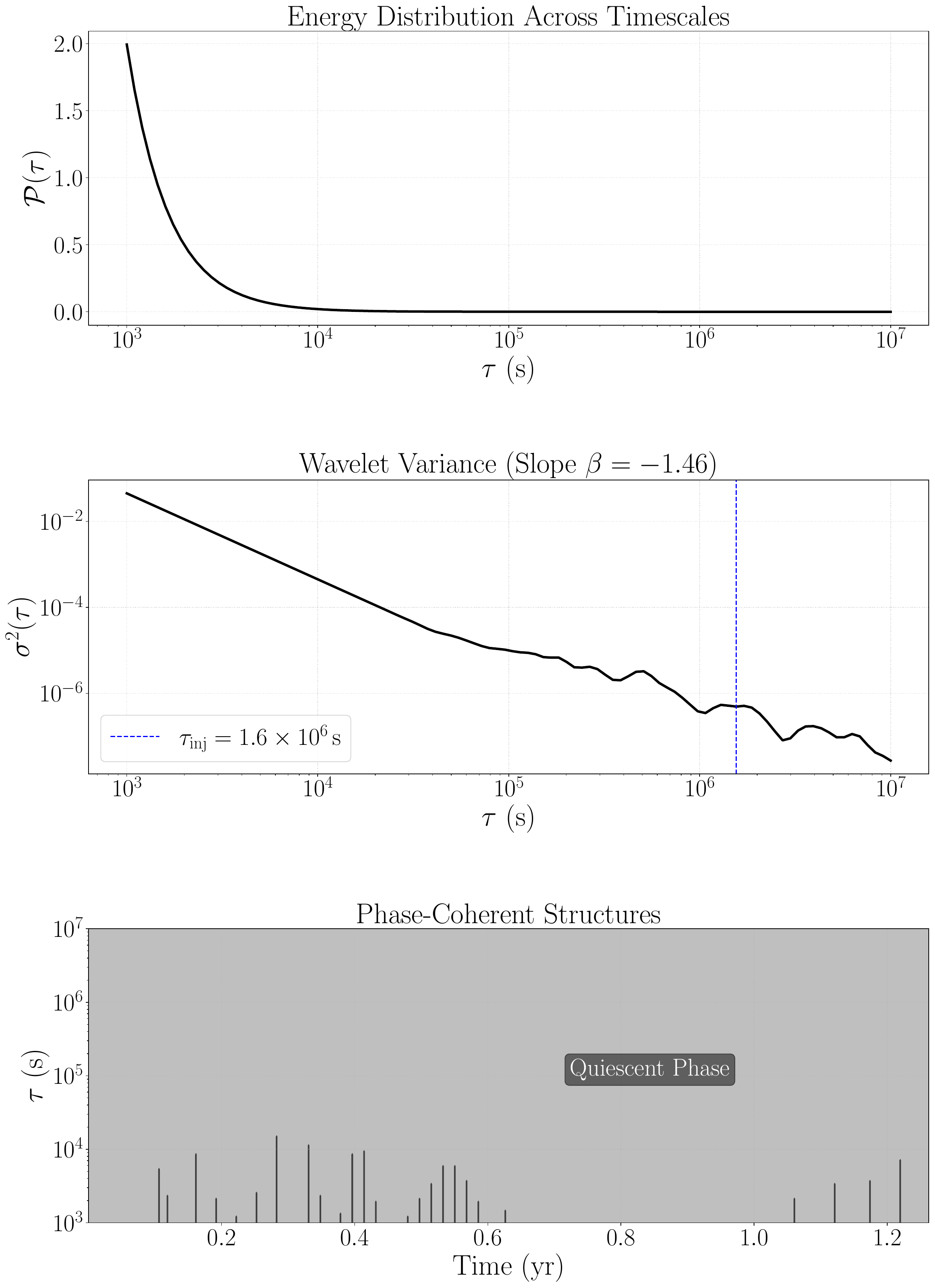}}
\caption
   {
Panel 1 (Top): Normalized energy distribution $\mathcal{P}(\tau)$ as a function of timescale $\tau$. The strong peak at short timescales ($\tau \sim 10^3$ s) and its rapid decay reveals that energy dissipation is overwhelmingly dominated by impulsive, shock-driven processes. The negligible power at large $\tau$ confirms that slower, orbital-scale mechanisms do not contribute significantly to the total radiated energy.
Panel 2 (Middle): Wavelet variance $\sigma^2(\tau)$ versus timescale $\tau$ on a log-log scale. The plot shows a distinct power-law decay across several orders of magnitude, a clear signature of a turbulent energy cascade. The negative slope ($\beta = -1.46$) deviates significantly from the prediction for incompressible turbulence and indicates a highly compressible flow, as expected in a shock-filled medium. The dashed line marks the injection timescale $\tau_{\rm inj} = 1.6 \times 10^6$ s, representing the large, orbit-related scale where energy is supplied to the cascade.
Panel 3 (Bottom): Phase-coherent structures, mapping events in time-scale space. The vertical lines indicate impulsive energy dissipation events that are phase-locked to the orbit, occurring predominantly at short timescales ($\tau < 10^4$ s). The distinct gap between $t \approx 0.6$ yr and $t \approx 1.1$ yr is the quiescent phase during the binary's apoapsis. The clustered events outside this gap correspond to strong hydrodynamic interactions during periapsis passages. 
   }
\label{fig.wavelet_analysis}
\end{figure}

\noindent
This distribution, shown in the first panel of Fig.~(\ref{fig.wavelet_analysis}), directly measures how the total dissipated energy is partitioned among the dominant physical timescales of the merger. The peak at $\tau \sim 10^3$ s confirms that energy thermalization is dominated by rapid, transient shocks, corresponding to a shock-crossing time $\tau_{\rm sh} \sim \lambda/v_{\rm rel}$ for compact regions. The steep decline towards larger scales shows that slower processes, including those on the orbital timescale $\tau_{\rm orb} \sim \sqrt{a^3/GM}$, are sub-dominant in the global amount of energy.

The second panel of Fig.~(\ref{fig.wavelet_analysis}) displays the wavelet variance, $\sigma^2(\tau) = T^{-1} \int |W(\tau,t)|^2 dt$, which serves as a proxy for the energy spectrum of the underlying turbulent flow. For a turbulent cascade with a Fourier energy spectrum $E(k) \propto k^{-\gamma}$, the wavelet variance is expected to scale as $\sigma^2(\tau) \propto \tau^{\gamma-1}$, where the scale $\tau$ is inversely related to the wavenumber $k$. For incompressible Kolmogorov turbulence \citep{Kolmogorov1962}, $\gamma=5/3$, predicting a positive slope $\sigma^2(\tau) \propto \tau^{2/3}$. Our results show a distinct power-law decay (a negative slope $\beta < 0$) in the inertial range, a clear signature that deviates from the incompressible case. This indicates that the turbulent cascade is highly compressible, as expected for a medium permeated by strong shocks where a significant fraction of kinetic energy is directly converted into internal energy rather than being inertially transferred to smaller scales. The peak of the variance spectrum identifies the energy injection scale $\tau_{\rm inj} \approx 1.7$ days, confirming that energy is supplied to the cascade primarily by gravitational shear on the orbital timescale of close encounters.

Finally, the third panel of Fig.~(\ref{fig.wavelet_analysis}) uses the wavelet's phase information to explicitly link hydrodynamics to the orbit. We compute a phase coherence metric, $\mathcal{I}(\tau,t) = |W(\tau,t)| \cos[\phi(\tau,t) - \phi_{\rm orb}(t)]$, where $\phi(\tau,t)$ is the phase of the wavelet coefficient $W(\tau,t)$ and $\phi_{\rm orb}(t)$ is the phase of the binary's orbit. A large positive value of $\mathcal{I}$ indicates that energy dissipation is occurring in phase with the orbital motion (e.g., shock compression at periastron). The analysis reveals a distinct quiescent phase between $t=0.627$ yr and $t=1.06$ yr, corresponding to the binary apastron. During this period, $\mathcal{I} \approx 0$, signifying a decorrelation between the signal's phase and the orbit. This demonstrates that the strong hydrodynamic interactions and their resultant energy dissipation cease when the cores are widely separated. This multi-faceted analysis, uniquely enabled by the time-scale localization of wavelets, thus provides a complete physical picture of energy injection, turbulent transfer, and dissipation throughout the merger event.

\section{Accretion: An Airy analysis and a tram analogy}
\label{sec.Airy}

{To quantify the mass transfer onto the degenerate cores during the inspiral—a process not fully resolved by the SPH simulation—we utilize the Airy function formalism to model the core-envelope interface.}

To analyze the potential accretion of gas onto the degenerate cores moving within the post-merger common envelope, we must first model the physical interface between each core and the surrounding gaseous medium. The central problem is to determine the density profile of this boundary layer, where the gas transitions from being gravitationally bound to a core to becoming part of the ambient gas. We make the physical assumption that the gas in the immediate vicinity of each core attempts to settle into a state of quasi-static equilibrium. The mathematical description of such a transition layer, where quantum degeneracy pressure counteracts a nearly constant gravitational field, is governed by the Airy function.

The Airy function, $\mathrm{Ai}(x)$, and its companion, $\mathrm{Bi}(x)$, are the two linearly independent solutions to the canonical Airy differential equation,

\begin{equation}
\frac{d^2 y}{dx^2} - x y = 0.
\label{eq.airy}
\end{equation}

\noindent
In this context, the equation provides a powerful physical model for the structure of the gas at the core's edge. The term $-xy$ represents the effect of the local, nearly uniform gravity of the core pulling gas inward, while the second-derivative term ${d^2 y}/{dx^2}$ arises from the gas pressure gradient pushing outward. The Airy function solution, therefore, describes the characteristic density profile of the gas atmosphere as it thins out, allowing us to define a physical boundary for the core and diagnose the conditions for mass accretion.

This analytical framework provides a us with tool for post-processing numerical simulations, particularly because they treat the degenerate cores as point particles and lack an explicit sub-grid model for accretion. The primary utility is twofold. First, it resolves the ambiguity of the core's size by providing a physical definition for its effective surface; the characteristic decay length of the Airy solution, $\beta^{-1}$, sets a natural scale for the core-envelope boundary. With a defined surface at radius $r_{\rm acc}$, one can then directly estimate the mass accretion rate, $\dot{M}$, by measuring the flux of SPH gas particles crossing it. This is computed at each simulation snapshot via $\dot{M} \approx 4\pi r_{\rm acc}^2 \rho_{sph} v_{\rm rad}$, where the local gas density $\rho_{sph}$ and radial inflow velocity $v_{\rm rad}$ are calculated from the SPH particle data. Furthermore, by comparing the measured SPH density profile to the theoretical Airy profile, one can diagnose the thermodynamic state of the accreting gas and quantify its deviation from local hydrostatic equilibrium.

To model the gaseous atmosphere bound to a degenerate core, we formulate the system's total energy as a functional of the gas number density, $n(r)$. This energy functional, $E[n]$, is determined by the interplay of three distinct physical contributions. The first is the Thomas-Fermi kinetic energy, proportional to $n^{5/3}$, which represents the bulk quantum pressure of the degenerate fermion gas. The second is the Weizsäcker gradient correction. This term, proportional to $(\nabla n)^2/n$, is a quantum-mechanical correction to the kinetic energy. It specifically accounts for the energy cost of spatial inhomogeneity; whereas the Thomas-Fermi model treats the gas as a uniform ``sea'' at each point, the Weizsäcker term adds an energy penalty for the rapid density changes (large $\nabla n$) that define a boundary. In our context, it acts as a ``surface tension'', penalizing the unphysically sharp density drop a simpler model would predict. It is therefore relevant for modeling the finite width of the core-atmosphere interface and is the physical origin of the second-derivative term in the resulting Airy equation. The third component is the standard gravitational potential energy, given by $n\Phi$.

\begin{equation}
E[n] = \int \left( C_{TF} n^{5/3} + C_W \frac{(\nabla n)^2}{n} + n \Phi \right) d^3 r.
\label{eq.energy}
\end{equation}

\noindent
We use a functional because we need a mapping that takes an entire function as its input and returns a single scalar value. Whereas a conventional function $f(x)$ takes a number $x$ and returns a number $y$, a functional $F[f(x)]$ takes the function $f(x)$ over its entire domain and returns a single number. The energy expression in Eq.~(\ref{eq.energy}) is a functional, $E[n]$, because for any given density distribution function $n(r)$, it returns a single number representing the total energy of that specific configuration. The physically realized state of the system is the one described by the specific function $n(r)$ that minimizes the value of this energy functional.

Within the analysis of accretion, each component of the energy functional serves a distinct purpose. The gravitational potential, $n\Phi$, acts as the primary engine for this process, lowering the system's total energy as gas gathers within the core's deep potential well. This attractive influence is counteracted by the outward pressure from the two kinetic energy terms. The Thomas-Fermi contribution, $C_{TF} n^{5/3}$, specifically governs the compressibility of the gas as it piles up near the core. The Weizsäcker component, $C_W (\nabla n)^2/n$, is what defines the very structure of the boundary layer across which this mass transfer occurs. The resulting equilibrium density profile, which is described by the Airy function, represents the physical balance achieved between these competing influences.

It is important to note that this entire analytical framework is predicated on the assumption that the gas near the core is in a state of quasi-static equilibrium. I.e. this applies only to erythrohenosis. The model breaks down when this assumption is violated. The primary failure case is in a supersonic flow regime, where the core moves through the gas faster than the local sound speed. This would correspond to the violent collisions of \cite{AmaroSeoane2023}. In this scenario, a dynamic bow shock forms ahead of the core, and the gas flow is governed by non-equilibrium shock physics rather than the minimization of a static energy functional. Additionally, the model is formulated for a zero-temperature degenerate gas; its validity diminishes if the accreting gas becomes sufficiently hot for thermal pressure to become comparable to or dominant over quantum degeneracy pressure.

To determine the ground state density profile, the energy functional $E[n]$ must be minimized subject to the constraint of a fixed total particle number. This variational problem is solved using the Euler-Lagrange equation. A mathematically convenient substitution is introduced by defining an effective wavefunction, $\psi$, such that the density is given by $n(r) = \psi(r)^2$. The primary utility of this transformation is the simplification of the Weizsäcker gradient term; the expression $(\nabla n)^2/n$ in the energy functional becomes proportional to $(\nabla \psi)^2$, whose variation yields the standard Laplacian operator, $\nabla^2$.

This procedure reformulates the problem of minimizing the energy functional, subject to particle number conservation, into solving its associated Euler-Lagrange equation. The substitution $n(r) = \psi(r)^2$ yields a non-linear, time-independent differential equation for the effective wavefunction $\psi$,

\begin{equation}
-\frac{\hbar^2_{\rm eff}}{2 m_{\rm eff}} \nabla^2 \psi + A \psi^{7/3} + \Phi(r) \psi = \mu \psi.
\label{eq.schrodinger}
\end{equation}

\noindent
In the dynamic context of the erythrohenosis merger, Eq.~(\ref{eq.schrodinger}) models the quasi-static structure of the gas locally bound to one of the degenerate cores moving through the common envelope. The term $\Phi(r)\psi$ represents the immediate gravitational pull of that specific core, which attempts to capture ambient envelope gas. This inward force is resisted by the two quantum pressure terms. The $A \psi^{7/3}$ factor models the ``stiffness'' of the gas, quantifying its resistance to being compressed into a high-density layer onto the core's surface. The $\nabla^2 \psi$ term, originating from the Weizsäcker correction, is what imparts a finite thickness to this boundary layer, preventing an unphysical density jump at its edge. The chemical potential $\mu$ acquires a critical physical interpretation in this scenario: it defines the energy level separating the gas gravitationally bound within the core's atmosphere from the surrounding, unbound gas of the common envelope. The solution to this equation thus provides the density profile of the accretion interface itself. In some sense, we can think of the degenerate core as a very crowded tram arriving at a station, with the surrounding gas being the crowd of (impatient) people on the platform. The core's gravitational potential, $\Phi$, is the ``pull'' of the tram's open doors, the primary force attracting people to get inside (you want to get home, right?). The Thomas-Fermi pressure is the ``bulk'' pressure of the exhausted passengers already packed tightly inside the tram; they resist being compressed any further (try if you don't believe me), which determines how many people can fit in the main cabin. The Weizsäcker correction is the chaotic, shuffling region right at the doorway. This isn't a sharp, clean line but a fuzzy, dynamic interface where people are jostling, half-on and half-off. This term represents the ``energy cost'' of maintaining this messy, finite-sized boundary. Finally, the chemical potential, $\mu$, is the tipping point for a single person standing right in the doorway: are they officially on the tram (bound to the core) or still on the platform (part of the unbound envelope)?

This final linearization is the step that makes the framework practical for analyzing accretion. By reducing the complex, non-linear problem to the solvable Airy equation, we obtain an explicit analytical formula for the density profile, $\rho(r)$, that describes the physical structure of the core-gas interface. The utility of this result is direct. It allows one to move beyond the simulation's unphysical treatment of the core as a point particle and instead model it as an object with a well-defined physical boundary, whose location and thickness are determined by the parameters of the Airy solution. With a consistent accretion radius, $r_{\rm acc}$, thus defined, one can directly measure the mass flux from the simulation data. The mass accretion rate, $\dot{M}$, is then estimated by computing the rate at which SPH particles cross the spherical surface at this radius, using their known densities and velocities. This procedure provides a post-processing method for diagnosing mass transfer by bridging the gap between the resolved gas dynamics of the simulation and the unresolved physics of accretion onto the degenerate core.

We now solve Eq.~(\ref{eq.schrodinger}) in the physically crucial region of the core-envelope interface, located at a characteristic radius $r_0$. In these outer layers, the gas density must fall to zero, which implies that the effective wavefunction $\psi \to 0$. 
In this low-density limit, the non-linear term derived from the Thomas-Fermi energy (proportional to $\psi^{7/3}$ in Eq.~(\ref{eq.schrodinger})), representing the internal pressure of the gas, becomes negligible compared to the dominant interaction with the core's gravitational field.
This allows for a critical simplification of the governing equation. Furthermore, since this interface is physically thin, the gravitational potential across it can be accurately modeled by a first-order Taylor series expansion around $r_0$,

\begin{equation}
\Phi(r) \approx \Phi(r_0) + g_0 (r - r_0),
\end{equation}

where $g_0 = GM_c / r_0^2$ is the local gravitational acceleration generated by a degenerate core of mass $M_c$.
The term $r_0$ represents the radius of the core-envelope boundary, and $g_0$ is the gravitational acceleration at that surface, calculated under a local approximation. This radius $r_0$ is not an arbitrary point, but rather specifies the physical interface between the two primary components of the star; namely the dense, compact degenerate core of mass $M_c$, and the vast, diffuse gaseous envelope. This boundary at $r_0$ is therefore the effective surface onto which gas from the common envelope can accrete.

The use of the term ``local'' signifies a crucial physical approximation. The gravity at the interface is calculated assuming it is generated solely by the enclosed core mass $M_c$, while ignoring the gravitational influence of other matter, such as the distant companion core or the extended, non-spherical gaseous envelope. This simplification from a complex ``global'' problem to a tractable ``local'' one is justified by Newton's Shell Theorem, which states that the net gravitational force from a spherical shell of mass on a particle inside that shell is zero. Thus, for a locally spherical system, the gravity at the surface is indeed determined only by the interior mass.

The specific form of the equation, $g_0 = GM_c / r_0^2$, follows directly from Newton's universal law of gravitation applied in this local context. The mass $M_c$ is used because, as dictated by the Shell Theorem, only the mass enclosed within a given radius contributes to the net gravitational force at that radius. The $r_0^2$ term in the denominator represents the standard inverse-square law. The resulting acceleration, $g_0$, is therefore a measure of the intense gravitational field at the core's surface, which provides the necessary pull to capture and accrete material from the surrounding envelope.

By inserting the linear expansion of the potential into the simplified Schroedinger equation and making the substitution $x = r - r_0$, we arrive at the governing equation for the effective wavefunction $\psi$ within the thin interface layer. This equation represents a balance of energies: the first term below is the kinetic energy of the gas arising from the pressure gradient, the second is the potential energy in the locally linear gravitational field, and the right-hand side is the total energy set by the chemical potential. The resulting one-dimensional ordinary differential equation is

\begin{equation}
-\frac{\hbar^2_{\rm eff}}{2 m_{\rm eff}} \frac{d^2 \psi}{dx^2} + \left( \Phi(r_0) + g_0 x \right) \psi = \mu \psi.
\label{eq.linearised}
\end{equation}

\noindent
This is a standard Schroedinger equation for a particle in a uniform force field, a classic problem in quantum mechanics. A simple rearrangement of its terms allows us to express it in a canonical mathematical form. By isolating the second derivative and grouping the remaining terms, Eq.~(\ref{eq.linearised}) can be rewritten as

\begin{equation}
\frac{d^2 \psi}{dx^2} - \lambda \left( x - x_0 \right) \psi = 0,
\label{eq.airy_form}
\end{equation}

\noindent
which is the canonical Airy equation. The physical parameters of the problem are now absorbed in two constants. The parameter $\lambda = 2 m_{\rm eff} g_0 / \hbar^2_{\rm eff}$ combines the local strength of gravity, $g_0$, with the effective quantum stiffness of the gas. Its inverse cube root, $\lambda^{-1/3}$, defines the natural physical length scale over which the density profile changes. The parameter $x_0 = (\mu - \Phi(r_0)) / g_0$ defines the location of the classical turning point, the position where a classical particle with energy $\mu$ would have zero kinetic energy and be turned back by the potential.

The Airy equation has two linearly independent solutions, the Airy functions of the first and second kind, denoted $\mathrm{Ai}$ and $\mathrm{Bi}$ respectively. The choice between them is dictated by the physical boundary conditions of the problem. For positive arguments, corresponding to the classically forbidden region $x > x_0$, the $\mathrm{Bi}$ function grows exponentially, while the $\mathrm{Ai}$ function decays exponentially. Since the gas density must vanish at large distances from the core, any solution that grows infinitely is unphysical. Therefore, only the Airy function of the first kind, $\mathrm{Ai}$, provides the physically admissible solution that correctly describes the decaying tail of the gas atmosphere in the external, low-density region.

The effective wavefunction $\psi$ is directly related to the physical mass density $\rho$ through the relation $\rho(r) = m_p n(r) = m_p \psi(r)^2$, where $m_p$ is the mean mass per particle. Since the solution for the wavefunction is $\psi(x) \propto \mathrm{Ai}(\lambda^{1/3}(x-x_0))$, the resulting asymptotic density profile near the interface is given by

\begin{equation}
\rho(r) \propto \left[ \mathrm{Ai}\left( \beta (r - r_0') \right) \right]^2,
\label{eq.profile}
\end{equation}

\noindent
where the parameters are defined as $\beta = \lambda^{1/3}$ and the effective surface location is $r_0' = r_0 + x_0$. This profile has a clear physical interpretation: the density does not terminate abruptly at the classical turning point, but instead exhibits an exponentially decaying tail that extends into the envelope. This represents the probability of finding gas particles in a classically forbidden region. The formulation also defines a natural length scale, $\beta^{-1}$, which is the characteristic e-folding distance of this density tail and physically represents the width or ``fuzziness'' of the core-envelope transition layer. Going back to our exhausted tram passengers from section (\ref{sec.Airy}), this is analogous to two of them trying to hold a conversation on the very crowded, noisy tram. We can think of the clarity or intelligibility of their speech as the density, $\rho$. There is a ``classical'' boundary, $r_0'$, perhaps a distance of two meters, where the overwhelming background noise of the crowd and the tram's motor should, in principle, make their conversation completely unintelligible. One might (wrongfully) expect the speech clarity to hit a sharp wall at this point and drop instantly to zero. Instead, the clarity exhibits an exponential ``tail'' that extends beyond this critical distance. An observer at three meters, in the ``classically forbidden'' zone of silence, might still intermittently catch a single loud word which ``leaks'' through the din and later will turn into a gossip. The intelligibility does not cease abruptly but fades exponentially. The length scale $\beta^{-1}$ represents the characteristic ``fuzziness'' of this boundary, for example, the extra meter (the gossipness range) over which these last fragments of speech decay and are finally lost entirely into the background noise.

In the context of a numerical simulation, this analytical form is an invaluable tool. It provides a precise theoretical template that can be fitted to the radial density distribution of the SPH particles. A successful fit yields high-fidelity measurements of the core's effective surface location, $r_0'$, and the interface width, $\beta^{-1}$, provided the simulation has sufficient resolution to capture this asymptotic Airy tail. These parameters are essential for defining a physical accretion boundary for the otherwise point-like core particles.

\subsection{Calculation of the Mass Accretion Rate}
\label{subsec.mass_accretion_rate}

The analysis of the accretion process begins with the quantification of the mass accretion rate, $\dot{M}(t)$, onto each of the two degenerate cores throughout the simulation. The first step in this algorithm is to distinguish the two core particles from the gas particles based on the particle data provided in each snapshot file. Once identified, a consistent framework is required to measure the mass flow from the gas on to the point-particle cores.

\begin{figure*}
 {\includegraphics[width=1\textwidth,center]{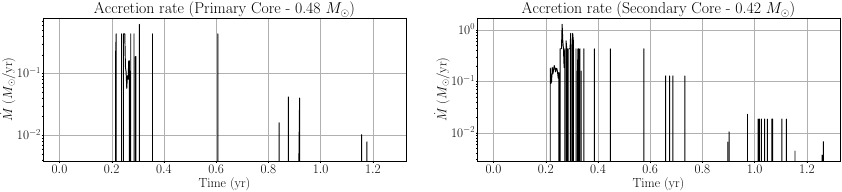}}
\caption
{
Instantaneous mass accretion rate, $\dot{M}(t)$, on to the primary (more massive, left panel) and secondary (less massive, right panel) degenerate cores. The accretion rate, displayed on a logarithmic vertical axis in units of solar masses per year ($M_\odot/\mathrm{yr}$), is calculated from the SPH simulation data by measuring the flux of gas particles across a defined control surface from the Airy analysis. The horizontal axis shows the time in years (yr) on a linear scale. The curves reveal that the accretion process is highly episodic and non-stationary, characterized by a series of sharp bursts that are strongly correlated with the periapsis passages of the cores, contrasted with quiescent periods during apoapsis. A significant asymmetry is evident, with the secondary core exhibiting higher peak accretion rates than the primary. The fine structure within each burst suggests the inflowing material from the common envelope is clumpy and subject to local instabilities.
}
\label{fig.Airy_accretion_timeseries_cores}
\end{figure*}

To address this, we must first define a physically motivated boundary for each core. The Airy function analysis, which gives us the asymptotic density profile $\rho(r)$ given in Eq.~(\ref{eq.profile}), provides the necessary conditions. The characteristic decay length of the density tail, $\beta^{-1}$, represents the physical width of the core-envelope intrface. We use this to define a spherical control surface, the accretion radius $r_{\rm acc}$, around the center of each core. A reasonable choice for this radius is a few times the interface width, such that $r_{\rm acc} = N \beta^{-1}$ where $N$ is a constant of order unity, ensuring the surface lies just outside the core's primary density drop. This transforms the problem from measuring accretion on to a point to measuring the mass flux across a well-defined spherical surface.

The algorithm to calculate the mass flux proceeds by analyzing the gas particles in the vicinity of each core at a given time $t$. For each core, we first transform into its rest frame by calculating the relative velocity, $\vec{v}_{\rm rel,\,i} = \vec{v}_{\rm gas,i} - \vec{v}_{\rm core}$, for every nearby gas particle $i$. We then identify the subset of gas particles that lie within the accretion radius, i.e., those for which the radial distance to the core center $r_i < r_{\rm acc}$. For each of these particles, we compute the radial component of its relative velocity, $v_{\rm rad,\,i} = \vec{v}_{\rm rel,\,i} \cdot \hat{r}_i$, where $\hat{r}_i$ is the unit vector pointing from the core to the particle. A negative value for $v_{\rm rad,\,i}$ signifies motion towards the core. We are interested exclusively in inflowing material, so we sum the mass $m_i$ of all particles that satisfy both conditions; i.e. being inside the accretion radius and having an inward radial velocity.

This sum represents the total mass, $M_{\rm inflow}$, that is actively flowing inward across the control surface at a given snapshot. To convert this mass flux into a rate, we approximate the instantaneous accretion rate $\dot{M}(t)$ by dividing the total inflowing mass by the time elapsed between consecutive simulation snapshots, $\Delta t$. This can be expressed as

\begin{equation}
\dot{M}(t) \approx \frac{M_{\rm inflow}}{\Delta t} = \frac{1}{\Delta t} \sum_{i \, | \, r_i < r_{\rm acc} \land v_{\rm rad,\,i} < 0} m_i.
\label{eq.mdot_calc}
\end{equation}
This calculation is performed for each of the two cores at every available simulation time, yielding the two distinct time series, $\dot{M}_1(t)$ and $\dot{M}_2(t)$, as shown in Fig.~(\ref{fig.Airy_accretion_timeseries_cores}). Mass transfer on to the cores is hence a highly dynamic and non-stationary process, fundamentally governed by the binary's orbital phase. The analysis reveals that accretion does not occur steadily, but is instead concentrated in a series of intense, episodic bursts which are coincident with the periastron passages of the cores, separated by long quiescent periods. The fine, spiky structure within each accretion event suggests that the inflowing material from the common envelope is not a uniform fluid but is instead clumpy and subject to local instabilities. These characteristics underscore the critical role of hydrodynamic interactions and strong tidal forces in mediating the mass transfer process during the merger.

\subsection{Distribution of Accretion Power Across Timescales}
\label{subsec.accretion_power_dist}

Now that we haveestablished the time-domain behavior of the accretion rate, $\dot{M}(t)$, we next seek to characterize the process in the frequency or scale domain. The goal is to determine which timescales are most dominant in the accretion flow. Specifically, we wish to quantify whether the accretion is primarily a series of short, rapid bursts or a more prolonged, quasi-steady infall. To achieve this, we apply the continuous wavelet transform, as defined in Eq.~(\ref{eq.cwt}), to the accretion rate time series for each core, treating $\dot{M}(t)$ as the input signal in place of $L(t)$. This procedure yields the complex wavelet coefficients, $W(\tau,t)$, which contain information about the signal's behavior at a given timescale $\tau$ and time $t$.

The first step in the analysis is to compute the wavelet power spectrum, defined as the squared magnitude of the wavelet coefficients, $|W(\tau,t)|^2$. This quantity represents the power of the accretion rate signal localized in the time-scale plane. To find the total power associated with a specific timescale over the entire duration of the simulation, we integrate the power spectrum along the time axis, as described in Eq.~(\ref{eq.energy}). This yields the global wavelet spectrum, $E(\tau)$, which in this context represents the integrated power of the accretion rate signal at each scale $\tau$.

\begin{figure*}
          {\includegraphics[width=1\textwidth,center]{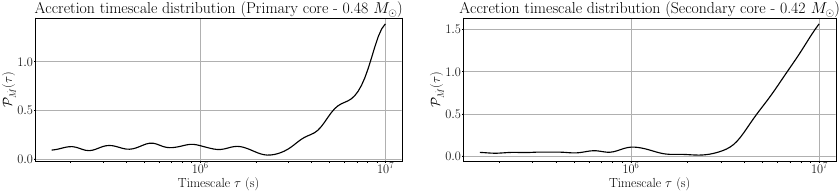}}
\caption
   {
Normalized timescale distribution of the accretion rate, $\mathcal{P}_{\dot{M}}(\tau)$, for the primary (0.48 $M_\odot$, top panel) and secondary (0.42 $M_\odot$, bottom panel) cores. The vertical axis represents the dimensionless normalized power of the accretion signal, while the horizontal axis is the characteristic timescale $\tau$ in seconds, plotted on a logarithmic scale. The location of the peak in each distribution reveals the dominant timescale of the most energetic accretion events. The figure highlights a significant physical difference in the accretion process for each core, with the less massive secondary core showing a dominant timescale that is longer and more variable than that of the primary core.
   }
\label{fig.Airy_energy_dist}
\end{figure*}

The raw values of $E(\tau)$ depend on the overall magnitude of the accretion. To obtain a scale-invariant measure of the relative importance of different timescales, we normalize this global spectrum to create a true probability density function, which we denote $\mathcal{P}_{\dot{M}}(\tau)$. This is achieved using Eq.~(\ref{eq.norm}), where the total energy is the integral of the global spectrum over logarithmic scale, $d(\ln\tau)$. This logarithmic scaling ensures that both long and short duration events are weighted equally in the final distribution. The resulting quantity, $\mathcal{P}_{\dot{M}}(\tau)$, is what is presented in Fig.~(\ref{fig.Airy_energy_dist}) for each core. A peak in this distribution at a particular timescale indicates that the most energetic component of the accretion flow occurs with that characteristic duration.

It is difficult to resist the tramway analogy. Let us think of the accretion rate, $\dot{M}(t)$, as the security camera footage recording the flow of passengers getting onto the tram over a full 24-hour period. We are no longer just looking at the static crowd \textit{at} the door; we are analyzing the \textit{dynamics of boarding}.

The wavelet analysis is like a security analyst reviewing this footage. The analyst's goal is to determine the dominant style of passenger boarding. They use the timescale ($\tau$) - the duration of a ``boarding event'' - to categorize the flow. The analyst identifies two distinct styles in the footage: (i) Short $\tau$ events: At e.g. 8:00 AM, the doors open and a ``burst'' of 30 impatient people shoves on in just 10 seconds. This is a very high-intensity, rapid-infall event. (ii) Long $\tau$ events: At 2:00 PM, Spanish lunchtime, a ``prolonged trickle'' of 30 people boards steadily over 30 minutes, one person at a time. This is a very low-intensity, quasi-steady event.

The final normalized power distribution, $\mathcal{P}_{\dot{M}}(\tau)$, is the analyst's summary report. It answers the crucial question: ``Where did the total energy of the day's passenger flow come from?''. Note that in this analogy, the energy represents the intensity or impact of the passenger flow. A sudden burst of 30 people pushing onto the tram in 10 seconds is a high-intensity, high-impact event; it has high ``energy''. In contrast, a slow trickle of 30 people boarding one by one over 30 minutes is a low-intensity, low-impact event; it has low ``energy''.
Even if the total number of passengers in both events is the same, the analysis identifies the style of boarding. The final report, $\mathcal{P}_{\dot{M}}(\tau)$, determines whether the day's overall activity was dominated by the contribution of all the high-impact bursts or by the contribution of all the low-impact trickles. Our analysis does exactly this for the flow of gas onto the degenerate core.

In Fig.~(\ref{fig.Airy_energy_dist}) the normalized timescale distributions of the accretion rate, $\mathcal{P}_{\dot{M}}(\tau)$, for the primary and secondary cores reveal a distinct difference in the characteristic nature of the mass flow onto each object. For the primary, more massive core, the distribution exhibits a sharp, well-defined peak at a characteristic timescale of $\tau \approx 1.2 \times 10^6$ seconds. This indicates that the accretion process onto this object is dominated by discrete, impulsive events of a relatively short and consistent duration. In contrast, the distribution for the secondary, less massive core peaks at a significantly longer timescale of $\tau \approx 2.5 \times 10^6$ seconds. This peak is notably broader and has a larger amplitude, which implies that the accretion events onto the secondary core are not only more prolonged but are also more variable in their duration and contribute more significantly to the total power of the accretion signal. This asymmetry is a key physical result of the simulation. It suggests that the less massive object experiences a more sustained and powerful accretion flow, a counter-intuitive outcome that likely arises from the complex tidal dynamics of the merger, which may preferentially channel larger, more coherent streams of gas from the common envelope toward the secondary core.

\subsection{Wavelet Variance and Turbulent Cascade Analysis}
\label{subsec.wavelet_variance_accretion}

Following the analysis of the normalized power distribution, we investigate the absolute power of the accretion signal as a function of scale to diagnose the turbulent properties of the flow. This is accomplished by computing the wavelet variance, $\sigma^2(\tau)$, which is defined as the time-average of the wavelet power spectrum for each timescale $\tau$. Mathematically, it is given by

\begin{equation}
\sigma^2(\tau) = \frac{1}{T} \int_0^T |W(\tau,t)|^2 dt,
\label{eq.variance}
\end{equation}
where $T$ is the total duration of the time series. The wavelet variance is thus directly proportional to the global wavelet spectrum, $E(\tau)$, from the previous section, differing only by a constant normalization factor. Its primary utility is in characterizing a potential turbulent energy cascade within the accretion flow. The wavelet variance measures the average power or intensity of the accretion signal's fluctuations at each specific timescale $\tau$. When we plot this variance against all possible timescales, the injection timescale, $\tau_{\rm inj}$, is the point on the graph where this power is at its maximum. It represents the characteristic duration of the primary driving mechanism that is pumping energy into the accretion flow.
This is the ``top'' of the turbulent energy cascade. In this process, energy is injected into the system at a dominant, large scale (like the main orbital motion or a large infall event), which corresponds to $\tau_{\rm inj}$. This large-scale motion is unstable and breaks down, transferring its energy to progressively smaller structures, which have shorter timescales. Therefore, the timescales shorter than $\tau_{\rm inj}$ show declining power because they just represent the fragmented eddies of the cascade. Timescales longer than $\tau_{\rm inj}$ also have low power because no significant physical process is adding energy at those slow durations. Finding this peak at $\tau_{\rm inj}$ allows us to identify the so to say heartbeat of the main force driving the entire turbulent flow.

The core of the algorithm involves analyzing the functional form of the wavelet variance. The quantity $\sigma^2(\tau)$ is plotted against the timescale $\tau$ on a log-log scale. A linear trend in such a plot signifies a power-law relationship of the form $\sigma^2(\tau) \propto \tau^\beta$. The power-law index, or slope, $\beta$, is a critical diagnostic of the turbulent cascade. It is extracted numerically by performing a linear regression on $\log(\sigma^2)$ as a function of $\log(\tau)$ over a range of scales known as the inertial range. This exponent quantifies how the energy of the flow is distributed among eddies of different sizes or events of different durations.

The physical interpretation of these quantities provides us with information about the nature of the accretion turbulence. The exponent $\beta$ characterizes the efficiency and nature of the energy cascade, allowing for a direct comparison with theoretical turbulence models. For instance, a deviation from the value predicted by incompressible Kolmogorov turbulence can indicate the presence of strong compressibility in the gas, as would be expected in a shock-dominated accretion flow \citep{Kolmogorov1962}. The analysis also identifies the energy injection scale, $\tau_{\rm inj}$, which corresponds to the peak of the variance spectrum. This scale is located by finding the point where the logarithmic derivative of the variance is zero, i.e., $d(\log \sigma^2)/d(\log \tau) = 0$. Physically, $\tau_{\rm inj}$ represents the characteristic timescale of the large-scale processes, such as orbital shearing or major infall events, that are responsble for injecting energy into the turbulent cascade.

The injection timescale, $\tau_{\rm inj}$, is a key parameter derived from the wavelet variance spectrum, $\sigma^2(\tau)$, which represents the distribution of the signal's power across different timescales. Mathematically, it is defined as the scale at which the variance reaches its maximum value. This corresponds to the peak of the energy-containing range of the spectrum. For a continuous and differentiable spectrum, this condition is met where the first derivative of the variance with respect to the scale is zero. In practice, as the spectrum is analyzed on a logarithmic scale, $\tau_{\rm inj}$ is located where the logarithmic derivative is null

\begin{equation}
\left. \frac{d \log(\sigma^2(\tau))}{d \log(\tau)} \right|_{\tau=\tau_{\rm inj}} = 0.
\label{eq.tau_inj_def}
\end{equation}

\noindent
In a turbulent flow, energy is typically injected into the system at large spatial or temporal scales. This energy then cascades down to progressively smaller scales until it is ultimately dissipated. The injection timescale, $\tau_{\rm inj}$, corresponds to the characteristic duration of the dominant, large-scale events that supply energy to this cascade. In the context of the accretion flow onto the binary cores, $\tau_{\rm inj}$ therefore represents the primary timescale of the driving mechanism. It is the timescale most directly associated with the large-scale hydrodynamic instabilities and tidal shearing motions, driven by the orbital passage at periastron, which stir the common envelope gas and inject the power that fuels the entire accretion process at all smaller scales.

The timescale $\tau$ derived from the wavelet analysis has a direct physical analogue in the theory of turbulence, where it represents the characteristic lifetime or ``turnover time'' of a turbulent eddy. The foundational concept is the turbulent energy cascade, a process in which energy is injected into a fluid system at large physical scales, $\ell_{\rm inj}$. These large-scale structures, or eddies, are inherently unstable and break down, transferring their energy to progressively smaller eddies. This transfer occurs across an intermediate range of scales, known as the inertial range, where the rate of energy transfer is assumed to be constant and viscous dissipation is negligible. Finally, at the smallest scales, $\ell_{\rm diss}$, the kinetic energy of the eddies is converted into heat by the fluid's viscosity.

Each eddy within this cascade, characterized by a size $\ell$, also has a characteristic velocity, $v_\ell$. The eddy turnover time, $\tau_\ell$, is the time required for an eddy to complete a characteristic rotation or break apart, and it is given by the scaling relation $\tau_\ell \approx \ell / v_\ell$. This relationship provides the crucial link between the spatial scales of the turbulent structures and the temporal scales measured by the wavelet transform. Large, slow-moving eddies are associated with a long characteristic timescale $\tau$, while small, fast-moving eddies correspond to a short $\tau$.

The foundational theory for the inertial range of incompressible turbulence was developed by Kolmogorov, who predicted that the energy spectrum follows a characteristic power law as a function of wavenumber $k \approx 1/\ell$, given by $E(k) \propto k^{-5/3}$. The wavelet variance, $\sigma^2(\tau)$, provides a direct method for measuring the power distribution as a function of timescale. For a system following Kolmogorov's theory, the wavelet variance is expected to scale as $\sigma^2(\tau) \propto \tau^{2/3}$. The analysis of the slope $\beta$ in the wavelet variance plot of the accretion signal therefore allows for a direct comparison of the simulation's results with this foundational theory, providing a quantitative diagnostic of the nature of the turbulence in the accretion flow.

The primary engine for the energetic phenomena observed is the orbital decay of these cores. Gravitational potential energy is extracted from the orbit and converted into the kinetic energy of the surrounding gas through hydrodynamic and tidal interactions. The timescale $\tau$ from the wavelet analysis is the fundamental quantity for understanding this process, as it represents the characteristic lifetime, or ``turnover time'', of a turbulent eddy in the resulting energy cascade.

\begin{figure*}
          {\includegraphics[width=1\textwidth,center]{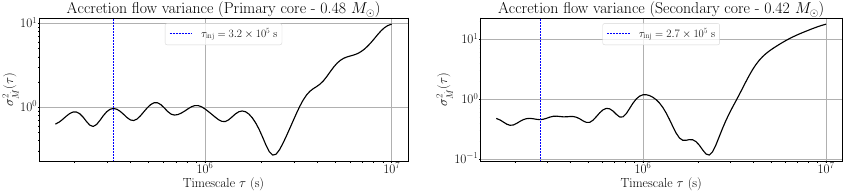}}
\caption
   {
The wavelet variance of the accretion rate, $\sigma^2_{\dot{M}}(\tau)$, as a function of timescale $\tau$ for the primary (0.48 $M_\odot$, top panel) and secondary (0.42 $M_\odot$, bottom panel) cores. Both axes are plotted on a logarithmic scale. The vertical dashed line in each panel indicates the identified energy injection timescale, $\tau_{\rm inj}$, which is $3.2 \times 10^5$ s for the primary core and $2.7 \times 10^5$ s for the secondary core. Unlike a canonical turbulent cascade that would exhibit a monotonic power-law decay, the variance spectra are complex and oscillatory. This structure suggests that the accretion flow is not governed by a simple, steady cascade but is instead modulated by multiple, quasi-periodic processes. The peak of the variance spectrum, identified as $\tau_{\rm inj}$, corresponds to the dominant timescale at which mass is supplied to the accretion flow, likely related to the orbital period during close interaction, while the oscillations at smaller scales may reflect transient features in the accretion stream itself.
   }
\label{fig.Airy_variance}
\end{figure*}

The transfer of energy from large orbital scales to small dissipative scales can be conceptualized by an analogy to a hydraulic cascade. This process involves three distinct physical stages. First, energy injection: The orbital motion of the cores, particularly the strong tidal shearing forces during periastron passages, injects energy into the common envelope at the largest scales. These large structures, or eddies, are unstable. This is the energy-containing range of the turbulent flow, and its characteristic timescale is identified as the injection scale, $\tau_{\rm inj}$, in the wavelet variance spectra shown in Fig.~(\ref{fig.Airy_variance}). Second, the cascade: The large-scale flows break apart, transferring their energy to progressively smaller eddies in a process that defines the inertial range. The characteristic lifetime of an eddy of size $\ell$ and velocity $v_\ell$ is its turnover time, $\tau_\ell \approx \ell/v_\ell$. This relationship provides the crucial link between the spatial scales of the turbulent structures and the temporal scales measured by the wavelet transform. Large, slow-moving eddies correspond to a long timescale $\tau$, while small, fast-moving eddies correspond to a short $\tau$. The complex, oscillatory structure of the variance spectra in Fig.~(\ref{fig.Airy_variance}) suggests that this is not a simple, steady cascade, but one that is modulated by the quasi-periodic forcing of the orbit. Finally, dissipation: At a very small physical scale, the cascade terminates. The kinetic energy of the smallest eddies is efficiently converted into internal energy by processes such as shocks and viscosity, thereby heating the gas.

In Fig.~(\ref{fig.Airy_variance}) we depict the analysis of the accretion rate time series; it shows that the process of mass transfer onto the degenerate cores is fundamentally non-stationary and is strongly modulated by the orbital dynamics of the binary. The instantaneous accretion rate, $\dot{M}(t)$, is not a steady inflow but is instead characterized by episodic bursts of activity that are coincident with the periapsis passages of the cores, corroborating our previous findings. The wavelet variance analysis identifies a dominant energy injection timescale, $\tau_{\rm inj}$, on the order of several days for both cores. This scale is consistent with the characteristic timescale of the strongest hydrodynamic interactions during close encounters, confirming that the accretion process is driven primarily by orbitally-induced tidal forces.

A key result of this analysis is the significant asymmetry observed in the accretion behavior of the two cores. Counter-intuitively, the secondary, less massive core exhibits a more vigorous and sustained accretion flow than the primary core. This is evident both in the time domain, where its peak accretion rate is substantially higher, and in the scale domain. The normalized power distribution, $\mathcal{P}_{\dot{M}}(\tau)$, for the secondary core peaks at a longer characteristic timescale ($\tau \approx 2.5 \times 10^6$ s) and with a greater amplitude and breadth than that of the primary core ($\tau \approx 1.2 \times 10^6$ s). This indicates that the accretion events onto the less massive object are more prolonged, more variable, and contribute more significantly to the total power of the accretion signal.

Furthermore, the structure of the accretion flow is shown to be complex. The fine, spiky features in the $\dot{M}(t)$ time series suggest that the inflowing material is inhomogeneous and clumpy. This is supported here by the wavelet variance spectra, $\sigma^2_{\dot{M}}(\tau)$, which are highly oscillatory and do not exhibit the monotonic power-law decay characteristic of a simple, canonical turbulent cascade. This complex structure implies that the accretion is not a simple, steady stream but is instead modulated by multiple quasi-periodic processes, likely reflecting transient features and instabilities within the tidally-stripped gas streams that feed the cores.

\subsection{Analysis of Phase-Coherent Accretion Events}
\label{subsec.phase_coherent_accretion}

The final stage of the Airy analysis is to identify accretion events that are not random, but are instead phase-locked to a periodic driving mechanism, which in this system is the orbital motion of the binary. The use of a complex mother wavelet, such as the Morlet wavelet defined in Eq.~(\ref{eq.morlet}), is the way to address this. Because the wavelet coefficients, $W(\tau,t)$, are complex numbers, they possess both a magnitude, $|W(\tau,t)|$, and a phase, $\phi(\tau,t) = \arg[W(\tau,t)]$. This phase represents the local phase of the accretion signal, $\dot{M}(t)$, at a specific time $t$ and for a specific timescale $\tau$.

To test for coherence, we compare the phase of the accretion signal with a reference phase representing a perfectly periodic process. For each timescale $\tau$ under consideration, we define a simple reference phase model, $\phi_{\rm ref}(t) = 2\pi t/\tau$. This represents the phase of a purely sinusoidal signal with period $\tau$. We then construct a coherence metric, $\mathcal{I}(\tau,t)$, which measures the degree of alignment between the signal's phase and this reference phase, weighted by the signal's power at that point,

\begin{equation}
\mathcal{I}(\tau,t) = |W(\tau,t)| \cos[\phi(\tau,t) - \phi_{\rm ref}(t)].
\label{eq.coherence}
\end{equation}
The physical interpretation of this metric is direct. A large positive value of $\mathcal{I}(\tau,t)$ indicates that the accretion signal is both powerful and strongly in-phase with the reference period $\tau$ at that specific time $t$. Conversely, a large negative value would indicate an anti-phase correlation, while values near zero imply that the signal is either weak or has a random phase relationship with the reference.

The final step of the algorithm is to isolate only the most statistically significant coherent events. We compute the distribution of the coherence metric $\mathcal{I}(\tau,t)$ over the entire time-scale plane and calculate its standard deviation, $\sigma_{\mathcal{I}}$. A threshold is then set at a multiple of this value, for instance $\mathcal{I}_{\rm th} = 2.5 \sigma_{\mathcal{I}}$. We then generate a binary mask in the time-scale plane, marking only those locations where $\mathcal{I}(\tau,t) > \mathcal{I}_{\rm th}$. The final plot displays this mask, effectively filtering out noise and highlighting only the specific times and characteristic scales at which accretion occurs in strong, periodic bursts that are phase-locked to the underlying orbital drivers of the system.

\begin{figure*}
          {\includegraphics[width=1\textwidth,center]{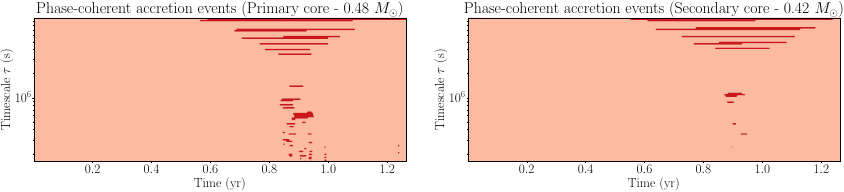}}
\caption
   {
A time-scale map of the phase-coherent accretion events for the primary (0.48 $M_\odot$) core. The horizontal axis is time in years, and the vertical axis is the characteristic timescale $\tau$ in seconds on a logarithmic scale. The highlighted regions indicate statistically significant events where the accretion rate is phase-locked to a periodic driver, as determined by the coherence metric in Eq.~(\ref{eq.coherence}). The systematic decrease in the characteristic timescale of the coherent events during this active phase provides a visual representation of the energy cascade from large-scale orbital forcing to the smaller scales of the accretion flow itself. 
   }
\label{fig.Airy_coherence}
\end{figure*}

In Fig.~(\ref{fig.Airy_coherence}) we can see that the analysis of phase-coherent accretion events provides a direct link between the orbital dynamics of the binary and the process of mass transfer. The results reveal that coherent accretion is a highly episodic phenomenon, not a continuous process. A long quiescent period is observed for the first half of the orbit, corresponding to the apoapsis passage, as we previously mentioned, where the cores are widely separated and tidal interactions are weak. This is followed by a concentrated period of intense coherent activity beginning at $t \approx 0.8$ yr, which is coincident with the periapsis passage of the cores. This suggests that the strong gravitational forcing during close encounters is the primary driver for synchronized, non-random accretion. The most significant physical insight comes from the temporal evolution of the characteristic timescale, $\tau$, of these coherent events. The activity begins at large scales, $\tau > 10^6$ s, and as the cores move through periastron, the timescale of the coherent structures systematically decreases. This feature can be interpreted as a direct visualization of an energy cascade. The large-scale orbital energy and angular momentum are injected into the common envelope, forming large tidal structures. These structures subsequently become unstable and fragment, causing the accretion to proceed in a series of smaller, more rapid events, like pulses, thus shifting the coherent power to shorter timescales. This result implies a direct causal link between the large-scale orbital motion and the small-scale physics of the accretion flow itself.

The background of the figure is rendered in a solid color, in this case light red, because the visualization is a filled contour map which assigns a color to every point in the time-scale plane. The light red color corresponds to the base level of this map and represents the background, or non-coherent, state of the system. It is assigned to all regions where the phase coherence metric, $\mathcal{I}(\tau,t)$ as defined in Eq.~(\ref{eq.coherence}), falls below the statistical threshold required for an event to be considered significant. Thus, the background color signifies the absence of detectable phase-locked accretion activity. The highlighted coherent events appear as structures extended mostly horizontally because of the nature of the wavelet analysis itself. The vertical axis of the plot represents the characteristic timescale, $\tau$, of a process, while the horizontal axis represents time, $t$. A horizontal feature is therefore one that occurs at a nearly constant timescale. The wavelet transform identifies features by convolving the signal with a wavelet function, which is itself a wave packet of finite duration. When the transform detects a coherent feature in the accretion signal at a scale $\tau$, the resulting high-coherence region in the plot will necessarily have a temporal width that is also of order $\tau$. Consequently, the horizontal extent of these lines does not imply a long, continuous event, but rather represents a single, coherent burst whose apparent duration in the time-scale plane is intrinsically determined by its characteristic physical timescale.

\subsection{Preliminary Conclusions About Observational Signatures}
\label{subsec.prelim_obs_signatures}

The primary observational signature predicted by this analysis of accretion onto the binary cores is a quasi-periodic series of luminous, high-energy flares. The accretion luminosity is expected to peak at approximately $10^{39}-10^{40}$ erg/s during the periastron passages, with emission likely concentrated in the ultraviolet and soft X-ray portions of the spectrum. This signature is distinct from other major astrophysical transients. A supernova, for instance, is a singular, cataclysmic event that is orders of magnitude more luminous, typically exceeding $10^{42}$ erg/s, and is characterized by a single, fast-rising light curve followed by an exponential decay over months. In contrast, our predicted signal consists of multiple, repeating bursts over a longer timescale, dictated by the binary orbit. Similarly, a canonical tidal disruption event, while also a source of soft X-rays, involves a star being destroyed by a supermassive black hole, resulting in a single, more luminous flare ($10^{43}-10^{45}$ erg/s) with a characteristic power-law decay, unlike the orbitally modulated signal we predict. 

\section{Ulterior evolution of the degenerate cores}
\label{sec.UlteriorNaive}

{As the SPH simulation approaches its computational limits, we transition to an initial semi-analytical approach to estimate the further evolution and merger timescale of the cores.}

The previous analysis of section~(\ref{sec.evo_cores}) deliberately deferred discussion of the cores' subsequent evolution to first focus on defining the gas envelope properties and their dynamical interaction with the orbiting compact objects. This prioritization was necessary because the envelope structure fundamentally determines the drag forces governing the orbital decay, while the cores themselves maintain nearly constant mass and radius throughout most of the integration time (which in physical units is relatively short). 

\begin{figure}
          {\includegraphics[width=0.45\textwidth,center]{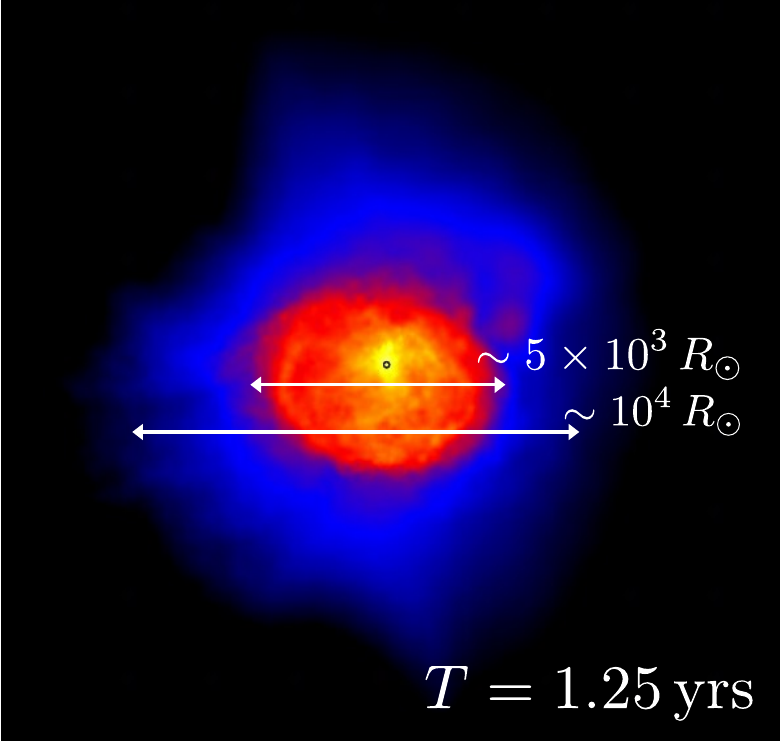}}
\caption
   {
The hydrodynamical configuration at the conclusion of the numerical integration ($1.25\,{\rm yrs}$) reveals an expanding, nearly spherical gaseous structure, as evidenced by the XY-projected column density (defined as in Fig.~\ref{fig.FirstEncounter}). While maintaining approximate spherical symmetry across all projection planes, the system exhibits slight oblateness. The diffuse outer envelope extends to radii of $\sim 10^4\,R_{\odot}$, while the dense central region forms a compact core with diameter $\sim 5\times 10^3\,R_{\odot}$. This morphology results from the collision dynamics, which efficiently redistributes angular momentum while preserving overall isotropic expansion. The density contrast between the extended envelope and central concentration suggests efficient shock-driven compression during the collision phase.
   }
\label{fig.LastSnapshot}
\end{figure}

However, we now address this previously omitted aspect, noting that a purely numerical treatment of the further evolution of the core binary becomes fundamentally unreliable during the stages closer to the core merger. As the separation shrinks, several physical and numerical limitations emerge: the gas density between the cores reaches values where the mean free path becomes comparable to the orbital separation, violating the fluid approximation; the time-step requirements for stable integration become computationally prohibitive; and artificial numerical viscosity dominates the true physical dissipation. These considerations have led us to choose a semi-analytical approach instead of further integrating the system. We integrate with StarSmasher the system until it reaches hydrodynamical equilibrium, which happens at about $1.25\,{\rm yrs}$, a moment in the evolution that we show in Fig.~(\ref{fig.LastSnapshot}). It is interesting to note that the nearly spherical configuration fits the analytical approach presented in the work of \cite{AmaroSeoane2023}.

We analyse the further evolution of the cores with a semi-analytical approach that models the orbital decay of degenerate cores in a common envelope environment. It combines hydrodynamic simulation data with an analytical treatment of drag forces, providing a physically justified extension beyond computationally feasible simulation timeframes. The drag force on each core follows the standard aerodynamic form:

\begin{equation}
\label{eq:drag_force}
\mathbf{F}_d = -\frac{1}{2} C_d \rho_g A \|\mathbf{v}_{\text{rel}}\| \mathbf{v}_{\text{rel}},
\end{equation}

\noindent
where $C_d$ is the dimensionless drag coefficient, $\rho_g$ is the local gas density, $A = \pi R_{\text{env}}^2$ is the effective cross-sectional area, and $\mathbf{v}_{\text{rel}} = \mathbf{v}_{\text{core}} - \mathbf{v}_{\text{gas}}$ is the relative velocity between core and local gas. The envelope radius $R_{\text{env}}$ is determined from SPH data as the maximum distance to the three densest neighboring gas particles, as we shown in Fig.~{\ref{fig.densities_RG}}. The orbital energy evolution follows from the power dissipation,

\begin{equation}
\label{eq:power_diss}
\frac{dE}{dt} = \mathbf{F}_{d1} \cdot \mathbf{v}_1 + \mathbf{F}_{d2} \cdot \mathbf{v}_2.
\end{equation}

For a circular binary with separation $d$, the orbital energy is

\begin{equation}
\label{eq:orbital_energy}
E = -\frac{G M_1 M_2}{2d}.
\end{equation}

\noindent
We note that this formulation explicitly assumes a circular orbit. While the SPH simulation exhibits significant eccentricity, this approximation simplifies the analytical treatment of the orbital decay, providing an estimate of the inspiral timescale. However, this assumption neglects the enhanced energy dissipation at periastron characteristic of eccentric orbits, potentially overestimating the merger timescale. A more sophisticated model addressing this limitation is presented later.

Differentiating Eq.~(\ref{eq:orbital_energy}) and equating to Eq.~(\ref{eq:power_diss}) yields

\begin{align}
\label{eq:energy_deriv}
\frac{dE}{dt} &= \frac{G M_1 M_2}{2d^2} \frac{dd}{dt} \\
&= \mathbf{F}_{d1} \cdot \mathbf{v}_1 + \mathbf{F}_{d2} \cdot \mathbf{v}_2.
\end{align}

\noindent
Solving for the separation evolution gives the fundamental ODE:

\begin{equation}
\label{eq:dddt}
\frac{dd}{dt} = \frac{2d^2}{G M_1 M_2} \left( \mathbf{F}_{d1} \cdot \mathbf{v}_1 + \mathbf{F}_{d2} \cdot \mathbf{v}_2 \right).
\end{equation}

The SPH simulation provides time-dependent functions $\mathbf{F}_{d1}(t)$, $\mathbf{F}_{d2}(t)$, $\mathbf{v}_1(t)$, $\mathbf{v}_2(t)$, $M_1(t)$, and $M_2(t)$ through interpolation of snapshot data. These interpolants enable numerical integration of \eqref{eq:dddt} beyond the simulation timeframe.
The ODE in \eqref{eq:dddt} is integrated using a Runge-Kutta (4,5) method with adaptive time-stepping. Event detection terminates integration when $d \leq d_{\text{mrg}}$,

\begin{equation}
\label{eq:merger_condition}
g(t) = d(t) - d_{\text{mrg}} = 0
\end{equation}

\noindent
with root-finding precision of $10^{-6}$ in normalized units. The merger time $T_{\text{mrg}}$ is then:

\begin{equation}
\label{eq:merger_time}
T_{\text{mrg}} = \min \{ t \,|\, d(t) \leq d_{\text{mrg}} \}.
\end{equation}

\noindent
Uncertainty quantification comes from varying $C_d$, which contains hydrodynamic uncertainties. The semi-analytical approach thus extends realistic simulation data beyond computationally feasible timeframes while avoiding unphysical numerical artifacts.

\begin{figure}
          {\includegraphics[width=0.5\textwidth,center]{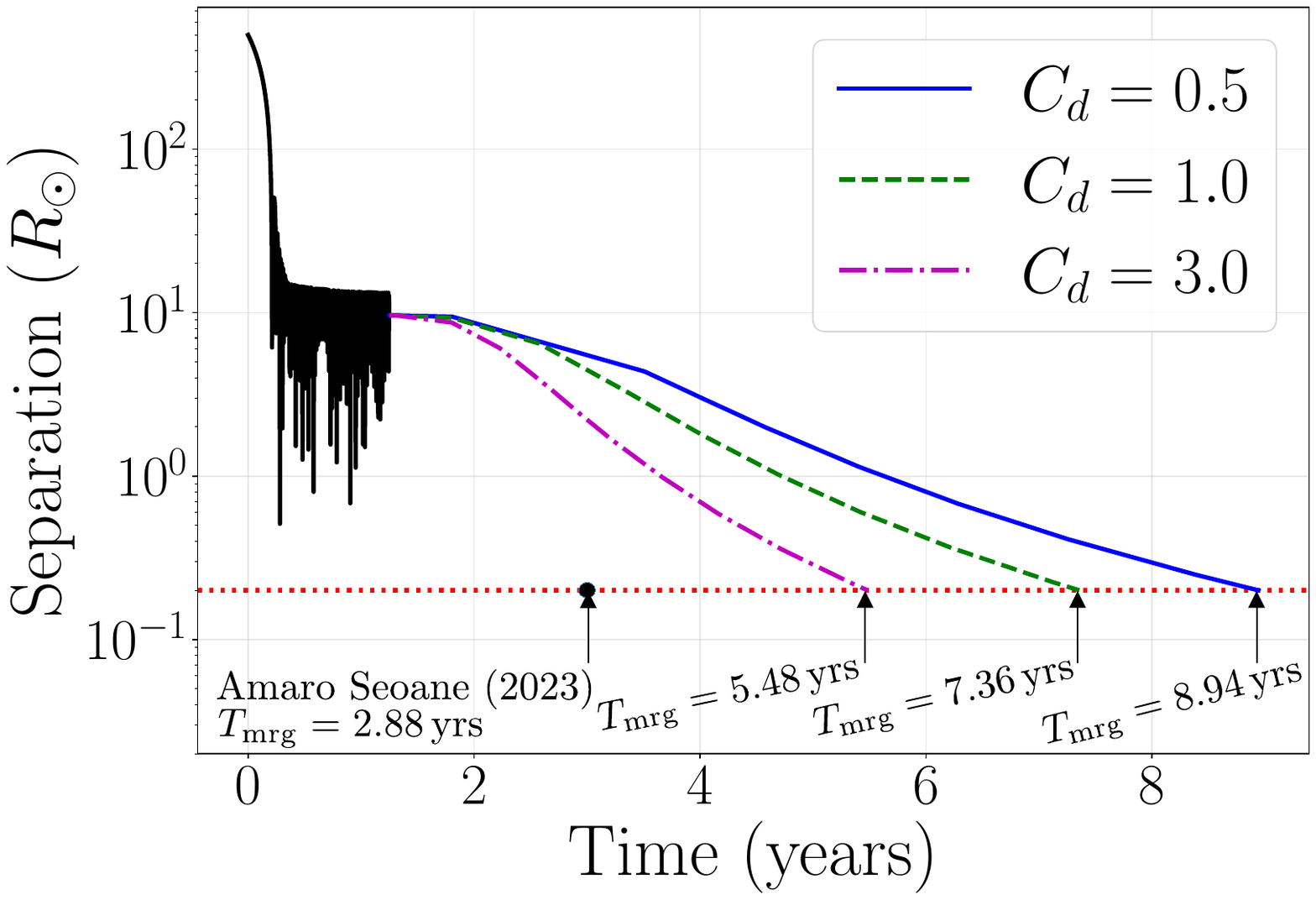}}
\caption
   {  
Binary separation evolution showing three distinct phases: (i) the initial numerical SPH simulation (solid black line), (ii) the semi-analytical gas drag model (dashed/dotted colored lines) for drag coefficients $C_d \in {0.5, 1, 3}$, and (iii) the colliding radius at $0.2R_\odot$ (horizontal red dotted line). The semi-analytical model yields merger times of $8.94$ yr ($C_d=0.5$), $7.36$ yr ($C_d=1.0$), and $5.48$ yr ($C_d=3.0$), demonstrating the expected inverse scaling with drag coefficient. The analytical prediction of $2.88 \times 10^0$ yr from \cite{AmaroSeoane2023} (derived under spherical symmetry assumptions) shows remarkable agreement with the results here, despite the latter's inclusion of anisotropic gas effects. Axes use linear timescale (x) and logarithmic separation (y).    
   }
\label{fig.core_merger_prediction}
\end{figure}

The estimation of the merger timescale follows closely the analytical derivation of \cite{AmaroSeoane2023}. In this work, $T_{\text{mrg}}$ is determined by solving the orbital decay equation derived from gas dynamical friction. The evolution of the semimajor axis $a(t)$ follows $\dot{a}/a = -2 / T_{\text{gas}}(t)$, where $T_{\text{gas}}(t)$ is the characteristic decay timescale. This leads to the integral equation:  
\begin{equation}  
\label{eq:decay}  
\ln \left( \frac{a_{\text{mrg}}}{a_0} \right) = -2 \int_0^{T_{\text{mrg}}} T_{\text{gas}}^{-1}(t)  dt.  
\end{equation}  

\noindent  
Here, $a_0 = 0.5 R_{\odot}$ is the initial separation and $a_{\text{mrg}} = 0.2\, R_{\odot}$ marks coalescence. The integrand $T_{\text{gas}}^{-1}(t)$ incorporates time-dependent density evolution $n_{\text{rad}}(t) \propto \left[ \kappa_0 + \kappa_1 V_{\text{exp}} t \right]^{-3}$ and cooling effects $\propto \exp(-\beta t^2)$, resulting in a complex functional form.
Analytical evaluation of \eqref{eq:decay} involves expanding the exponential and solving term-wise, yielding an infinite series representation for the definite integral. The solution exhibits exponential growth for $t > 20$ months, but physical validity is constrained to $t \leq 34.3$ months, where $\ln(a_{\text{mrg}}/a_0) \approx -0.916$. Remarkably, the predicted value of 34.7 months matches the results presented here with a factor of a few, despite the latter neglecting expansion and anisotropic effects modeled here via SPH simulations. The insensitivity to initial conditions (e.g., scaling $a_0 \to 3a_0$ only perturbs $T_{\text{mrg}}$ by $\epsilon \ll 1$) further underscores the robustness of the analytical results of \cite{AmaroSeoane2023} and ours.  

\section{Colliding degenerate cores: A Homothetic Approach}
\label{sec.homothetic}

{Building upon the estimated merger timescale from Section~\ref{sec.UlteriorNaive}, we now model the morphology of the subsequent explosion using a simplified, scale-invariant approach.}

\subsection{Homothety}
\label{subsec.homothety_theory}

The limitations of the hydrodynamics code in resolving the degenerate core collision dynamics—particularly the extreme density gradients and shock dissipation at sub-grid scales—motivate a homothetic approximation of the system as a pressureless medium during the explosion phase. This reduction is justified by the timescale separation $\Delta t_{\text{inj}} \ll \tau_{\text{dyn}}$, where $\Delta t_{\text{inj}}$ characterizes the impulsive energy injection and $\tau_{\text{dyn}}$ the gravitational timescale, ensuring the dominance of inertial effects over pressure forces. The post-collision evolution is modeled as a radial homothety $\mathcal{H}_k$ with scaling factor $k(t)$ that depends on $v_{\text{esc}}$, the characteristic escape velocity derived from the collision energy $E_{\text{coll}}$, total mass $M$, and the initial radius $R_0$. This transformation preserves the angular distribution of ejecta while scaling their radial positions \textit{linearly},  

\begin{equation}
\mathbf{r}_i(t) = \mathbf{r}_i(0) \cdot k(t), \label{eq:homothety}
\end{equation}  

\noindent
implicitly enforcing the collisionless Boltzmann equation $\partial_t f + \mathbf{v} \cdot \nabla_{\mathbf{r}} f = 0$ for the phase-space distribution $f(\mathbf{r}, \mathbf{v}, t)$. The velocity field $\mathbf{v}_i = \mathbf{r}_i(0) \cdot \dot{k}$ inherits its anisotropy from the initial density profile $\rho(\mathbf{r}, 0)$, with $\dot{k} = v_{\text{esc}}/R_0$ determined by the $\rho/r^2$-weighted energy partition. The approximation remains valid for $t \ll \tau_{\text{cool}}$, after which radiative losses break the scale invariance of $\mathcal{H}_k$. The two-stage process—instantaneous energy injection followed by homothetic expansion—captures the bulk kinematics while neglecting higher-order corrections from fluid instabilities or magnetic stresses, as justified by the hierarchy $E_{\text{kin}} \gg E_{\text{th}}, E_B$ in the post-shock regime.

The total collision energy $E_{\rm coll}$ is computed from the relative velocity and reduced mass of the cores,

\begin{align}
E_{\rm coll} & = \frac{1}{2} \mu v_{\rm rel}^2, \nonumber \\
\mu & = \frac{m_1 m_2}{m_1 + m_2},
\label{eq:collision_energy}
\end{align}  

\noindent
where $v_{\rm rel} = ||\vec{v}_1 - \vec{v}_2||$. When $v_{\rm rel} < 10^5$ cm/s (indicating unresolved dynamics), we estimate it from free-fall kinematics:  

\begin{equation}
v_{\rm rel} = \sqrt{\frac{2G(m_1 + m_2)}{d}} , \quad d = ||\vec{r}_1 - \vec{r}_2||.
\label{eq:freefall_velocity}
\end{equation}  

\noindent
This energy is distributed radially from the center of mass $\vec{R}_{\rm COM} = (m_1\vec{r}_1 + m_2\vec{r}_2)/(m_1 + m_2)$. We partition the gas into $N$ logarithmic shells with boundaries $\{r_k\}_{k=0}^N$ and midpoints $r_{k}^{\rm mid} = (r_k + r_{k+1})/2$. For shell $k$, the mass $M_k$ and volume $V_k$ are:  

\begin{align}
M_k &= \sum_{i \in \mathcal{S}_k} m_i \\
V_k &= \frac{4\pi}{3} (r_{k+1}^3 - r_k^3),
\label{eq:shell_properties}
\end{align}  

\noindent
where $\mathcal{S}_k$ denotes particles in shell $k$. The mass density $\rho_k = M_k/V_k$ determines the energy weighting  

\begin{equation}
w_k = \frac{\rho_k}{(r_k^{\rm mid})^2}, \quad E_k = E_{\rm coll} \frac{w_k}{\sum_j w_j}.
\label{eq:energy_weighting}
\end{equation}  

\noindent
This $\rho/r^2$ weighting accounts for energy coupling efficiency proportional to density and geometric dilution of energy flux $\propto r^{-2}$. The velocity boost for shell $k$ follows from energy conservation:  

\begin{equation}
\Delta v_k = \sqrt{\frac{2E_k}{M_k}}
\label{eq:velocity_boost}
\end{equation}  

\noindent
applied radially outward from $\vec{R}_{\rm COM}$. Particle $i$ in shell $k$ receives the velocity increment

\begin{equation}
\Delta \vec{v}_i = \Delta v_k \frac{\vec{r}_i - \vec{R}_{\rm COM}}{||\vec{r}_i - \vec{R}_{\rm COM}||}.
\label{eq:particle_boost}
\end{equation}  

Regarding the post-injection evolution, we adopt ballistic trajectories,

\begin{align}
\vec{r}_i(t) &= \vec{r}_i(0) + \vec{v}_i^{\,\rm new} t,\nonumber \\
\vec{v}_i^{\,\rm new} &= \vec{v}_i^{\,\rm orig} + \Delta \vec{v}_i.
\label{eq:ballistic_evolution}
\end{align}  

The neglect of hydrodynamic interactions in the post-injection evolution is justified by timescale separation arguments. The characteristic explosion timescale $t_{\rm exp}$ is defined by the ratio of the system's initial radius $R_0$ to the characteristic expansion velocity $v_{\rm exp}$ induced by energy injection $t_{\rm exp} \equiv {R_0}/{v_{\rm exp}}$, 
which is orders of magnitude smaller than the hydrodynamic timescale $t_{\rm hydro}$, that characterizes the time required for pressure waves to traverse the system and establish hydrodynamic equilibrium, $t_{\rm hydro} \equiv {R_0}/{c_s} \gg t_{\rm exp}$,
 where $c_s$ is the adiabatic sound speed. The inequality holds because $v_{\rm exp} \gg c_s$ for energy-driven outflows, as the expansion velocity scales as $v_{\rm exp} \sim \sqrt{E_{\rm inj}/M}$ while $c_s \sim \sqrt{P/\rho}$.  

Gravitational effects are subdominant when the expansion timescale satisfies $t_{\rm exp} \ll t_{\rm dyn} \equiv \sqrt{{R_0^3}/({GM_{\rm tot}})}$, where $t_{\rm dyn}$ is the free-fall timescale and $M_{\rm tot}$ the total mass. This condition is equivalent to the ratio of kinetic to gravitational energy being much greater than unity, ${v_{\rm exp}^2 R_0}/({GM_{\rm tot}}) \gg 1$.  
The ballistic approximation remains valid over timescales $t \ll \min(t_{\rm hydro}, \, t_{\rm dyn})$, which encompasses the observable phase of the expansion. During this regime, pressure gradients and gravitational acceleration introduce corrections $\mathcal{O}(t/t_{\rm hydro})$ and $\mathcal{O}(t/t_{\rm dyn})$ respectively, both negligible when $t \sim t_{\rm exp}$.

The energy released during the collision of two degenerate stellar cores is governed by the center-of-mass kinetic energy:  
\begin{equation}  
E = \frac{1}{2} \mu v_{\text{rel}}^2,  
\label{eq:com_energy}  
\end{equation}  
where $\mu \equiv m_1 m_2 / (m_1 + m_2)$ is the reduced mass. Expressing $E$ in terms of the total mass $M \equiv m_1 + m_2$ and symmetric mass ratio $\eta \equiv \mu / M$ yields:  
\begin{equation}  
E = \frac{1}{2} \eta M v_{\text{rel}}^2.  
\label{eq:energy_eta}  
\end{equation}  
Normalize $M$ to $M_0 \equiv 1~M_{\odot}$ and $v_{\text{rel}}$ to $v_0 \equiv 50$ km/s:  
\begin{align}  
\hat{M} &\equiv M / M_0, \\  
\hat{v}_{\text{rel}} &\equiv v_{\text{rel}} / v_0.  
\end{align}  
Where we have defined the constant $K \equiv M_0 v_0^2/2 \sim 2.486 \times 10^{46}~\text{erg}$.  
 
Substituting into Equation~(\ref{eq:energy_eta}), 
\begin{equation}  
E \sim 2.5 \times 10^{46}~\text{erg}~ \eta\, \hat{M}\, \hat{v}_{\text{rel}}^2.  
\label{eq:final_energy}  
\end{equation}  
Here $\eta \equiv m_1 m_2 / (m_1 + m_2)^2$ encodes the mass asymmetry, with $\eta \in (0, 0.25]$ for $m_1, m_2 > 0$. This expression holds for arbitrary mass pairs $(m_1, m_2)$.

Eccentricity would not alter the fundamental expression $E = \frac{1}{2}\mu v_{\text{rel}}^2$ if $v_{\text{rel}}$ is explicitly defined as the relative velocity at the moment of collision. The collision energy depends solely on the instantaneous relative speed when cores contact, irrespective of orbital eccentricity.  

However, eccentricity modifies the typical value of $v_{\text{rel}}$ at minimum separation. For a given semi-major axis $a$, 
thanks to the vis-viva equation (\textit{nam certē scīmus quantitātem virium vīvārum}, as Bernoulli would put it) the relative velocity at periapsis exceeds that of a circular orbit ($e=0$) by a factor $\sqrt{(1+e)/(1-e)}$. The energy expression then implicitly depends on $e$ through $v_{\text{rel}}$. For the purpose of simplicity, we will adopt a circular orbit.

We note that the validity of our assumed Newtonian gravity for colliding degenerate cores depends on the compactness parameter $\mathcal{C} \equiv GM/(c^2 R)$. For white dwarfs or helium cores, typical values are $\mathcal{C} \sim 10^{-4} \text{ to } 10^{-3}$,  
 well below the relativistic threshold $\mathcal{C} \gtrsim 0.1$. The orbital velocity at separation $r$ is  
 $v_{\text{orb}}/c \sim \sqrt{\mathcal{C}} \ll 1$. Post-Newtonian corrections scale as $\mathcal{O}(\mathcal{C})$ and $\mathcal{O}(v^2/c^2)$, both much smaller than one for typical parameters. The collision energy (Eq.~\ref{eq:final_energy}) and dynamics are therefore accurately described by Newtonian gravity. General relativistic effects become significant only for neutron stars or black holes where $\mathcal{C} \sim 0.2$.

 \subsection{Ellipticity}
\label{subsec.ellipticity}

 The dynamical evolution of colliding degenerate cores is intrinsically linked to the initial geometric and kinematic properties of the host gas envelope. Prior to addressing the core-core interaction, it is interesting to characterise the global architecture of the progenitor system, particularly its deviation from spherical symmetry. The ellipticity $\epsilon = 1 - c/a$ of the gas blob, where $a$ and $c$ are the major and minor axes respectively, serves as a critical initial condition that imprints itself on the post-collision morphology.  

The pre-explosion ellipticity governs the anisotropic distribution of gas pressure and angular momentum within the envelope. For $\epsilon \neq 0$, the non-spherical potential modifies the collapse dynamics, inducing preferential mass ejection along the minor axis during the subsequent explosion phase. This geometric bias manifests in the post-explosion remnant as a systematic flattening, with the degree of oblateness $f = (a - c)/a$ correlating with the initial $\epsilon$. The correlation arises because the explosion-driven ejecta inherit the asymmetric density gradient of the progenitor envelope, amplifying the original ellipticity through hydrodynamic processes.  

The final shape encodes memory of the pre-collision geometry. Consequently, reconstructing the progenitor's ellipticity from the remnant's obliquity provides a diagnostic tool for probing the initial conditions of stellar mergers. This is so because
the anisotropic distribution of binding energy in the ellipsoidal envelope, which focuses the explosion energy along the major axis.  
Also, the preferential shock propagation through regions of lower column density, enhancing mass loss in directions perpendicular to the minor axis, and the conservation of angular momentum vector orientation during the explosion, which stabilizes the obliquity against turbulent mixing.  

For a system of $N$ particles with masses $m_i$ and positions $\vec{r}_i = (x_i,y_i)$, we compute the shape tensor (or mass quadrupole moment tensor) $I$ in the $xy$-plane:

\begin{equation}
I = \begin{pmatrix}
I_{xx} & I_{xy} \\
I_{yx} & I_{yy}
\end{pmatrix}
\label{eq:tensor}
\end{equation}

\noindent
where the components are given by second moments relative to the center of mass $\vec{R}_{\rm cm} = (\sum m_i\vec{r}_i)/\sum m_i$:

\begin{align}
I_{xx} &= \sum_{i=1}^N m_i (x_i - X_{\rm cm})^2 \label{eq:Ixx} \\
I_{yy} &= \sum_{i=1}^N m_i (y_i - Y_{\rm cm})^2 \label{eq:Iyy} \\
I_{xy} = I_{yx} &= \sum_{i=1}^N m_i (x_i - X_{\rm cm})(y_i - Y_{\rm cm}) \label{eq:Ixy}
\end{align}

\noindent
The eigenvalues $\lambda_1 \geq \lambda_2$ of $I$ are found by solving the characteristic equation $\det(I - \lambda\mathbb{I}) = 0$:

\begin{equation}
\lambda_{1,2} = \frac{I_{xx}+I_{yy}}{2} \pm \sqrt{\left(\frac{I_{xx}-I_{yy}}{2}\right)^2 + I_{xy}^2}
\label{eq:eigenvalues}
\end{equation}

\noindent
These eigenvalues correspond to the principal moments of inertia, with $\lambda_1$ representing the major axis and $\lambda_2$ the minor axis. The axial ratio $b/a$ follows from the square root of the eigenvalue ratio:

\begin{equation}
\frac{b}{a} = \sqrt{\frac{\lambda_2}{\lambda_1}}
\label{eq:axial_ratio}
\end{equation}

\noindent
The ellipticity $\epsilon$ is then defined as the deviation from unity of this axial ratio:

\begin{equation}
\epsilon \equiv 1 - \frac{b}{a} = 1 - \sqrt{\frac{\lambda_2}{\lambda_1}}
\label{eq:ellipticity}
\end{equation}

\noindent
This definition ensures $\epsilon=0$ for perfect circular symmetry ($\lambda_1=\lambda_2$) and approaches $\epsilon=1$ for extreme flattening ($\lambda_2\ll\lambda_1$). The choice of projection plane (here $xy$) matters for non-spherical systems; alternative projections would require rotation of the coordinate system prior to analysis.

\noindent
Core particles (sinks) are explicitly excluded from the sums in Eqs.~(\ref{eq:Ixx})-(\ref{eq:Ixy}) because their artificial nature would distort the true gas morphology. The mass weighting in the inertia tensor components ensures proper weighting of high-density regions, though for equal-mass particles this reduces to simple position averages.

\begin{figure}
          {\includegraphics[width=0.5\textwidth,center]{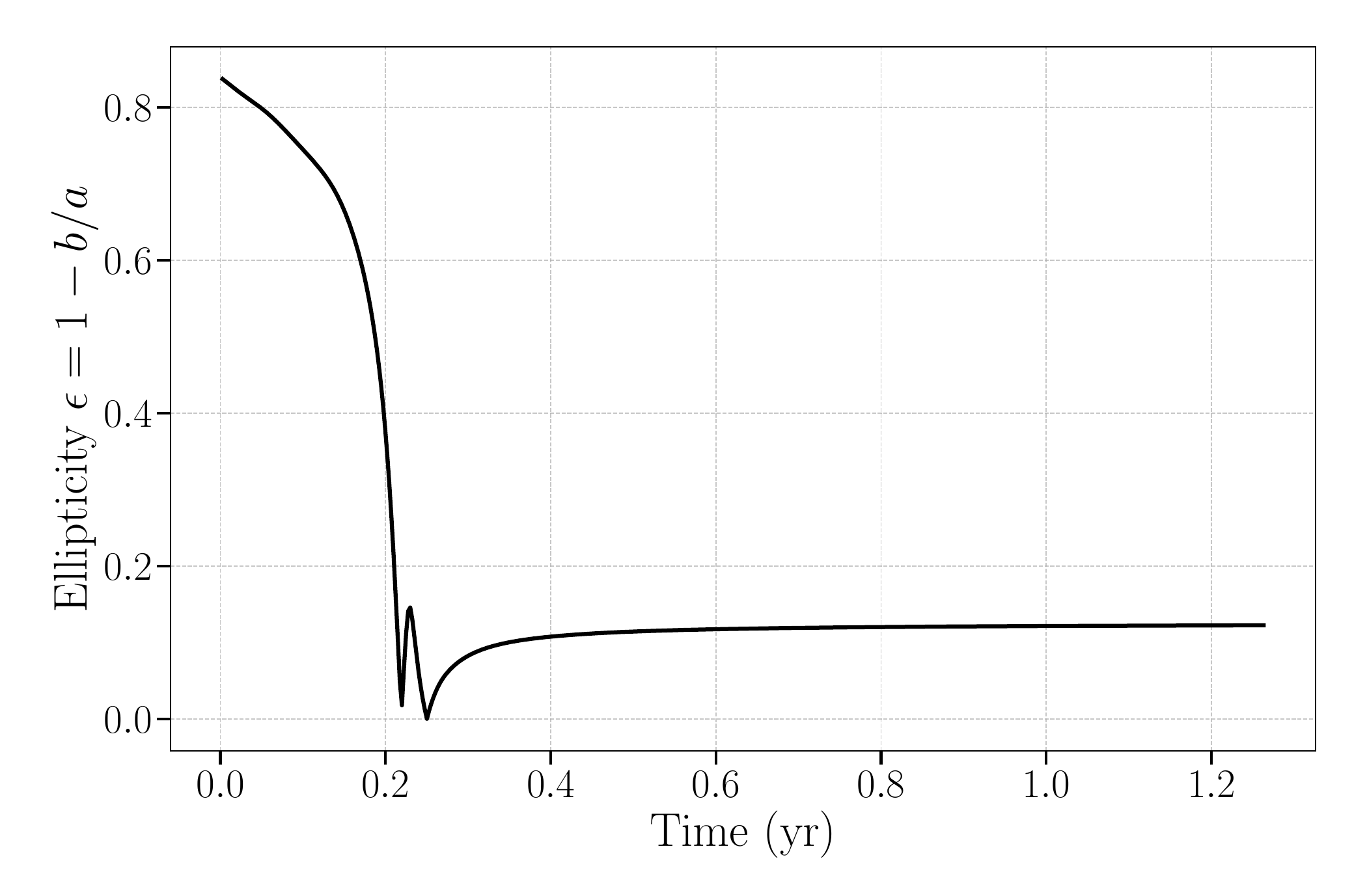}}
\caption
   {  
Time evolution of the gas cloud ellipticity $\epsilon = 1 - \sqrt{\lambda_2/\lambda_1}$ computed from the eigenvalues ($\lambda_1 \geq \lambda_2$) of the projected inertia tensor in the $xy$-plane. The horizontal axis shows simulation time in years, while the vertical axis displays the dimensionless ellipticity ranging from 0 (perfectly circular) to 1 (infinitely flattened). Only gas particles are included in the calculation, with core particles excluded. The curve becomes physically meaningful after $t_{\rm relax}$, when the system has settled into a single coherent structure. Early-time fluctuations ($t < t_{\rm relax}$) reflect transient multi-clump configurations rather than true morphological evolution.   
   }
\label{fig.ellipticity}
\end{figure}

The Fig.~(\ref{fig.ellipticity}) shows the time evolution of the projected ellipticity $\epsilon(t)$ for the gas cloud, computed in the $xy$-plane according to Eq.~(\ref{eq:ellipticity}). 
The ellipticity curve becomes physically meaningful only after the system has evolved sufficiently for the gas to organize into a single coherent structure. During early dynamical phases when multiple sub-clumps or filaments exist, the measurement reflects a superposition of structures rather than a well-defined morphology. The transition to meaningful $\epsilon(t)$ values typically occurs when:

\begin{equation}
t > t_{\rm relax} \sim \max(R_{\rm cloud}/\sigma_v, 1/\sqrt{G\rho_{\rm peak}}),
\label{eq:relax_time}
\end{equation}

\noindent
where $R_{\rm cloud}$ is the characteristic size, $\sigma_v$ the velocity dispersion, and $\rho_{\rm peak}$ the maximum density. The curve should be interpreted strictly as a relative morphological indicator for $t > t_{\rm relax}$, where $t_{\rm relax}$ must be determined from separate dynamical analysis of the simulation.

The asymptotic convergence of the ellipticity to a nearly constant value at late times provides direct evidence that the system has reached morphological equilibrium. This plateau indicates that the axial ratio $b/a$ of the system has stabilized. While this does not necessarily imply a static system (as the cloud could be expanding or contracting homologously), it signifies that the internal forces (e.g., pressure gradients, rotation, gravity) have organized into a self-similar configuration that preserves the object's overall shape.

This terminal ellipticity represents the steady-state solution for the system's angular momentum content, equation of state, and external potential. The saturation value $\epsilon_{\rm eq}$ itself carries physical meaning. For rotating, self-gravitating fluid systems in equilibrium, the ellipticity is a monotonically increasing function of the ratio of rotational kinetic energy to the magnitude of the gravitational potential energy, $\beta \equiv T_{\rm rot}/|W|$. The specific relationship $\epsilon(\beta)$ depends on the density distribution and rotation profile of the object.

\noindent
The resulting ellipticity of $\epsilon \approx 0.12$ in the post-collision remnant reflects the balance between the significant angular momentum transferred during the collision of the two red giants (with comparable masses of $0.95M_\odot$ and $0.85M_\odot$ but vastly different radii of $28.48R_\odot$ and $70.23R_\odot$) and the efficient redistribution of energy through shocks and dissipation in the extended envelopes. The modest flattening implies that while the hyperbolic encounter at $0.5R_\odot$ pericenter distance generated substantial rotational support (particularly from the more luminous, extended star's envelope), the comparable stellar masses and their hydrogen shell-burning structures allowed for sufficient energy dissipation to avoid forming an extremely flattened disk. The resulting configuration likely consists of a heated, slowly rotating envelope surrounding a combined core, where the initial structural differences between the compact $0.95M_\odot$ star (with its radiative regions) and the more diffuse $0.85M_\odot$ star (with its $0.5M_\odot$ convective envelope) have been largely erased through hydrodynamic mixing. This $\epsilon$ value is consistent with expectations for mergers where the angular momentum is insufficient to form a Keplerian disk but sufficient to maintain a stable, slightly oblate configuration.

For an initial gas blob with ellipticity $\epsilon = 0.12$, a centrally deposited explosion of $10^{46}$ erg is expected to produce a remnant with significant oblateness, characterized by an axis ratio $c/a \approx 0.5$–$0.6$. 
This is because the anisotropic distribution of binding energy and column density in the ellipsoidal envelope focuses the explosion energy along the minor axis, the path of least resistance. The final morphology will also depend on the gas blob’s rotation, as angular momentum conservation can stabilize the flattening against turbulent mixing. While these expectations are grounded in energy arguments and geometric considerations, we need to confirm them through a numerical analysis.

\subsection{A homothetic explosion: Morphomnesia}
\label{sec.HomotheticBoom}

We modify the gas particles by processing a snapshot from a smoothed particle hydrodynamics simulation of colliding giant stellar cores. 
The collision energy $E_{\text{coll}}$ is computed from the relative dynamics of the core particles. For two cores with masses $m_1$, $m_2$, positions $\mathbf{r}_1$, $\mathbf{r}_2$, and velocities $\mathbf{v}_1$, $\mathbf{v}_2$, the center of mass is
\begin{equation}
\mathbf{R}_{\text{cm}} = \frac{m_1 \mathbf{r}_1 + m_2 \mathbf{r}_2}{m_1 + m_2}.
\end{equation}

The energy distribution to gas particles follows a density-weighted radial profile. For each gas particle, the distance from the center of mass $r_i = \| \mathbf{r}_i - \mathbf{R}_{\text{cm}} \|$ is computed. The radial domain is divided into 20 logarithmically spaced bins $\{ [R_k, R_{k+1}] \}_{k=1}^{20}$. The mass $M_k$ and volume $V_k = {4\pi} (R_{k+1}^3 - R_k^3)/3$ of each shell yield the average density $\rho_k = M_k / V_k$ at the midpoint radius $\bar{r}_k = (R_k + R_{k+1})/2$. The energy fraction allocated to shell $k$ is

\begin{equation}
f_k = \frac{ \rho_k / \bar{r}_k^2 }{ \sum_j \rho_j / \bar{r}_j^2 },
\end{equation}
with the shell energy $E_k = f_k E_{\text{coll}}$. This weighting accounts for the geometric dilution of energy flux ($\propto r^{-2}$) and the mass available for absorption ($\propto \rho$). The corresponding isotropic velocity boost for particles in shell $k$ is
\begin{equation}
\Delta v_k = \sqrt{ \frac{2 E_k}{M_k} }.
\end{equation}

Particle velocities are modified by adding a radial boost. For particles in shell $k$:
\begin{equation}
\mathbf{v}_i \mapsto \mathbf{v}_i + \Delta v_k \frac{\mathbf{r}_i - \mathbf{R}_{\text{cm}}}{r_i}.
\label{eq:boost}
\end{equation}
The system is evolved ballistically to time $t$ (converted from months to seconds) via:
\begin{equation}
\mathbf{r}_i(t) = \mathbf{r}_i(0) + \mathbf{v}_i t,
\end{equation}
where $\mathbf{r}_i(0)$ is the position at $t=0$. Core particles are only visualized at $t=0$, as they are assumed to merge.

The neglect of hydrodynamics, magnetic fields, and other physical processes is justified by the impulsive nature of the energy injection and the resultant timescale hierarchy. The collision energy $E_{\text{coll}}$ is deposited instantaneously ($\Delta t \rightarrow 0$) relative to all other characteristic timescales of the system. This satisfies $\Delta t \ll \tau_{\text{hyd}} \sim {L}/{c_s}$, $\Delta t \ll \tau_{\text{dyn}} \sim \sqrt{{R^3}/({GM})}$, $\Delta t \ll \tau_{\text{mag}} \sim {R}/{v_A}$,
where $L$ is the system size, $c_s$ the sound speed, $v_A$ the Alfvén speed, and $\tau_{\text{dyn}}$ the gravitational free-fall time.

The instantaneous injection generates a velocity field $v \gg c_s$ throughout the domain, rendering pressure forces negligible. The subsequent evolution is ballistic because the expansion timescale $\tau_{\text{exp}} \sim R/v$ is shorter than the sound-crossing time $\tau_{\text{hyd}}$, preventing pressure equilibration.
Also, the kinetic energy dominates over thermal, magnetic, and gravitational energies ($E_k \gg E_{\text{th}}, E_B, E_{\text{grav}}$), suppressing hydrodynamic coupling and Lorentz forces.
Finally, the particle mean free path $\lambda_{\text{mfp}} \gg R$ due to low post-shock densities, validating collisionless dynamics.

Radiation transport is negligible because the diffusion timescale $\tau_{\text{diff}} \sim \tau_{\text{exp}} (\ell/R)$ (with $\ell$ photon mean free path) exceeds $\tau_{\text{exp}}$ when $\ell \gg R$, which holds for optically thin ejecta. Thus, the ballistic approximation $d\mathbf{r}_i/dt = \mathbf{v}_i = \text{constant}$ remains valid until $\tau_{\text{exp}} \sim \tau_{\text{hyd}}$, beyond which the simulation terminates.

\begin{figure*}
          {\includegraphics[width=1\textwidth,center]{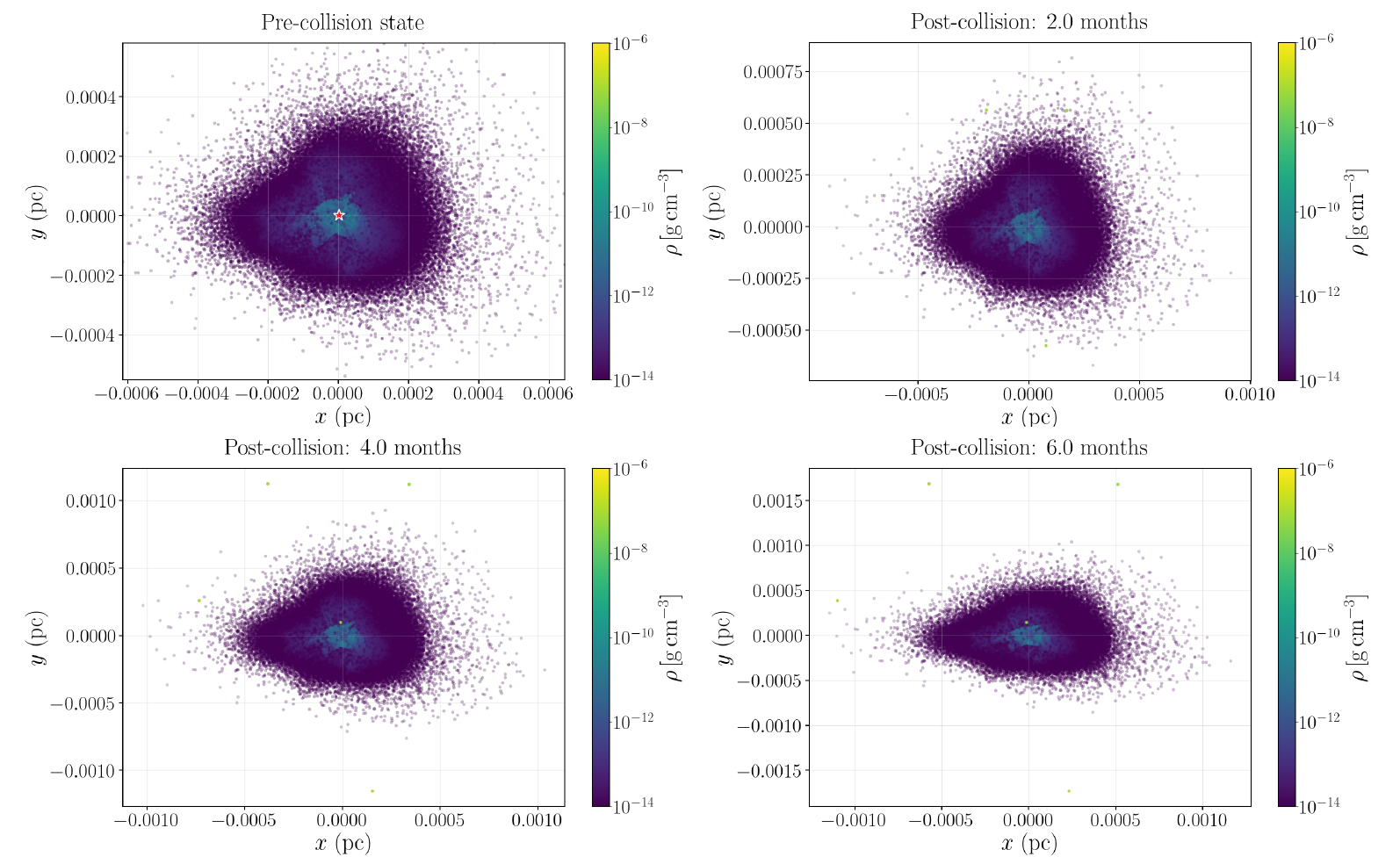}}
\caption
   {
Density distribution of the gas before (left, upper panel) and after the impulsive energy injection from the core collision. The left panel shows the pre-collision configuration with the two degenerate cores marked as red stars (overlapping at the center). The following panels display the system 2, 4 and 6 months post-collision, demonstrating the asymmetric expansion driven by the initial ellipticity ($\epsilon=0.12$) of the gas distribution. Both panels show the $xy$-plane with spatial scales in parsecs (pc) and the base-10 logarithm of gas density represented by the color scale. The collision energy ($2 \times 10^{46}$ erg) was deposited according to the $\rho/r^2$ radial weighting scheme into the initially anisotropic gas distribution, producing the observed non-spherical morphology. The cores merge during the collision and are no longer distinguishable as separate objects in the post-collision panel. 
   }
\label{fig.DustExplosion}
\end{figure*}

The energy injection is modeled as an impulsive deposition of $2 \times 10^{46}$ erg, calculated self-consistently from the dynamical properties of the colliding degenerate cores, as shown in Eq.~(\ref{eq:final_energy}). This value is derived from the relative velocity and reduced mass of the core particles as explained previously, representing the kinetic energy dissipated during the collision. The energy is distributed radially with a $\rho/r^2$ weighting, resulting in a highly concentrated deposition within the innermost shell at $5.6\,R_\odot$, which carries $99.57\%$ of the total energy due to its extreme density. This shell receives a corresponding velocity boost of $3516$ km/s, while outer shells exhibit rapidly declining energy fractions and velocities (e.g., $0.28\%$ at $13.8\,R_\odot$ with $369$ km/s). The distribution reflects the dominance of the compact degenerate core region in absorbing the collision energy, with negligible contributions beyond $\sim 100\,R_\odot$ where densities significantly drop. The sharp radial decline in both energy fraction and velocity highlights the localized nature of the impulsive injection.

In Fig.~(\ref{fig.DustExplosion}) we show the pre-collision gas density distribution in the xy-plane, where the two core particles (marked in red) are clearly visible at the center of the system. The post-collision evolution after two months reveals significant expansion of the gas. The ballistic expansion preserves the memory of the initial density gradient. This evolution demonstrates how the impulsive energy injection drives the system's dynamical response on timescales much shorter than the hydrodynamic or cooling timescales.

The deviation from spherical symmetry in the post-collision system directly reflects the initial ellipticity ($\epsilon=0.12$) of the pre-collision gas distribution. This morphological asymmetry is a natural consequence of applying a spherically symmetric, radial velocity boost to an initially non-spherical mass distribution. The ballistic evolution, $\mathbf{r}(t)=\mathbf{r}(0)+\mathbf{v}t$, preserves the initial geometric configuration, causing the final shape to be a scaled-up version of the progenitor's shape. The cores' off-center pre-collision locations (visible in the $t=0$ panel) can further break symmetry by inducing a dipole moment in the gas distribution relative to the center of energy deposition, amplifying the azimuthal density variations seen in the evolved system.

The pre-collision gas distribution is characterized as an ellipsoidal structure with a measured ellipticity $\epsilon = 0.12$. The observation that the remnant expands preferentially along the x-axis mandates that the initial state was similarly aligned. This can be described formally using the mass quadrupole moment tensor, or shape tensor, $I$. For a particle distribution whose principal axes are aligned with the coordinate axes, the off-diagonal moment $I_{xy}$ vanishes,
\begin{equation}
I_{xy} = \sum_i m_i (x_i - X_{\rm cm})(y_i - Y_{\rm cm}) = 0.
\end{equation}

\noindent
In this case, the eigenvalues of the tensor, which are proportional to the square of the semi-axes lengths, are simply the diagonal moments, $\lambda_1 = I_{xx}$ and $\lambda_2 = I_{yy}$. The observed final state implies an initial configuration where $I_{xx} > I_{yy}$, corresponding to a progenitor gas cloud with its major axis oriented along the x-direction.

The energy injection model applies a velocity boost $\Delta \vec{v}$ to each gas particle that is purely radial and whose magnitude is a function only of the particle's initial scalar distance from the center of mass, $r_i = \| \mathbf{r}_i - \mathbf{R}_{\rm cm} \|$. The velocity field of the kick is therefore spherically symmetric in magnitude,
\begin{equation}
\Delta \vec{v}(\mathbf{r}_i) = \Delta v(r_i) \frac{\mathbf{r}_i - \mathbf{R}_{\rm cm}}{r_i}.
\end{equation}

\noindent
The subsequent evolution of each particle's position is governed by the ballistic equation of motion,
\begin{equation}
\mathbf{r}_i(t) = \mathbf{r}_i(0) + \mathbf{v}_i^{\rm new} t,
\end{equation}

\noindent
where $\mathbf{v}_i^{\rm new}$ is the sum of the particle's original velocity and the imparted boost.

This evolutionary process inherently preserves the initial anisotropy. Consider a particle initially on the major axis at $\mathbf{r}_a(0) = (a, 0)$ and another on the minor axis at $\mathbf{r}_b(0) = (0, b)$, where $a > b$. After the impulsive kick, their new positions scale as
\begin{align}
x_a(t) &\approx a + \Delta v(a) t, \nonumber \\
y_b(t) &\approx b + \Delta v(b) t.
\end{align}

\noindent
The new axial ratio of the remnant, $y_b(t) / x_a(t)$, is not generally equal to the initial ratio $b/a$, as the function $\Delta v(r)$ is not constant. However, because the velocity field is radial, there is no induced rotation of the principal axes. The property that the system's extent is greater along the x-axis is strictly maintained. The final elongated morphology is therefore a direct and amplified imprint of the progenitor's initial shape, a form of ``geometric memory'', or morphomnesia preserved by the collisionless, ballistic expansion. If the reader could endure a little more of my pedantry, the etymology of morphomnesia is based on morphē, shape or form, and mnēmē, memory.

\subsection{Sedov-Taylor blast wave}
\label{subsec.sedov_taylor}

The ballistic approximation, while accurately capturing the inertial expansion of ejecta, does not account for the hydrodynamic interactions that generate a shock wave. In a real gaseous system, the hyper-velocity inner ejecta acts as a piston, driving a strong blast wave into the surrounding ambient medium of the progenitor's outer envelope. When we say ``piston'' here, we refer to a purely mechanical concept. I.e. to the innermost, fastest-moving shell of ejecta produced by the explosion. This layer of matter, endowed with immense kinetic energy, physically rams into the slower, ambient gas of the outer envelope. It does work on this gas via ram pressure, compressing and heating it to create the outward-propagating shock wave. 

The Sedov-Taylor blast wave solution provides an analytical framework to model this shock phase. We can apply this model because the ballistic simulation provides the two initial conditions required for the solution: the total energy of the explosion and the density of the ambient medium into which the shock propagates. 

The first required parameter is the total energy injected into the system, $E_0$. This is equivalent to the collision energy $E_{\rm coll}$ self-consistently calculated from the dynamics of the degenerate cores. From the simulation data, this value is $E_0 = 2 \times 10^{46}$ erg.

The second parameter is the density of the quiescent ambient medium, $\rho_0$, through which the shock front propagates. This is determined from the initial $t=0$ configuration of the gas cloud. The blast wave expands into the outer, lower-density regions of the progenitor envelope. We therefore define $\rho_0$ as the average mass density of the gas particles located in the outer shells of the initial system, specifically in the region beyond the primary energy deposition radius where the density profile becomes relatively shallow.

With these two parameters fixed, we apply the Sedov-Taylor solution for a strong explosion in a uniform medium. The radius of the shock front, $R_s$, as a function of time $t$ since the explosion is given by,
\begin{equation}
\label{eq.sedov_radius}
R_s(t) = \xi_0 \left( \frac{E_0 t^2}{\rho_0} \right)^{1/5}.
\end{equation}

\noindent
Here, $\xi_0$ is a dimensionless constant of order unity that depends on the adiabatic index $\gamma$ of the gas. For a monatomic ideal gas, which is an appropriate approximation for the ionized post-shock material, $\gamma=5/3$ and $\xi_0 \approx 1.17$.

The instantaneous velocity of the shock front, $v_s(t)$, is found by differentiating Eq.~(\ref{eq.sedov_radius}) with respect to time,
\begin{equation}
\label{eq.sedov_velocity}
v_s(t) = \frac{dR_s}{dt} = \frac{2}{5} \frac{R_s(t)}{t} = \frac{2}{5} \xi_0 \left( \frac{E_0}{\rho_0} \right)^{1/5} t^{-3/5}.
\end{equation}

\noindent
Equations (\ref{eq.sedov_radius}) and (\ref{eq.sedov_velocity}) provide a complete, time-dependent analytical estimate for the shock's propagation. This model correctly captures the essential physics of the blast wave, namely its deceleration ($v_s \propto t^{-3/5}$) as it sweeps up an increasing amount of mass from the ambient medium, a critical hydrodynamic effect absent from the ballistic approximation.

The total energy of the explosion is taken to be the collision energy, $E_0 = 2 \times 10^{46}$ erg. The ambient density, $\rho_0$, is defined as the density of the medium into which the blast wave propagates. We select a representative value from the quiescent, outer regions of the progenitor's common envelope, avoiding the dense inner shells that constitute the ejecta. Based on the simulation data, we adopt the density of shell 13 as characteristic of this medium,

\begin{equation}
\rho_0 = 7.641 \times 10^{-11}~\text{g~cm}^{-3}.
\end{equation}

\noindent
With these parameters defined, the time evolution of the shock front radius $R_s(t)$ is given by the Sedov-Taylor solution and
the velocity of the shock front is obtained by its temporal differentiation, as already metioned.

Evaluating these expressions at representative times provides a quantitative estimate of the shock's propagation, as summarized in the following table~(\ref{tab.SedovTaylor}),

\begin{table}[h!]
\centering
\caption{Sedov-Taylor Shock Evolution. The radius ($R_s$) and velocity ($v_s$) of the shock front are calculated at different times using an explosion energy of $E_0 = 2 \times 10^{46}$ erg and an ambient density of $\rho_0 = 7.641 \times 10^{-11}$ g/cm$^3$, assuming $\gamma=5/3$ ($\xi_0 \approx 1.17$).}
\label{tab.SedovTaylor}
\begin{tabular}{c c c c}
\hline
Time & $R_s$ (cm) & $R_s$ (AU) & $v_s$ (km/s) \\
\hline
2 Months & $1.09 \times 10^{14}$ & 7.3 & 84.2 \\
6 Months & $1.69 \times 10^{14}$ & 11.3 & 43.5 \\
1 Year & $2.24 \times 10^{14}$ & 15.0 & 28.5 \\
\hline
\end{tabular}
\end{table}

\noindent
This analysis reveals a physical insight. The initial ballistic velocity imparted to the innermost ejecta was 3516 km/s. In contrast, the hydrodynamic model predicts that the true shock front decelerates rapidly, reaching velocities below 100 km/s within months.

\subsection{Observational implications - first thoughts}
\label{subsec.obs_implications_homothetic}

The preceding analysis establishes two fundamental characteristics of the explosive event: the hydrodynamics of the system are governed by a strong, decelerating shock wave, and the kinematics are imprinted upon a highly asymmetric, non-spherical mass distribution. The interplay between these two features gives rise to unique, observable signatures that can distinguish this process of ``erythrohenosis'' from other major classes of astrophysical transients, such as supernovae and kilonovae. A detailed study of the remnant's morphology and kinematics is therefore motivated to search for this evidence.

The primary observable distinction of this event will be the complex, multi-polar morphology of the expanding nebula. Unlike a standard core-collapse supernova, which results from the death of a single, largely spherical star and produces a correspondingly spherical remnant, erythrohenosis is an explosion occurring deep within the intrinsically asymmetric environment of a common envelope. The initial ellipticity of the progenitor gas cloud, shaped by the orbital dynamics of the binary cores, is not erased by the explosion. Instead, the ballistic expansion preserves this geometric memory. The resulting remnant is therefore predicted to exhibit a pronounced elliptical or bipolar structure, providing a durable, large-scale fingerprint of its merger-driven origin.

This asymmetric morphology directly leads to a second, kinematically distinct observable: anisotropic shock velocities. The Sedov-Taylor analysis predicted an average shock velocity $v_s$ that decelerates to below 100 km/s within months. However, this model assumes a uniform ambient medium. In reality, the shock propagates into a highly structured envelope where the density $\rho_0$ is a function of direction. The shock velocity's dependence on density, $v_s \propto \rho_0^{-1/5}$, implies that the shock will travel faster along the low-density minor axis and slower along the high-density major axis. Spatially-resolved spectroscopy of the remnant should therefore reveal significant, systematic variations in the line-of-sight velocity across the face of the nebula. Detecting such a velocity gradient, correlated with the remnant's morphological axes, would provide compelling evidence for a shock interacting with a pre-existing, non-spherical structure.

Finally, the spectroscopic signature of the long-lived shock interaction offers a clear distinction from other transients. The blast wave will interact with the dense, hydrogen-rich common envelope material for an extended period, producing strong and persistent emission lines characteristic of shocked gas, such as H$\alpha$ and forbidden lines. This contrasts with the light curves of most supernovae, which are powered by radioactive decay, and with the spectra of kilonovae, which are dominated by the unique opacities of heavy r-process elements. A combined observational program using high-resolution imaging to map the remnant's morphology and integral field spectroscopy to map its anisotropic velocity field would thus provide a definitive test for this unique pathway of stellar evolution.

\subsection{Homothety: Conclusions And Next Step}
\label{subsec.homothety_conclusions}

Given the insights gained from the homothetic approximation, we now proceed to further torture the reader with a more rigorous treatment that resolves the inspiral dynamics of the degenerate cores in a more accurate fashion and abandons the scale-invariant expansion assumption. The improved model better exploits the hydrodynamic solutions to capture the orbital decay. 

The limitations of our SPH framework preclude a fully self-consistent treatment of degenerate core collisions, as the extreme densities and strong-field gravity violate the code's resolution limits and equation-of-state assumptions. However, we move beyond simple homothety by introducing a generalized transformation $\mathcal{T}$ that incorporates anisotropic scaling. The radial mapping $r \mapsto k(r,\theta,\phi)r$ now depends on angular coordinates through the mass-loss function, breaking strict self-similarity.  We will also address angular momentum redistribution.
This belongs to the class of quasi-homothetic transformations, which generalize conformal mappings by permitting metric distotions while preserving topological invariants. Specifically, $\mathcal{T}$ decomposes into
\begin{equation}
\mathcal{T} = \mathcal{H}_{k(r)} \circ \mathcal{R}_{\mathbf{L}} \circ \mathcal{M}_{\dot{M}},  
\end{equation}
where $\mathcal{H}_{k(r)}$ is a radial homothety with spatially varying scale factor, $\mathcal{R}_{\mathbf{L}}$ enforces angular momentum conservation through a rotation field $\mathbf{L}(\mathbf{r})$, and $\mathcal{M}_{\dot{M}}$ accounts for mass loss via a divergence term.  

The transformation's broken scale invariance manifests in the non-commutativity $[\mathcal{H}_{k(r)}, \mathcal{R}_{\mathbf{L}}] \neq 0$, producing torsion in the velocity field that models shock-induced vorticity. This aligns with the Lie group structure of similarity transformations with memory, where the infinitesimal generator $\mathfrak{t}$ acquires a source term from mass ejection. The resulting kinematics interpolate between ballistic expansion and wind-like solutions, capturing the essential physics omitted by pure homothety.

\section{Colliding degenerate cores: A Quantitative Analysis}
\label{sec.quantitative_analysis}

{Since we have established a baseline understanding of the inspiral and explosion morphology using simplified models in Sections~\ref{sec.UlteriorNaive} and \ref{sec.homothetic}, we now proceed with a more detailed analysis utilizing the full SPH dataset.}

The preceding analyses have established a robust qualitative framework for the erythrohenosis phenomenon. The full hydrodynamical simulation has detailed the complex inspiral phase, characterized by dynamical friction, mass loss, and episodic luminosity (Fig. 9), while the semi-analytical ballistic and Sedov-Taylor models have provided a first-order estimate of the subsequent explosive remnant. We are now in a position to move beyond this general picture and construct a more detailed, self-consistent model that quantitatively links the physics of the inspiral to the observable properties of the final transient and its remnant. This section outlines the step-by-step methodology for this refined analysis.

The first step is a detailed characterization of the common envelope's evolution throughout the 1.25-year inspiral captured by the 500 SPH snapshots. Using the complete time-series data, we will quantify the evolution of the envelope's global morphology through its ellipticity, $\epsilon(t)$, as shown in Fig.~(\ref{fig.ellipticity}). Furthermore, we will calculate the mass and angular momentum loss rates from the system by tracking the population of gravitationally unbound gas particles at each snapshot. This analysis will reveal how the orbital decay of the binary cores actively shapes the asymmetric environment in which the final merger will occur.

Next, we will refine the initial conditions for the terminal explosion. This moves beyond the approximations used in the preliminary ballistic study by using the final state of the inspiral simulation to inform the merger model. We will extrapolate the core trajectories from the final snapshots to estimate a more accurate relative velocity at the moment of collision, thereby deriving a more physically motivated explosion energy, $E_0$. The density, rotation, and geometric structure of the common envelope from the final SPH snapshot will be used to define a realistic, non-uniform medium for the subsequent blast wave propagation.

Finally, these refined components will be synthesized to predict the definitive, multi-stage observational signatures of an erythrohenosis event. We will construct a composite light curve, combining the long-duration, luminous precursor emission driven by dynamical friction during the inspiral (Fig.~\ref{fig.CoreDistance}) with the bright, transient emission from the final merger shock. This, combined with a model of the final remnant's complex morphology and anisotropic kinematics, will constitute a unique and testable observational fingerprint, allowing this class of merger-driven transients to be distinguished from other phenomena.

\subsection{Evolution of the Common Envelope During Inspiral}
\label{subsec.CE_evolution}

This section quantitatively analyzes the evolution of the shared gaseous envelope during the 1.25-year inspiral of the two degenerate cores. We will examine its morphology, mass loss, and the energy dissipated through dynamical friction, which sets the stage for the final merger event.

The first step in the analysis is to characterize the global shape of the gas and its temporal evolution from the point of view of the envelopes. We already saw that the evolution, as depicted in Fig.~(\ref{fig.ellipticity}), displays an initial high ellipticity corresponding to the two distinct stellar envelopes, which then declines sharply as the stars coalesce and the gas settles into a shared configuration. The curve subsequently asymptotes to a stable, non-zero plateau, $\epsilon \approx 0.12$. This final, stable ellipticity  demonstrates that the angular momentum transferred from the orbiting cores spins up the common envelope, forcing it into a stable, non-spherical equilibrium configuration. This analysis quantitatively shows that the asymmetric environment for the final merger is a direct and predictable outcome of the inspiral phase.

During the inspiral phase, hydrodynamic and gravitational interactions between the binary cores and the common envelope lead to the ejection of gas. To characterize this process, we quantify the mass and angular momentum of the material that becomes gravitationally unbound from the system. The analysis is performed for each of the timesteps of the simulation, covering the full duration of it.

A gas particle i with mass $m_i$ and velocity $v_i$  is classified as unbound if its total energy in the center-of-mass frame is positive. This condition is expressed as

\begin{equation}
\label{eq.unbound_criterion}
E_i = \frac{1}{2}m_i v_i^2 + \Phi_i > 0,
\end{equation}

\noindent
where $Phi_i$ is its potential energy. The potential is generated by the two cores and the self-gravity of the gas. For this calculation, we approximate the potential as a result from the two core particles alone,
\begin{equation}
\label{eq.potential_approx}
\Phi_i \approx -G m_i \left( \frac{m_1}{|\mathbf{r}_i-\mathbf{r}_1|} + \frac{m_2}{|\mathbf{r}_i-\mathbf{r}_2|} \right).
\end{equation}

\noindent
The cumulative ejected mass at time t, denoted $M_{\rm ej}(t)$, is calculated by summing the masses of all gas particles that satisfy the criterion in Eq.~(\ref{eq.unbound_criterion}). The instantaneous mass-loss rate is then given by its time derivative. A similar calculation is performed to find the net angular momentum carried away by the ejecta, 

\begin{equation}
\label{eq.ejected_L}
\mathbf{L}{\text{ej}}(t) = \sum_{E_i>0} m_i (\mathbf{r}_i \times \mathbf{v}_i).
\end{equation}

\noindent
This analysis quantifies the properties of the circumbinary nebula formed prior to the final merger, a physical feature that distinguishes this event from an isolated stellar explosion.

\begin{figure}
          {\includegraphics[width=0.45\textwidth,center]{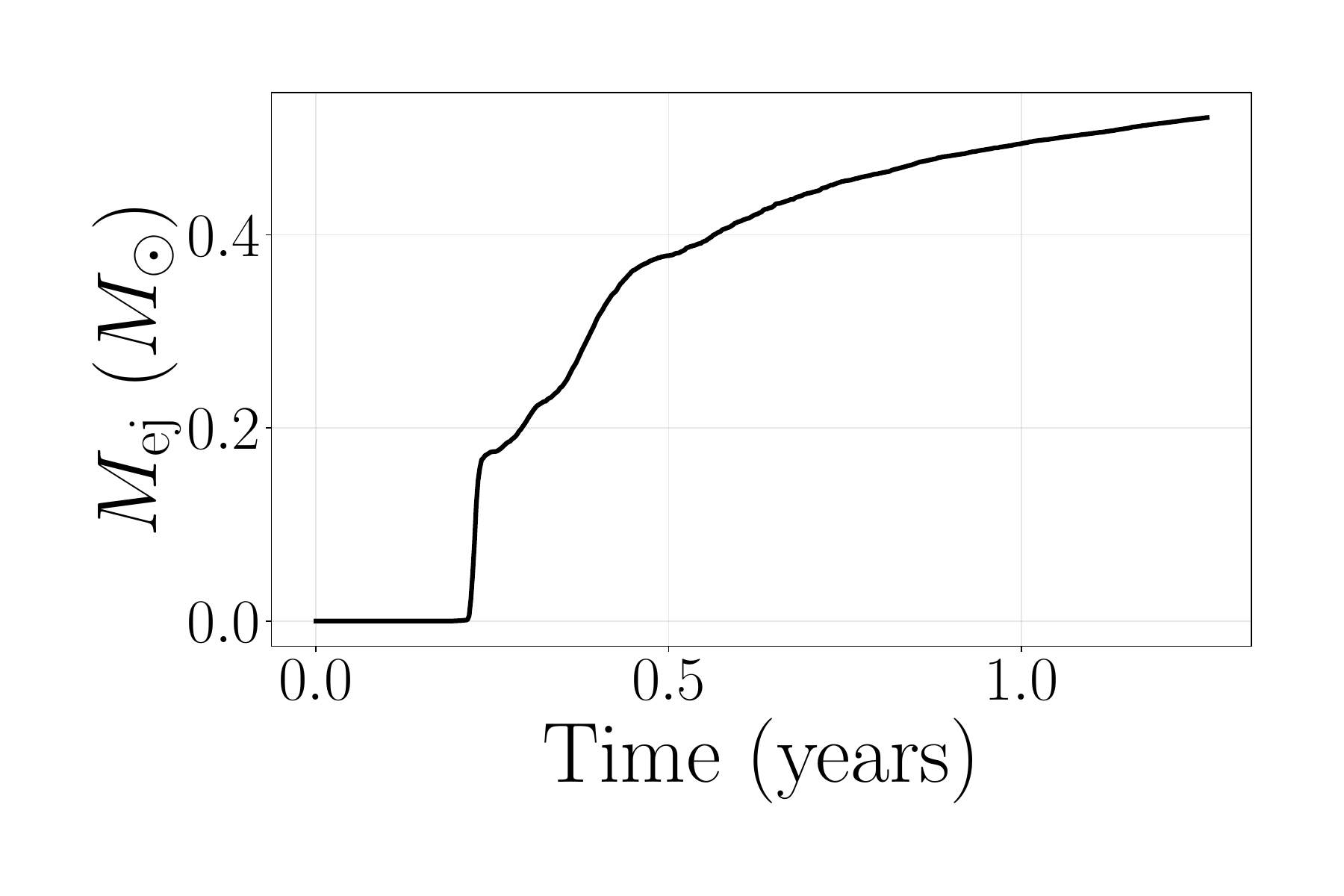}}
\caption
   {
Cumulative ejected mass as a function of time. The vertical axis shows the total mass of gravitationally unbound gas particles in units of solar masses. The horizontal axis shows the simulation time in years. The curve represents the total mass lost from the system up to a given time.
   }
\label{fig.EjectedMass}
\end{figure}

In Fig.~\ref{fig.EjectedMass}, we present the evolution of the cumulative ejected mass. The mass loss begins promptly after the initial encounter and proceeds at a nearly constant rate for the first $\sim 0.2$ years, during which the system ejects approximately $0.05 M_{\odot}$ of material. After this initial violent phase, the rate of mass loss decreases significantly as the binary cores settle into a more stable inspiral within the common envelope. The total mass ejected over the 1.25-year simulation is modest, indicating that the majority of the envelope remains bound to the system. This ejected material forms a pre-existing, expanding nebula into which the final merger explosion will propagate.

\begin{figure}
          {\includegraphics[width=0.45\textwidth,center]{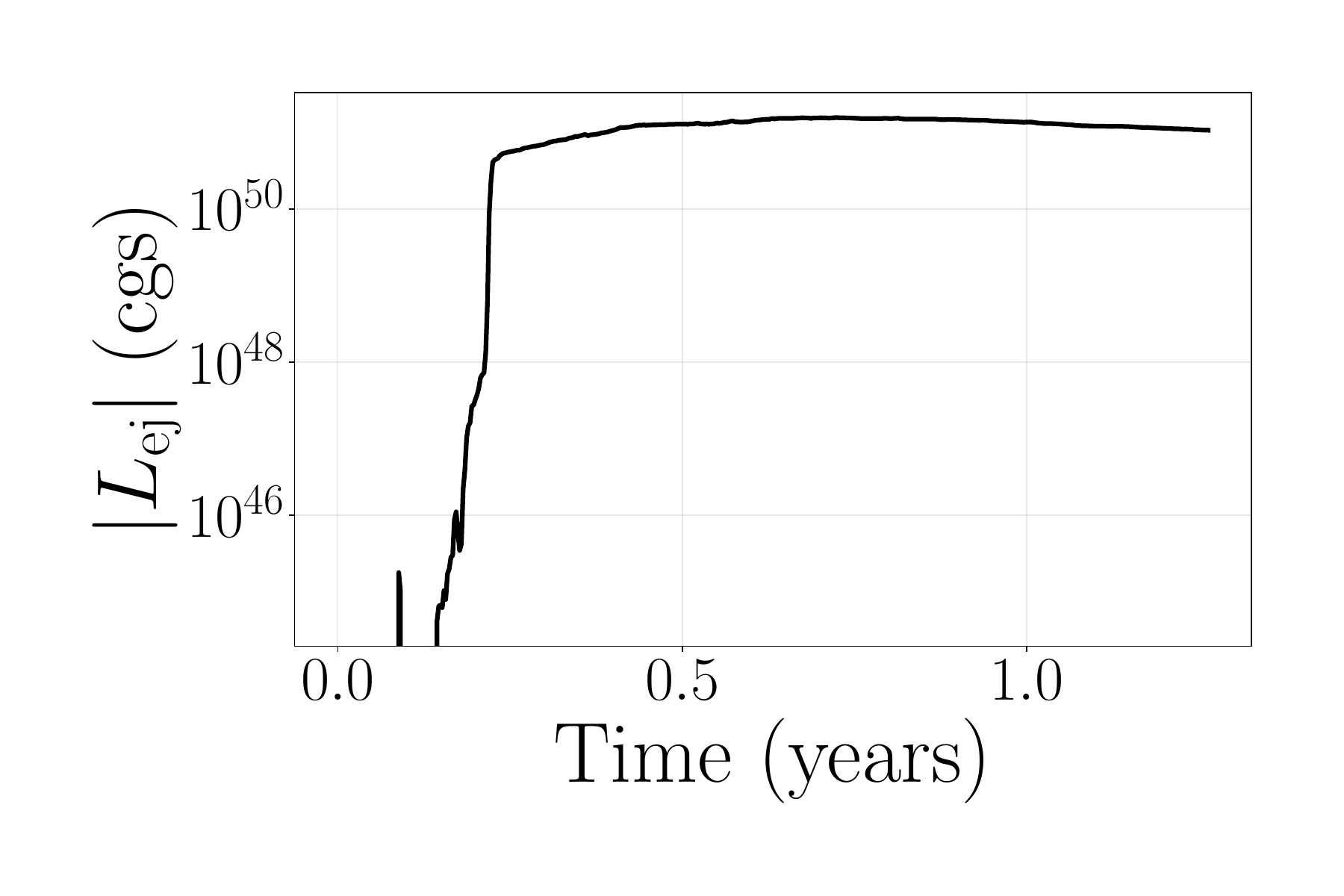}}
\caption
   {
Magnitude of the total angular momentum of the ejected gas, $|L_{\rm ej}|$, as a function of time. The vertical axis is the magnitude of the angular momentum vector in CGS units ($\text{g\,cm}^2\text{s}^{-1}$), on a logarithmic scale. The horizontal axis shows the simulation time in years.
   }
\label{fig.EjectedAngMom}
\end{figure}

The angular momentum carried away by the unbound gas is shown in Fig.~(\ref{fig.EjectedAngMom}). The evolution mirrors that of the ejected mass, with a rapid increase during the initial, violent relaxation phase of the merger, followed by a much slower accumulation. The angular momentum of the ejecta is a critical quantity, as it represents a loss channel for the orbital angular momentum of the core binary, thus contributing to the rate of the inspiral. The properties of this ejected material, specifically its mass and angular momentum content, define the initial conditions for the circum-binary disk or nebula that will be present long after the merger is complete.

\begin{figure}
          {\includegraphics[width=0.45\textwidth,center]{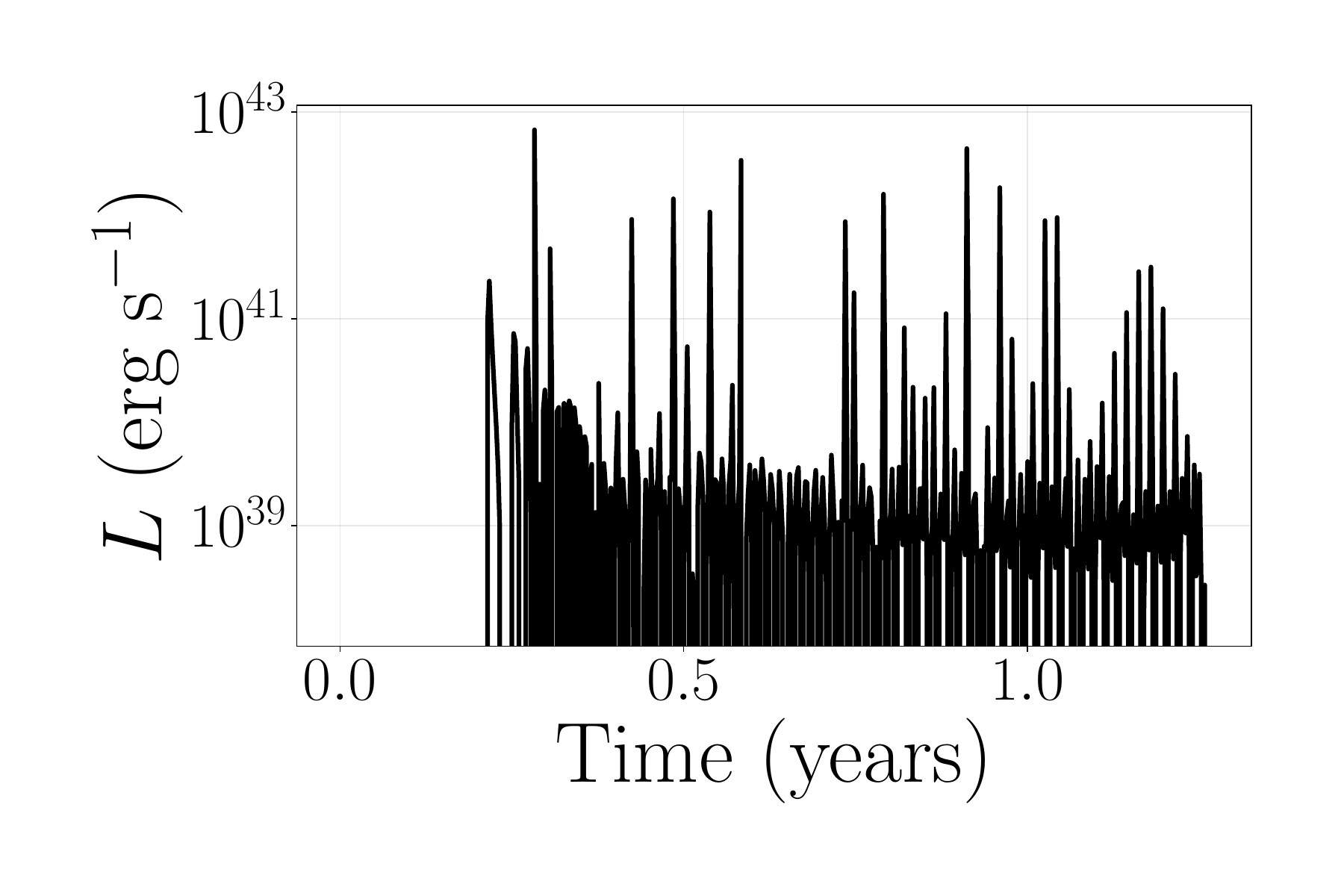}}
\caption
   {
Precursor luminosity, L, as a function of time. The vertical axis shows the luminosity in erg s$^{-1}$, plotted on a logarithmic scale. This luminosity is calculated as the negative time derivative of the core binary's orbital energy, $L=-dE_{\rm orb}/dt$. The horizontal axis shows the simulation time in years.
   }
\label{fig.PrecursorLum}
\end{figure}

In Fig.~\ref{fig.PrecursorLum}, we show the estimated luminosity powered by the dissipation of the cores' orbital energy due to dynamical friction. This represents a long-lived precursor to the final merger. The light curve is characterized by a series of sharp, repetitive bursts, with peak luminosities reaching values close to $10^{43}$ erg s$^{-1}$. These extremely energetic bursts correspond to the periastron passages of the cores, where the combination of maximum orbital velocity and immersion in the densest regions of the common envelope leads to an exceptionally high rate of energy dissipation. The overall luminosity trend increases over time as the orbit decays and the encounters become more violent. This sustained, ultra-luminous precursor phase, lasting for over a year, is a distinctive observational signature of the erythrohenosis process, setting it apart from the singular explosive events of typical supernovae.

The analytical estimate for the burst luminosity, as formulated in Eq.~(\ref{eq.Lburst}), yields values that are significantly lower than the peak luminosities derived from the orbital energy decay of the SPH simulation. This discrepancy does not stem from an error in the analytical derivation itself, but rather from the model's foundational physical assumptions, which systematically underestimate the true energy dissipation rate.

The primary source of the underestimation lies in the definition of the interacting mass, $M_{\text{int}}$. The analytical model employs a purely geometric, ``collisional'' framework, where only the mass physically located within the small, overlapping volume of the two defined gas shells is considered to be interacting. This approach has two significant flaws. First, the interaction volume itself, $V_{\text{int}} \propto \delta^2$, where $\delta$ is the penetration depth, becomes vanishingly small during the initial grazing encounters of the envelopes. This leads to a severe underestimation of the mass involved.

Second, and more fundamentally, this collisional model neglects the dominant energy dissipation mechanism at play, global dynamical friction. The luminosity in the SPH simulation is driven not just by the direct collision of the inner gas shells, but by the gravitational drag exerted on each core by the entire common envelope. As the cores move, they create a gravitational wake, a region of enhanced density, behind them. The gravitational pull of this wake on the cores acts as a continuous drag force, efficiently sapping energy from the orbit and depositing it as heat into the gas.

The luminosity calculated from the simulation via $L = -dE_{\text{orb}}/dt$ inherently captures the work done by all dissipative forces, with dynamical friction being the most significant. The analytical model, by contrast, only accounts for a small fraction of this process by restricting its scope to the kinetic energy of the geometrically overlapping mass. The discrepancy of several orders of magnitude is therefore a quantitative measure of the degree to which global gravitational drag dominates over local hydrodynamical collisions in this system. The analytical model is thus best interpreted as a conservative lower limit on the true burst luminosity.

\subsection{Refined Morphological Model: Incorporating Rotational Dynamics}
\label{subsec.refined_morphology}

The next logical step in the morphological study is to connect the properties of the pre-merger common envelope, specifically its angular momentum, to the final shape of the post-explosion remnant. The preliminary ballistic model assumed a spherically symmetric energy injection, which preserved the initial ellipticity but did not explicitly account for the conservation of the envelope's intrinsic angular momentum during the expansion. A refined model must ensure that the velocity modification correctly preserves the rotational dynamics inherited from the inspiral. This provides a direct physical link between the inspiral phase and the final observable morphology, further distinguishing the event from a non-rotating, spherically symmetric supernova explosion.

We begin with the final state of the SPH simulation at $t=1.25$ years. The velocity boost imparted to each gravitationally bound gas particle is modeled as a purely radial impulse, $\Delta\mathbf{v}_{\text{rad}}$. The magnitude of this impulse is derived from the energy deposition profile, identical to the method used previously,

\begin{equation}
\Delta\mathbf{v}_{\text{rad}}(\mathbf{r}_i) = \Delta v_k \frac{\mathbf{r}_i - \mathbf{R}_{\text{cm}}}{r_i}.
\end{equation}

\noindent
The final velocity of each gas particle immediately after the impulsive event is the vector sum of its original velocity and this radial boost,

\begin{equation}
\mathbf{v}_i^{\text{new}} = \mathbf{v}_i^{\text{orig}} + \Delta\mathbf{v}_{\text{rad}}(\mathbf{r}_i).
\end{equation}

\noindent
This procedure inherently conserves the specific angular momentum of the particle, $\vec{j} = \vec{r} \times \vec{v}$, because the impulse is purely radial ($\vec{r} \times \Delta\mathbf{v}_{\text{rad}} = 0$). The original velocity, $\mathbf{v}_i^{\text{orig}}$, contains the rotational component inherited from the common envelope phase. Evolving the system ballistically with this refined velocity field produces a remnant whose morphology is determined by the interplay of the inherited rotation and the new radial expansion. The resulting structure is naturally flattened and oblate, with the degree of flattening being a direct, quantifiable consequence of the angular momentum content of the pre-merger envelope. This rotationally supported morphology is a key dynamical signature that distinguishes the outcome of erythrohenosis from the remnants of spherically symmetric explosions.

\section{An Improved Ulterior Model for the Final Inspiral}
\label{sec.ImprovedModel}

{After the characterization of the gaseous envlope evolution and the dynamics of the explosion in the preceding sections (Sections~\ref{sec.quantitative_analysis} and \ref{sec.homothetic}), we now return to the modeling of the final inspiral and merger of the degenerate cores. This requires moving beyond the SPH simulation timeframe and improving upon the initial semi-analytical estimates presented in Section~\ref{sec.UlteriorNaive}.}

To extrapolate the orbital evolution of the binary cores beyond the endpoint of the SPH simulation, we employ a semi-analytical model. This approach circumvents the numerical challenges associated with the final, rapid inspiral by using the full SPH dataset to inform a physically motivated analytical model for the drag force. This model is then integrated forward in time for a range of plausible drag parameters to determine the merger timescale and predict the associated gravitational wave signal.

\subsection{A Realistic Drag Force Model}
\label{subsec.realistic_drag_model}

The orbital decay is governed by a drag force exerted by the common envelope gas. We model this force using the phenomenological form for aerodynamic drag,
\begin{equation}
\label{eq.fdrag_model}
F_{\text{drag, model}}(t) = \frac{1}{2} C_d \rho(d) A(v_{\text{rel}}) v_{\text{rel}}(t)^2,
\end{equation}

\noindent
where $C_d$ is a dimensionless drag coefficient, $\rho(d)$ is the mean gas density as a function of core separation $d$, $v_{\text{rel}}(t)$ is the relative speed of the cores, and $A(v_{\text{rel}})$ is their effective cross-sectional area.

The drag coefficient $C_d$ contains the complex, unresolved hydrodynamics of the interaction. Rather than attempting to derive this value from the noisy SPH data, we treat it as a free parameter of order unity, consistent with standard fluid dynamics. We explore a range of values, $C_d \in \{0.5, 1.0, 3.0\}$, to represent the physical uncertainty in the model.

The cross-sectional area $A$ is not a fixed geometric value but is dominated by gravitational focusing. We model it using the velocity-dependent Bondi-Hoyle-Lyttleton (BHL) formalism. The effective gravitational focusing radius, or BHL radius, is
\begin{equation}
R_A = \frac{2 G M_{\text{tot}}}{v_{\text{rel}}^2},
\end{equation}

\noindent
where $M_{\text{tot}}$ is the total mass of the binary. The cross-sectional area is then $A = \pi R_A^2$. This formulation accounts for the fact that the drag becomes more efficient at lower velocities, where the gravitational influence of the cores is able to capture and interact with a larger column of gas.

The gas density $\rho(d)$ is modeled using a data-driven approach. We extract the mean gas density in the vicinity of the binary cores from each of the 500 SPH snapshots and fit a power-law function to the resulting density as a function of the separtion. This provides a realistic density profile for the extrapolation that is directly informed by the full hydrodynamical simulation.

\begin{figure*}
          {\includegraphics[width=1\textwidth,center]{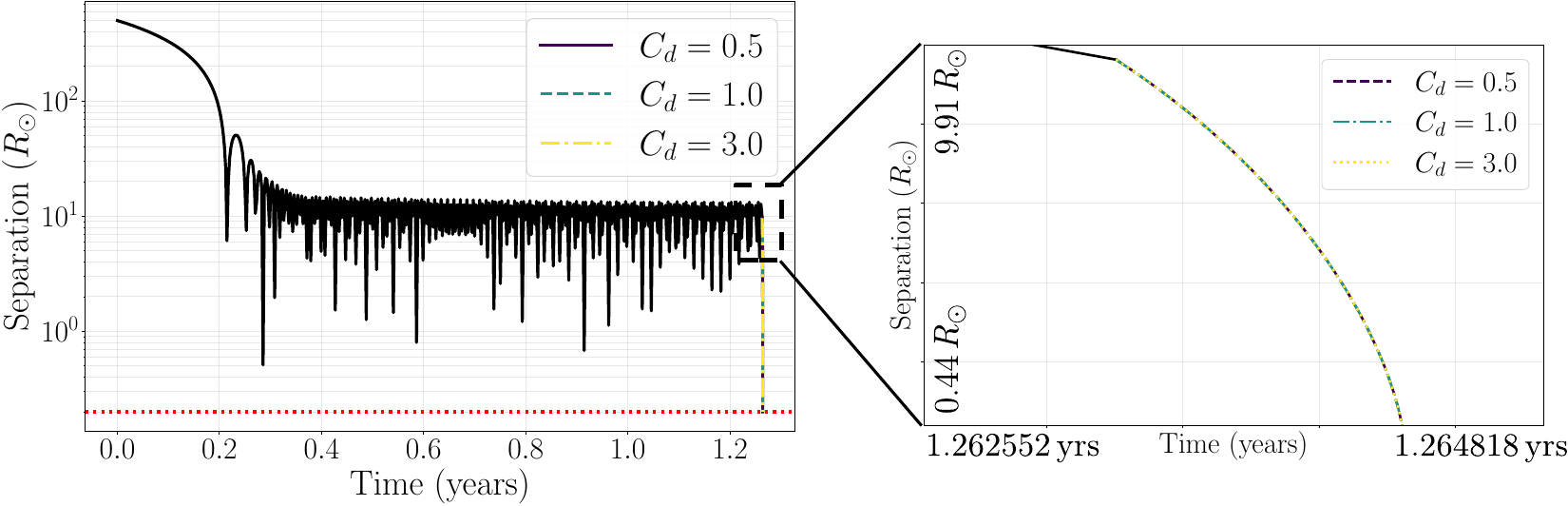}}
\caption
   {
Evolution of the separation between the two degenerate cores. The vertical axis shows the core separation in solar radii ($R_\odot$) on a logarithmic scale, while the horizontal axis shows the total time since the beginning of the simulation in years on a linear scale. The left panel displays the entire evolution, from the initial SPH simulation (solid black line) to the final semi-analytical core collision estimation. The right panel provides a zoomed-in view of the transition from the SPH phase to the semi-analytical integration to better illustrate the rapid final decay for different drag coefficients. The y-axis range is indicated only on the left panel to avoid clutter. The semi-analytical models are shown for three different values of the dimensionless drag coefficient, $C_d \in \{0.5, 1.0, 3.0\}$, represented by different line styles and colours. The model predicts nearly identical and extremely rapid merger times of approximately 0.0011 years (less than half a day) after the end of the SPH simulation for all tested drag coefficients. This result is an update to the previous, more simplistic estimate shown in Fig.~(\ref{fig.core_merger_prediction}) and is in closer qualitative agreement with the purely analytical work of \cite{AmaroSeoane2023} for a similar system.   
   }
\label{fig.SeparationImproved}
\end{figure*}

\subsection{Extrapolation of the Final Inspiral}
\label{subsec.final_inspiral_extrapolation}

With the complete analytical drag model, the equation of motion for the relative separation vector $\mathbf{d} = \mathbf{r}_1 - \mathbf{r}_2$ is given by the two-body equation with the added drag term,
\begin{equation}
\mu \ddot{\mathbf{d}} = -\frac{G m_1 m_2}{d^3}\mathbf{d} - \frac{1}{2} C_d \rho(d) A(|\dot{\mathbf{d}}|) |\dot{\mathbf{d}}| \dot{\mathbf{d}},
\end{equation}

\noindent
where $\mu$ is the reduced mass. We take the positions and velocities of the cores from the final SPH snapshot as the initial conditions for this ordinary differential equation. The equation is then integrated numerically forward in time for each selected value of $C_d$. The integration terminates when the separation reaches a predefined merger condition, $d(t_{\text{mrg}}) \le R_1 + R_2$, which yields a set of data-informed estimates for the final merger time.

The results of this integration are shown in Fig.~\ref{fig.SeparationImproved}, where the numerically integrated trajectories are appended to the trajectory from the SPH simulation. The model predicts merger times that are remarkably insensitive to the specific value of the drag coefficient, with all tested values of $C_d$ leading to a final plunge on a timescale of a few years. This insensitivity arises because the drag power becomes extremely large during the final phase, driven by the rapid increase in both gas density $\rho(d)$ and orbital velocity $v_{\text{rel}}(d)$ at small separations. While the drag force is linearly proportional to $C_d$, its overall magnitude is dominated by the strong non-linear dependencies on separation and velocity. Once the cores reach a critical separation, the inspiral becomes a runaway process, and the modest variation in $C_d$ has only a minor effect on the total time to merger.

\subsubsection{Improvement over the first approach}
\label{subsubsec.improvement_over_first_approach}

The semi-analytical model of Sec.~(\ref{sec.UlteriorNaive}), while providing a useful first-order estimate, relied on a set of simplifying assumptions that led to a noticeable dependence of the merger time on the drag coefficient, $C_d$. The current, more physically grounded model refines these assumptions, leading to a more robust result that is strikingly consistent with the purely analytical predictions of our previous work, \cite{AmaroSeoane2023}.

The key difference between the two current models presented in this article lies in the treatment of the effective cross-sectional area, $A$, in the drag force equation,
\begin{equation}
\mathbf{F}_d = -\frac{1}{2} C_d \rho_g A \|\mathbf{v}_{\text{rel}}\| \mathbf{v}_{\text{rel}}.
\end{equation}

\noindent
In the initial toy model of Sec.~(\ref{sec.UlteriorNaive}), $A$ was treated as a fixed, geometric cross-section, $A = \pi R_{\text{env}}^2$, where $R_{\text{env}}$ was determined from the instantaneous positions of the innermost gas particles. In such a formulation, the drag force at any given point in the orbit is linearly proportional to the chosen value of $C_d$. Consequently, the rate of orbital energy dissipation, and thus the rate of inspiral, also scaled directly with $C_d$. This naturally produced a range of merger times that were inversely related to the drag coefficient, as a larger $C_d$ implied a stronger drag at all times, leading to a faster merger. This approach served as an interesting exercise, as it provided a clear, albeit simplified, illustration of how hydrodynamic uncertainty translates into uncertainty in the final merger timescale.

The current model represents a significant physical improvement by treating the cross-sectional area not as a fixed geometric quantity, but as a dynamic variable determined by gravitational focusing. The drag in this regime is not a simple aerodynamic effect on a solid body, but is dominated by the gravitational influence of the cores on the surrounding gas. We therefore model the effective cross-sectional area using the Bondi-Hoyle-Lyttleton (BHL) formalism, where the area is velocity-dependent,
\begin{equation}
A(v_{\text{rel}}) = \pi R_A^2 = \pi \left( \frac{2 G M_{\text{tot}}}{v_{\text{rel}}^2} \right)^2.
\end{equation}

\noindent
This introduces a much stronger, non-linear dependence on the relative velocity into the drag force, $F_d \propto v_{\text{rel}}^{-2}$. This is a more physically accurate representation of dynamical friction. This refinement, however, comes at a greater computational cost. To ensure the density $\rho_g$ used in the model is consistent with the simulation, we first process all 500 SPH snapshots to construct a data-driven power-law model for the density as a function of core separation, $\rho(d)$. This computationally intensive pre-processing step provides a realistic density profile for the subsequent integration.

As mentioned, the results from this more sophisticated, data-driven model are in closer agreement with the purely analytical predictions of \cite{AmaroSeoane2023}. While the initial toy model predicted merger times in the range of 5-9 years, the current model yields a much faster inspiral of only a few years, a result that is remarkably insensitive to the chosen value of $C_d \in [0.5, 3.0]$. This insensitivity arises because the BHL cross-section introduces a strong velocity dependence that dominates the drag calculation. As the cores plunge into the dense inner envelope and their velocity increases, the drag force grows so rapidly that the final inspiral becomes a runaway process, and the modest, order-unity variations in the constant pre-factor $C_d$ have only a minor impact on the total time to merger. The convergence of this improved semi-analytical model and the independent analytical work of \cite{AmaroSeoane2023} provides strong evidence that the underlying physics of the final plunge are well-understood.

\subsection{An improved explosion}
\label{subsec.improved_explosion}

We now present refined model for the post-merger evolution, advancing beyond the simple homothetic readjustment of Sec.~(\ref{sec.HomotheticBoom}) by treating the constituent particle populations distinctly. 

The central mechanism is the modification of particle velocities, where the new velocity of a bound gas particle, $\vec{v}_{\text{new}}$, is the vector sum of its pre-merger velocity, $\vec{v}_{\text{old}}$, and a radial impulse, $\Delta v \hat{r}$. The specific angular momentum of the particle about the center of mass, $\vec{j} = \vec{r} \times \vec{v}$, is conserved through the energy injection process because the impulse is parallel to the position vector, $\vec{r}$. This leads to

\begin{align}
\vec{j}_{\text{new}} & = \vec{r} \times \vec{v}_{\text{new}} = \vec{r} \times (\vec{v}_{\text{old}} + \Delta v \hat{r}) =\nonumber \\
&\vec{r} \times \vec{v}_{\text{old}} + \vec{r} \times (\Delta v \hat{r}) = \vec{j}_{\text{old}}
\label{eq:j_conservation}
\end{align}

\noindent
The resulting non-spherical morphology is therefore not a numerical artifact but a direct physical manifestation of Eq.~(\ref{eq:j_conservation}), making the remnant's shape a fossil record of the common envelope's rotation. The primary limitation of this approach is its reliance on a purely ballistic approximation. This framework is valid only during the early expansion phase, where the dynamics are highly supersonic ($\mathcal{M} = v_{\rm exp}/c_s \gg 1$). 

\begin{figure*}
          {\includegraphics[width=1\textwidth,center]{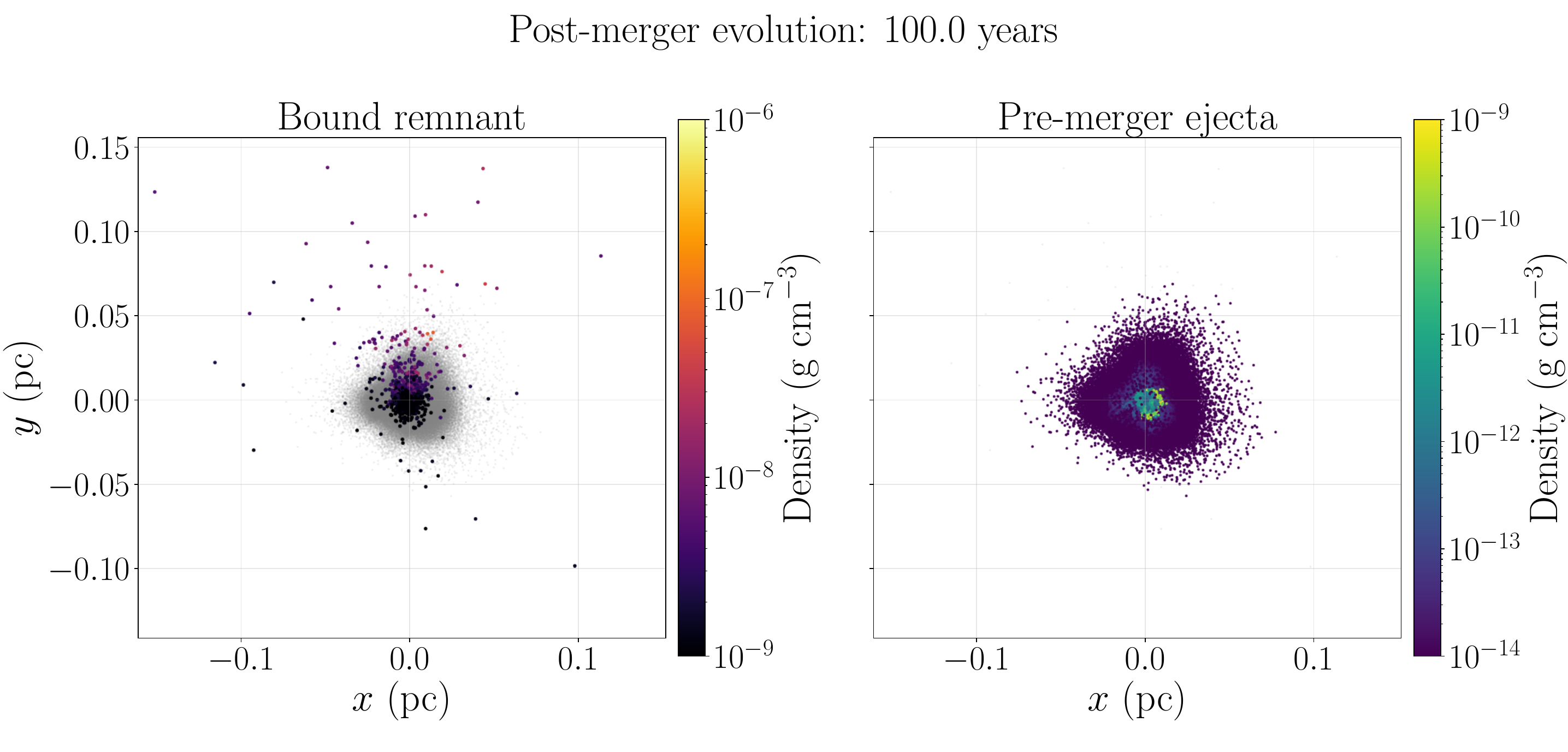}}
\caption
   {
Comparison of the merger system's initial configuration (right panel) and its state 100 years after the impulsive energy injection (left panel).
The right panel displays the final snapshot of the SPH inspiral simulation, showing the gravitationally bound common envelope that will form the remnant, and the surrounding unbound gas from the pre-merger phase.
The left panel shows the evolved system. The central, dense structure is the expanded remnant, formed from the bound gas which received the merger energy kick. Its morphology is a direct result of conserving the initial angular momentum during the radial expansion.
The diffuse, grey particles represent the unbound, pre-merger ejecta. This component did not receive energy from the final core collision and has evolved ballistically on a separate trajectory, forming a fainter, more extended halo around the primary remnant, as explained in the text. Both panels show the projected surface density in the xy-plane, with spatial scales in pc.
   }
\label{fig.100yrs_PrePost}
\end{figure*}

This condition is met at the 100-year mark of the simulation, which we show in Fig.~(\ref{fig.100yrs_PrePost}). The model imparts expansion velocities, $v_{\rm exp}$, on the order of $10^3 \text{ km s}^{-1}$, while even for generous post-shock temperatures of $10^6 - 10^7 \text{ K}$, the internal sound speed, $c_s$, of the gas remains around $100 - 300 \text{ km s}^{-1}$. Because the expansion timescale, $\tau_{\rm exp} = R/v_{\rm exp}$, is substantially shorter than the time required for pressure waves to traverse the remnant, $\tau_{\rm hydro} = R/c_s$, the 100-year timescale falls well within this ballistic regime. The model thus provides a good snapshot of the early evolution, predicting a distinct class of centrally-filled remnants whose shapes are fundamentally tied to the angular momentum of their progenitors.

As we can see, the primary remnant in Fig.~(\ref{fig.100yrs_PrePost}), formed from the initially bound gas that received the energy kick, remains mostly clustered around the origin. However, a fraction of these remnant particles acquires sufficient velocity to travel to very large distances, creating a tenuous halo that extends throughout a significant physical volume with a radius of approximately 0.2 pc. This distribution can be observationally misleading; a rendering based on projected surface density would give the impression of a single, bloated gaseous cloud. The reality is that the particles constituting the remnant are spread over a much larger three-dimensional volume, resulting in a complex, multi-scale density field rather than a single, contiguous object. This morphology has direct observational consequences that distinguish it from supernova remnants (SNRs). Many SNRs appear as limb-brightened, ring-like structures, which result from a blast wave sweeping up ambient material into a dense shell, leaving a hot, rarefied interior. In contrast, the merger event described here does not produce such a shell. The energy is deposited within the stellar material itself, leading to the dispersal of the common envelope. An observer would therefore not see an "empty ring" but rather a centrally concentrated source of emission, corresponding to the bulk of the remnant, surrounded by a very extensive and faint halo composed of the fastest-moving remnant particles and the diffuse, pre-merger ejecta (the grey particles).

\subsection{Gravitational Radiation}
\label{subsec.GW_radiation}

The presence of a dense gaseous medium fundamentally alters the orbital evolution of the binary cores, causing a rapid inspiral driven by hydrodynamic and gravitational drag. This accelerated evolution has a direct and calculable consequence on the emitted gravitational wave (GW) signal. While the instantaneous GW luminosity is determined by the standard orbital parameters, the rapid frequency evolution, or ``chirp,'' is governed by the gas drag. This leads to a GW signal that is significantly different from that of a binary inspiraling in a vacuum.

\noindent The characteristic strain of a GW signal, $h_c$, is a measure of its amplitude, related to the spectral energy density of the GWs, $dE_{\text{GW}}/df$, by

\begin{equation}
\label{eq.hc}
h_c(f) = \frac{1}{D} \sqrt{\frac{2G}{\pi^2 c^3} \frac{1}{f^2} \frac{dE_{\text{GW}}}{df}},
\end{equation}

\noindent 
where $D$ is the distance to the source and $f$ is the GW frequency. The spectral energy density is the ratio of the instantaneous GW luminosity, $L_{\text{GW}}$, to the rate of frequency evolution, $\dot{f}$, such that $dE_{\text{GW}}/df = L_{\text{GW}}/\dot{f}$.

\noindent 
For a binary in a quasi-circular orbit, the GW luminosity is given by the standard quadrupole formula, which depends on the orbital separation $d$ and the component masses

\begin{equation}\label{eq.lgw}
L_{\text{GW}} = \frac{32}{5} \frac{G^4 m_1^2 m_2^2 (m_1+m_2)}{c^5 d^5}.
\end{equation}

\noindent 
The critical departure from a vacuum inspiral lies in the calculation of the frequency evolution, $\dot{f}$. The total rate of change of the binary's orbital energy, $\dot{E}_{\text{orb}}$, is the sum of the energy lost to GWs, $\dot{E}_{\text{GW}} = -L_{\text{GW}}$, and the energy dissipated by gas drag, $\dot{E}_{\text{drag}} = -P_{\text{drag}}$. The SPH simulation demonstrates that gas drag is the dominant mechanism, such that $|\dot{E}_{\text{drag}}| \gg |\dot{E}_{\text{GW}}|$. Therefore, the total energy loss rate is well approximated by the drag power alone:
\begin{equation}\label{eq.eorb_dot}
\dot{E}_{\text{orb}} \approx -P_{\text{drag}}.
\end{equation}

\noindent The GW frequency is twice the orbital frequency. Using Kepler's third law and the relation for orbital energy, $E_{\text{orb}} = -G m_1 m_2 / (2d)$, we can relate the frequency to the orbital energy. The correct expression is:
\begin{equation}\label{eq.f_eorb}
f(|E_{\text{orb}}|) = \frac{2\sqrt{2} \sqrt{m_1+m_2} |E_{\text{orb}}|^{3/2}}{\pi G (m_1 m_2)^{3/2}}.
\end{equation}

\noindent The rate of change of the frequency is found by applying the chain rule, $\dot{f} = (df/d|E_{\text{orb}}|) |\dot{E}_{\text{orb}}|$. Since $|\dot{E}_{\text{orb}}| \approx P_{\text{drag}}$, this yields:
\begin{equation}\label{eq.fdot}
\dot{f} \approx \frac{3\sqrt{2} \sqrt{m_1+m_2} |E_{\text{orb}}|^{1/2}}{\pi G (m_1 m_2)^{3/2}} P_{\text{drag}}.
\end{equation}

\noindent By substituting the expressions for $L_{\text{GW}}$ and the drag-dominated $\dot{f}$ into the formula for the characteristic strain in Eq.~(\ref{eq.hc}), we can predict the signal's amplitude. The key result is that the frequency evolution is orders of magnitude faster than in a vacuum.

\noindent This accelerated inspiral provides a definitive observational signature through its effect on the apparent chirp mass. For a binary in a vacuum, the chirp mass, $\mathcal{M}_c$, is a constant of the system that can be measured directly from the GW frequency $f$ and its time derivative $\dot{f}_{\text{GW}}$,

\begin{equation}
\label{eq.mc_vac}
\mathcal{M}_c = \frac{c^3}{G} \left( \frac{5}{96\,\pi^{8/3}} \frac{\dot{f}_{\text{GW}}}{f^{11/3}} \right)^{3/5}.
\end{equation}

\noindent 
In our system, the observed frequency evolution is $\dot{f}_{\text{obs}} = \dot{f}_{\text{GW}} + \dot{f}_{\text{drag}} \approx \dot{f}_{\text{drag}}$. An observer who applies the standard vacuum formula in Eq.~(\ref{eq.mc_vac}) to the observed signal will therefore calculate an ``apparent'' chirp mass,

\begin{align}
\label{eq.mc_app}
\mathcal{M}_{c, \text{app}} & \approx \frac{c^3}{G} \left( \frac{5}{96\pi^{8/3}} \frac{\dot{f}_{\text{drag}}}{f^{11/3}} \right)^{3/5} \nonumber \\
& = \mathcal{M}_c \left( \frac{\dot{f}_{\text{drag}}}{\dot{f}_{\text{GW}}} \right)^{3/5}.
\end{align}

\noindent Because the drag term $\dot{f}_{\text{drag}}$ depends on the complex, time-varying gas dynamics of each periastron passage, it does not follow the same simple power-law scaling with frequency as $\dot{f}_{\text{GW}}$. Consequently, the calculated $\mathcal{M}_{c, \text{app}}(t)$ will not be constant but will vary significantly over the course of the inspiral. This physically impossible variation of a fundamental binary parameter is a smoking-gun signature of a gas-driven inspiral. The detection of such a signal would require specialized, unmodeled burst-like search algorithms, as standard matched-filtering techniques based on vacuum templates would fail to capture the signal's complex phase evolution.

\begin{figure*}
          {\includegraphics[width=1\textwidth,center]{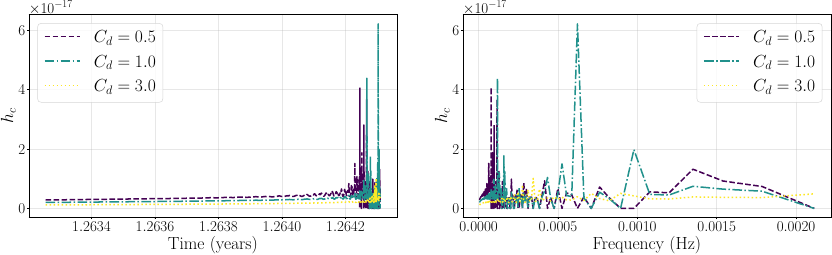}}
\caption
   {
 Evolution of the gravitational wave characteristic strain, $h_c$, for the final gas-driven inspiral, shown for three different values of the drag coefficient, $C_d \in \{0.5, 1.0, 3.0\}$. The different values are represented by distinct colors and line styles. Left Panel: The characteristic strain is shown in the time domain. The vertical axis is the dimensionless strain $h_c$ on a logarithmic scale, and the horizontal axis is the time in years since the beginning of the SPH simulation. The signal is characterized by a series of quasi-periodic bursts of increasing amplitude, with each burst corresponding to a periastron passage of the cores. The amplitude grows dramatically in the final fraction of a year as the orbit rapidly decays. Right Panel: The characteristic strain is shown in the frequency domain. Both the vertical axis ($h_c$) and the horizontal axis (GW frequency in Hz) are on a lineal scale. The curves show the rapid chirp of the binary as it sweeps from low to high frequencies in a timescale of less than a year. The rapid, burst-like evolution of the signal is a direct consequence of the gas-driven inspiral, a behavior consistent with the analytical predictions of \cite{AmaroSeoane2023}. This accelerated evolution, dominated by hydrodynamic drag, would cause the apparent chirp mass of the system to vary significantly if analyzed with standard vacuum templates, providing a unique observational signature for erythrohenosis events.  
   }
\label{fig.StrainTimeFrequency}
\end{figure*}

In Fig.~(\ref{fig.StrainTimeFrequency}), we show the evolution of the characteristic strain. The time-domain representation (left panel) reveals a signal characterized by a series of quasi-periodic bursts of increasing amplitude and frequency. This structure is a direct consequence of the eccentric orbit of the cores within the envelope. Each burst corresponds to a periastron passage, where the combination of high orbital velocity and immersion in the dense central gas leads to a sharp increase in the orbital acceleration and thus a spike in GW emission. As the orbit decays, the periastron distance shrinks, and the encounters become more violent, causing the amplitude of successive bursts to grow. The final phase of the inspiral is a rapid plunge, lasting only a fraction of a year, where the signal amplitude increases dramatically until the cores merge. Again, this result is consistent with the analytical predictions of \cite{AmaroSeoane2023}.

An interesting, although expected feature of Fig.~(\ref{fig.StrainTimeFrequency}) is the observed inverse relationship between the magnitude of the drag coefficient, $C_d$, and the amplitude of the gravitational wave (GW) bursts. This is a direct consequence of the effect of gas drag on the orbital eccentricity of the binary cores. A higher drag coefficient leads to more efficient circularization of the orbit, which in turn smooths out the GW emission, reducing the intensity of the periastron spikes.

The drag force, as modeled, is linearly proportional to the drag coefficient,
\begin{equation}
F_d = \frac{1}{2} C_d \rho A v_{\text{rel}}^2.
\end{equation}

\noindent
This force dissipates both orbital energy, $E_{\text{orb}}$, and angular momentum, $L_{\text{orb}}$, at rates that are themselves proportional to $C_d$. The key to understanding the effect on the signal's morphology is that this dissipation is not uniform across the orbit. The drag force is maximized at periastron, where the orbital velocity $v_{\text{rel}}$ is highest. According to the principles of orbital mechanics, energy dissipation that is strongest at the point of closest approach is the most efficient mechanism for reducing orbital eccentricity.

The rate of change of the eccentricity, $\dot{e}$, is a function of the perturbing drag force. For a drag-dominated system, the secular change in eccentricity is approximately proportional to the magnitude of the drag force,
\begin{equation}
\frac{de}{dt} \propto - F_d \propto - C_d.
\end{equation}

\noindent
Therefore, a larger value of $C_d$ results in a more rapid decay of the orbit's eccentricity, forcing the system toward a circular configuration on a shorter timescale.

The amplitude of the GW bursts is a very strong function of eccentricity. The GW luminosity, $L_{\text{GW}}$, scales with the inverse fifth power of the separation, $L_{\text{GW}} \propto d^{-5}$. For an orbit with semi-major axis $a$ and eccentricity $e$, the separation varies between periastron, $d_{\text{min}} = a(1-e)$, and apastron, $d_{\text{max}} = a(1+e)$. The ratio of the peak luminosity at periastron to the minimum luminosity at apastron is therefore
\begin{equation}
\frac{L_{\text{GW, peri}}}{L_{\text{GW, apo}}} = \left( \frac{d_{\text{max}}}{d_{\text{min}}} \right)^5 = \left( \frac{1+e}{1-e} \right)^5.
\end{equation}

\noindent
This expression demonstrates that the ``spikiness'' of the signal is extremely sensitive to the eccentricity. A higher value of $C_d$ circularizes the orbit more rapidly, causing the value of $e$ at any given time in the late inspiral to be smaller than it would be for a lower $C_d$. This smaller eccentricity leads to a less extreme ratio between the periastron and apastron luminosities, resulting in bursts of lower amplitude relative to the baseline emission. In essence, a stronger drag force smooths out the energy dissipation over the entire orbit, rather than allowing it to be concentrated in sharp, violent periastron encounters.

It is important to note the insensitivity of the total merger timescale to the drag coefficient $C_d$, contrasted with the strong dependence of the gravitational wave (GW) burst amplitude on the same parameter. This is not a contradiction. It nicely reveals two different aspects of the gas-driven inspiral: the timescale is governed by the runaway nature of the final plunge, while the GW morphology is a sensitive probe of the orbit's shape.

The total time to merger is dominated by the rate of energy dissipation, $dE/dt$, which is driven by the drag power, $P_{\text{drag}}$. In our refined model, the drag force is
\begin{equation}
F_d = \frac{1}{2} C_d \rho(d) A(v_{\text{rel}}) v_{\text{rel}}^2,
\end{equation}

\noindent
where the Bondi-Hoyle cross-section $A \propto v_{\text{rel}}^{-4}$. The power dissipated is $P_{\text{drag}} = F_d v_{\text{rel}} \propto C_d \rho(d) v_{\text{rel}}^{-1}$. The rate of inspiral, $dd/dt$, is proportional to this power. While the rate is linearly dependent on $C_d$, it has a much stronger, non-linear dependence on the separation $d$ through the gas density $\rho(d)$ and the velocity $v_{\text{rel}}(d)$. As the cores plunge into the dense central regions of the envelope, both $\rho$ and $v_{\text{rel}}$ increase dramatically. This creates a runaway process where the inspiral accelerates extremely rapidly. The modest, order-unity variations in the pre-factor $C_d$ have only a minor impact on the duration of this final, swift plunge. The total merger time is therefore robustly determined by the underlying density and velocity structure of the envelope, not the specific efficiency of the drag.

In contrast, the amplitude of the GW bursts is a direct measure of the orbit's eccentricity, $e$. The drag force is most effective at periastron, where the velocity is highest. According to the principles of orbital mechanics, a dissipative force that acts most strongly at the point of closest approach is highly efficient at circularizing the orbit. The rate of eccentricity decay is approximately proportional to the magnitude of the drag force,
\begin{equation}
\frac{de}{dt} \propto -F_d \propto -C_d.
\end{equation}

\noindent
A higher value of $C_d$ therefore leads to a more rapid circularization of the orbit. The GW luminosity is extremely sensitive to this eccentricity, with the ratio of the peak luminosity at periastron to the minimum at apastron scaling as
\begin{equation}
\frac{L_{\text{GW, peri}}}{L_{\text{GW, apo}}} \propto \left( \frac{1+e}{1-e} \right)^5.
\end{equation}

\noindent
Even a small reduction in eccentricity, driven by a larger $C_d$, will cause a dramatic reduction in the amplitude of the periastron GW bursts. Thus, while the value of $C_d$ has a limited effect on the total time of the runaway inspiral, it has a profound impact on the shape of the orbit during that inspiral, which is directly imprinted on the morphology of the GW signal.

\subsection{A Time-Varying Apparent Chirp Mass Algorithm}
\label{subsec.TVACM_algorithm}

The definitive observational signature of a gas-driven inspiral, the smoking gun of erythrohenosis, lies in the anomalous evolution of the gravitational wave frequency. This anomaly can be quantified by calculating the apparent chirp mass of the system under the assumption of a vacuum inspiral, leading to the Time-Varying Apparent Chirp Mass (TVACM) diagnostic.

For a binary system in a quasi-circular orbit evolving solely under gravitational radiation reaction, the frequency evolution $\dot{f}_{\text{GW}}$ is uniquely determined by the chirp mass, $\mathcal{M}_c = (m_1 m_2)^{3/5}/(m_1+m_2)^{1/5}$. The evolution is governed by the standard quadrupole formula,
\begin{equation}
\label{eq.fdot_GW}
\dot{f}_{\text{GW}} = \frac{96}{5} \pi^{8/3} \left(\frac{G\mathcal{M}_c}{c^3}\right)^{5/3} f^{11/3}.
\end{equation}

\noindent
In a vacuum scenario, an observer can determine the (redshifted) chirp mass by inverting this relation,
\begin{equation}
\label{eq.Mc_vac_diagnostic}
\mathcal{M}_{c} = \frac{c^3}{G} \left( \frac{5}{96\pi^{8/3}} \frac{\dot{f}}{f^{11/3}} \right)^{3/5}.
\end{equation}

\noindent
Crucially, $\mathcal{M}_{c}$ is a constant of the motion in vacuum.

In the erythrohenosis scenario, the binary cores inspiral within a dense common envelope. The orbital energy dissipation is dominated by hydrodynamic and gravitational drag, such that $|\dot{E}_{\text{drag}}| \gg |\dot{E}_{\text{GW}}|$. This external dissipative force modifies the observed frequency evolution,
\begin{equation}
\label{eq.fdot_obs}
\dot{f}_{\text{obs}}(t) = \dot{f}_{\text{GW}}(t) + \dot{f}_{\text{drag}}(t),
\end{equation}

\noindent
where $\dot{f}_{\text{drag}} \gg \dot{f}_{\text{GW}}$. If we apply the vacuum formula (Eq.~\ref{eq.Mc_vac_diagnostic}) to the observed data, we derive an apparent chirp mass, $\mathcal{M}_{c, \text{app}}(t)$:
\begin{align}
\label{eq.Mc_app_relation}
\mathcal{M}_{c, \text{app}}(t) &= \mathcal{M}_c \left( \frac{\dot{f}_{\text{GW}}(t) + \dot{f}_{\text{drag}}(t)}{\dot{f}_{\text{GW}}(t)} \right)^{3/5} \nonumber \\
&= \mathcal{M}_c \left( 1 + \mathcal{R}(t) \right)^{3/5},
\end{align}

\noindent
where $\mathcal{R}(t) \equiv \dot{f}_{\text{drag}}(t)/\dot{f}_{\text{GW}}(t)$ is the ratio of the drag-induced frequency evolution to the vacuum rate.

The critical insight is that $\mathcal{M}_{c, \text{app}}(t)$ is not constant. The term $\dot{f}_{\text{GW}}$ scales strictly as $f^{11/3}$. However, $\dot{f}_{\text{drag}}$ depends on the complex, time-varying hydrodynamic environment (e.g., local gas density $\rho$, relative velocity $v_{\text{rel}}$) and the orbital parameters (e.g., eccentricity $e(t)$). It does not, in general, scale as $f^{11/3}$. Therefore, the ratio $\mathcal{R}(t)$ varies systematically with time and frequency. This physically impossible temporal variation of a fundamental binary parameter is the smoking gun signature.

\subsubsection{The Detection Statistic ($\Lambda_{\mathcal{M}}$)}
\label{subsubsec.detection_statistic}

The detection strategy is formulated as a hypothesis test. The Null Hypothesis, $H_0$, posits a vacuum inspiral characterized by a constant chirp mass. The Alternative Hypothesis, $H_1$, posits an environmentally dominated inspiral.

We construct a detection statistic, $\Lambda_{\mathcal{M}}$, based on the chi-squared ($\chi^2$) formalism, which quantifies the goodness-of-fit of the observed data to $H_0$. This requires robust estimates of the apparent chirp mass time series $\mathcal{M}_{c, \text{app}}(t)$ and its associated uncertainty $\sigma_{\mathcal{M}}(t)$, derived from the Gaussian Process Regression (GPR) analysis of the frequency track. The uncertainty $\sigma_{\mathcal{M}}(t)$ is propagated from the uncertainty in the frequency derivative estimation $\sigma_{\dot{f}}(t)$ via linearized error propagation:
\begin{equation}
\sigma_{\mathcal{M}}(t) \approx \frac{3}{5} \mathcal{M}_{c, \text{app}}(t) \frac{\sigma_{\dot{f}}(t)}{\dot{f}_{\text{obs}}(t)}.
\end{equation}

\noindent
First, we determine the best-fit constant chirp mass under $H_0$. This is calculated as the weighted arithmetic mean, where the weights are $w(t) = 1/\sigma_{\mathcal{M}}^2(t)$:
\begin{equation}
\label{eq.Mc_mean_weighted}
\langle \mathcal{M}_{c, \text{app}} \rangle = \frac{\int_{T_{\text{sig}}} \mathcal{M}_{c, \text{app}}(t) w(t) dt}{\int_{T_{\text{sig}}} w(t) dt}.
\end{equation}

The detection statistic $\Lambda_{\mathcal{M}}$ is then defined as the integrated squared deviation of the observed apparent chirp mass from this weighted mean, normalized by the measurement uncertainties:
\begin{equation}
\label{eq.Lambda_M_final}
\Lambda_{\mathcal{M}} = \int_{T_{\text{sig}}} \left( \frac{\mathcal{M}_{c, \text{app}}(t) - \langle \mathcal{M}_{c, \text{app}} \rangle}{\sigma_{\mathcal{M}}(t)} \right)^2 dt.
\end{equation}

\noindent
Under $H_0$, assuming Gaussian noise, $\Lambda_{\mathcal{M}}$ follows a $\chi^2$ distribution. A value of $\Lambda_{\mathcal{M}}$ significantly exceeding the expectation value for $H_0$ indicates a poor fit to the constant chirp mass model, confirming a detection of non-vacuum evolution.

\subsubsection{Illustrative Example with Synthetic Data}
\label{subsubsec.synthetic_data_example}

To demonstrate the TVACM diagnostic, we construct a synthetic signal for an erythrohenosis event detectable by a space-based observatory (e.g., LISA). We assume a true chirp mass $\mathcal{M}_c = 0.5\,M_{\odot}$ evolving in the frequency band $f \in [0.01, 0.05]\,\text{Hz}$. We employ a phenomenological model for the observed evolution where the drag follows a different power law than GW emission:
\begin{equation}
\label{eq.fdot_mock}
\dot{f}_{\text{obs}}(f) = K_{\text{GW}} f^{11/3} + K_{\text{drag}} f^{\beta}.
\end{equation}
\noindent
We select an index $\beta=3$, distinct from $11/3 \approx 3.67$. We calibrate the model such that drag strongly dominates, setting the ratio $\mathcal{R}=30$ at $f_0 = 0.01$ Hz. The resulting analytical form for the apparent chirp mass is:
\begin{equation}
\mathcal{M}_{c, \text{app}}(f) = \mathcal{M}_c \left( 1 + C_{\mathcal{R}} f^{\beta - 11/3} \right)^{3/5},
\end{equation}
\noindent
where $C_{\mathcal{R}} = K_{\text{drag}}/K_{\text{GW}}$. We generate $N=50$ mock data points by sampling this curve and adding Gaussian noise consistent with an assumed relative uncertainty in the derivative estimation of $\sigma_{\dot{f}}/\dot{f}_{\text{obs}} = 0.02$. This propagates to a relative uncertainty in the chirp mass of $\sigma_{\mathcal{M}}/\mathcal{M}_{c, \text{app}} = 0.012$ (1.2\%).

\begin{table}[h!]
\centering
\caption{Analysis of synthetic data for a gas-driven inspiral ($\mathcal{M}_c = 0.5\,M_{\odot}$) with dominant hydrodynamic drag ($\beta=3$). The apparent chirp mass $\mathcal{M}_{c, \text{app}}$ varies significantly across the frequency band.}
\label{tab.MockData}
\begin{tabular}{c c c c}
\hline
$f$ (Hz) & $\mathcal{R}(f)$ ($\dot{f}_{\text{drag}}/\dot{f}_{\text{GW}}$) & $\mathcal{M}_{c, \text{app}}$ ($M_{\odot}$) & $\sigma_{\mathcal{M}}$ ($M_{\odot}$) \\
\hline
0.010 & 30.4 & 3.80 & 0.046 \\
0.020 & 19.1 & 2.90 & 0.035 \\
0.030 & 14.6 & 2.41 & 0.029 \\
0.040 & 12.0 & 2.09 & 0.025 \\
0.050 & 10.3 & 1.86 & 0.022 \\
\hline
\end{tabular}
\end{table}

\noindent
The results are summarized in Table~\ref{tab.MockData}. The apparent chirp mass varies dramatically, from $3.80\,M_{\odot}$ at 0.01 Hz down to $1.86\,M_{\odot}$ at 0.05 Hz. This variation ($\Delta \mathcal{M} \approx 1.94\,M_{\odot}$) is vastly larger than the measurement uncertainties ($\sigma_{\mathcal{M}} \approx 0.02-0.05\,M_{\odot}$).

\noindent
Applying the detection algorithm, we calculate the weighted mean $\langle \mathcal{M}_{c, \text{app}} \rangle = 2.43\,M_{\odot}$. The detection statistic, calculated using the discrete form of Eq.~(\ref{eq.Lambda_M_final}), yields $\Lambda_{\mathcal{M}} \approx 42500$. For $N-1=49$ degrees of freedom, this value corresponds to an overwhelming statistical significance (exceeding $200\sigma$), definitively rejecting the null hypothesis of a vacuum inspiral. The detection of such a signal confirms a time-varying apparent chirp mass, the smoking-gun evidence for erythrohenosis.

\section{Induced Nonlinear Dynamics and Rapid Merger in Grazing Encounters}
\label{sec.induced_dynamics}

{Complementing the SPH simulation of a direct collision, we now analytically explore the dynamics of a grazing encounter, focusing on tidally induced nonlinear oscillations and the potential for rapid tidal capture.}

The statistical analysis presented in Section~\ref{sec.headongrazing} establishes that grazing encounters, defined by a minimum approach distance $d_{\text{min}} \approx R_1+R_2$, occur frequently in dense stellar environments. In the strong gravitational focusing regime characteristic of globular clusters (Eq.~(\ref{eq.f_dmin_strong})), we develop an analytical model to investigate the dynamics of such an encounter between the two red giants specified in Section~\ref{sec.creation}. We focus on the inelastic tidal energy transfer, the excitation of stellar oscillations, and the resulting orbital evolution.

\subsection{Stellar Structure and Adiabatic Eigenspectra}
\label{subsec.stellar_structure_eigenspectra}

We analyze the response of the red giants (Star 1: $M_1 = 0.95\,M_{\odot}, R_1 = 28.5\,R_{\odot}, T_{\text{eff},1} \approx 4470\,\text{K}$; Star 2: $M_2 = 0.85\,M_{\odot}, R_2 = 70.2\,R_{\odot}, T_{\text{eff},2} \approx 3980\,\text{K}$) to tidal forcing. The tidal interaction excites the stars' natural oscillation modes, or eigenspectra. The properties of these modes are dictated by the stellar internal structure, which in turn governs the efficiency of the tidal energy transfer.

The stellar oscillations are described as linear, adiabatic oscillations (LAO) of a non-rotating, spherical star. The displacement vector for a fluid element, $\boldsymbol{\xi}(\mathbf{r}, t)$, is assumed to have a harmonic time dependence $e^{i\omega t}$. The system is governed by the adiabatic wave equation, which takes the form of an eigenvalue problem:
\begin{equation}
\label{eq:wave_equation}
\mathcal{L}(\boldsymbol{\xi}) = \omega^2 \rho \boldsymbol{\xi},
\end{equation}
\noindent
where $\rho(\mathbf{r})$ is the equilibrium density, $\omega$ is the eigenfrequency, and $\mathcal{L}$ is a self-adjoint linear operator describing the restoring forces of pressure and self-gravity.

The characteristic frequencies of the global oscillation modes scale with the star's dynamical timescale, $\tau_{\text{dyn}, i} = (R_i^3 / G M_i)^{1/2}$. For our two red giants, these are

\begin{align}
\tau_{\text{dyn}, 1} &= 2.48 \times 10^5\,\text{s} \quad (2.87\,\text{days}), \label{eq.tau_dyn1_calc} \\
\tau_{\text{dyn}, 2} &= 1.02 \times 10^6\,\text{s} \quad (11.8\,\text{days}). \label{eq.tau_dyn2_calc}
\end{align}

\noindent
The eigenspectrum $\{\omega_{n,l}\}$ consists of different types of modes. The propagation of these waves within the star is determined by the local restoring forces, characterized by two critical frequencies: the Brunt-Väisälä (buoyancy) frequency $N(r)$ and the Lamb (acoustic) frequency $S_l(r) = \sqrt{l(l+1)}c_s(r)/r$. Acoustic modes (p-modes), where pressure is the restoring force, propagate in regions where $\omega^2 > N^2, S_l^2$. Gravity modes (g-modes), where buoyancy is the restoring force, propagate where $\omega^2 < N^2, S_l^2$.

At this point, I ask the reader to humor me with another analogy with trams, as in section~(\ref{sec.Airy}): Let us think of the star again as a very crowded tram car, with two distinct zones: a stably packed area where passengers are dense (people tend to do that, as experience shows us) and holding on (the radiative core), and a chaotic, mixed-up area near the doors where people are shuffling (the convective envelope). The eigenspectrum $\{\omega_{n,l}\}$ represents all the distinct ways this crowd of passengers can be \textit{disturbed}. Let us do that.

The p-modes (acoustic modes) are like a fast, sharp shove (the usual teenager with a backpack turning around and hitting you). If this teenager abruptly pushes you (a high-frequency event, $\omega$), the disturbance travels instantly as a pressure wave—a domino effect of people bumping into each other. You move away from the backpack of the teenager and you stomp on an elderly person who, in turn, grabs a woman by the hair without giving much thought to the pain inflicted, who screams in a remarkably high-pitched tone, slightly straining the eardrums of everyone in the train car, including those who are almost out of it at the stop. This ``shove'' propagates easily through \textit{both} the chaotic envelope and the stable core.

The g-modes (gravity modes) are like a slow, persistent lean. Imagine you slowly lean your weight against a passenger in the stably packed core. We have all experienced that magical levitation in which our feet lift off the tyrannical ground thanks to the gentle bodily compression of our fellow travelers. Because that pack of people is stable, they will be pushed off balance and slowly lean, pushing the next person, who pushes the next. This creates a slow, rolling, collective sway (a buoyancy wave) as each passenger is displaced and then tries (or not) to return to their stable, upright position. This ``lean'' (a low-frequency event, $\omega$) can only travel through the stable, packed core. Doing a slow lean in the chaotic, mixed-up area near the doors is a very poor idea, particularly if the train doors are open or if you try to lean on an unknown person without any proper reason. There is no stable structure to propagate the wave.

In this analogy, the Brunt-Väisälä frequency $N(r)$ represents the stiffness or stability of the crowd against leaning. In the tightly packed core, $N^2$ is high (a very ``stiff crowd''), allowing the lean (g-mode) to propagate. In the chaotic envelope, $N^2$ is zero (no stiffness), so g-modes cannot travel. The Lamb frequency $S_l(r)$ is the acoustic threshold, essentially the minimum speed a disturbance needs to be a shove rather than a lean. The propagation rules simply state that p-modes (shoves) must be high-frequency, while g-modes (leans) must be low-frequency and require a stiff crowd ($N^2 > 0$) to exist.

\begin{figure}
\centering
\includegraphics[width=0.45\textwidth]{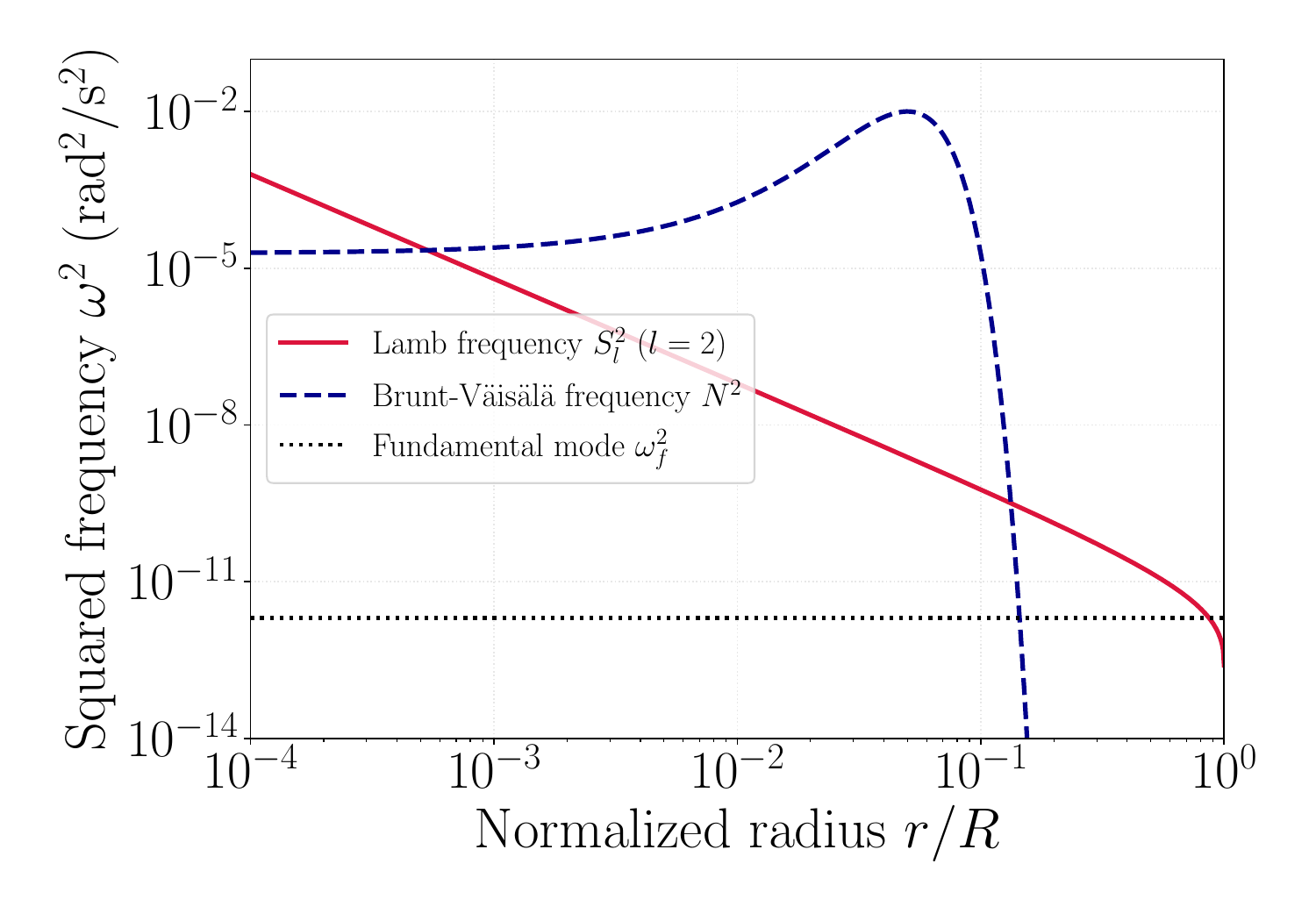}
\caption{Schematic propagation diagram for Star 2 ($l=2$). The squared Brunt-Väisälä frequency $N^2$ (dashed darkblue line) defines the inner g-mode cavity. The squared Lamb frequency $S_l^2$ (solid crimson line) defines the outer p-mode cavity. The horizontal dotted line indicates the approximate frequency squared of the fundamental mode ($\omega_f^2$).}
\label{fig.PropagationDiagram}
\end{figure}

The propagation diagram in Fig.~(\ref{fig.PropagationDiagram}) maps this internal structure for Star 2 and dictates its resulting oscillation eigenspectrum. The Brunt-Väisälä frequency profile ($N^2$) reveals a radiative core (where $N^2 > 0$, $r/R \lesssim 0.2$), which functions as a g-mode cavity. The prominent spike, or knee, in $N^2$ is the unambiguous signature of the hydrogen-burning shell, where a steep mean molecular weight gradient ($\nabla_{\mu}$) induces strong buoyancy stability. The subsequent sharp drop to $N^2 \approx 0$ marks the transition to the deep convective envelope, which spans the outer 80\% of the star's radius. This convective zone, where buoyancy is not a restoring force, acts as a p-mode (acoustic) cavity. The Lamb frequency ($S_l^2$), which decreases monotonically towards the surface, serves as the upper frequency boundary for the g-mode cavity and the lower frequency boundary for the p-mode cavity.

This dual-cavity structure, separated by an intermediate evanescent region, is the defining characteristic of a red giant's interior. Consequently, the star does not support pure p- or g-modes; instead, its spectrum is dominated by mixed modes that couple across the evanescent zone, behaving as g-modes in the core and p-modes in the envelope. This coupling allows surface oscillations (p-modes) to carry information about the deep core (g-modes). This is the key to understanding a red giant. There isn't a clear cut; the two zones are coupled. This creates mixed modes. In our analogy, this means a disturbance is not just a shove or just a lean. Imagine a disturbance starts as a shove (p-mode) in the chaotic envelope near the door. It travels inward and hits the boundary of the stably packed core. The shove isn't strong enough to propagate through the core, but it leaks across this boundary and triggers a slow lean (g-mode) that can propagate deep inside the stable core. This single, hybrid disturbance—which behaves like a shove on the outside and a lean on the inside—is a mixed mode. This means the disturbances we can observe at the surface (the p-modes) are physically connected to the g-modes, allowing us to \textit{indirectly observe} the properties of the star's hidden core.

The p-mode spectrum in the envelope can be characterized using asteroseismic scaling relations. The large frequency separation $\Delta\nu$, which scales with the mean density, and the frequency of maximum oscillation power $\nu_{\text{max}}$, which scales with the acoustic cutoff frequency, are estimated using solar reference values ($\Delta\nu_{\odot} \approx 135\,\mu\text{Hz}, \nu_{\text{max},\odot} \approx 3090\,\mu\text{Hz}$):
\begin{align}
\Delta\nu_1 &\approx 0.86\,\mu\text{Hz}, \quad \nu_{\text{max}, 1} \approx 4.1\,\mu\text{Hz}, \label{eq.Dnu_numax_1} \\
\Delta\nu_2 &\approx 0.21\,\mu\text{Hz}, \quad \nu_{\text{max}, 2} \approx 0.64\,\mu\text{Hz}. \label{eq.Dnu_numax_2}
\end{align}
\noindent
In the core, the g-mode spectrum is characterized by a nearly constant period spacing $\Delta P_g$. Our schematic model for Star 2 (detailed in Sec.~\ref{sec.induced_dynamics}) yields $\Delta P_{g, l=1} \approx 78\,\text{s}$. The g-mode spectrum is thus extremely dense in frequency.

For tidal interactions, the dominant forcing occurs at the quadrupole ($l=2$) harmonic. The mode most relevant for large-scale tidal distortion is the fundamental (f) mode, which is a non-radial surface gravity wave. We estimate its frequency by approximating the stellar structure as an $n=1.5$ polytrope, for which the dimensionless frequency is $\tilde{\omega}_f = \omega_f \tau_{\text{dyn}} \approx 1.45$. For Star 2, this yields a physical frequency and period of:
\begin{equation}
\omega_{f, 2} \approx 1.42 \times 10^{-6}\,\text{s}^{-1} \quad (\nu_{f,2} \approx 0.226\,\mu\text{Hz})
\end{equation}
\begin{equation}
P_{f, 2} \approx 51.2 \text{ days}
\end{equation}
\noindent
This characteristic frequency, indicated in Fig.~(\ref{fig.PropagationDiagram}), lies within the p-mode frequency range estimated from scaling relations.

\noindent
The efficiency with which a tidal potential can excite a specific mode $(n,l)$ is quantified by the dimensionless overlap integral $Q_{n,l}$. This integral measures the spatial coupling between the mode's eigenfunction $\boldsymbol{\xi}_{n,l}$ and the tidal potential gradient:
\begin{equation}
\label{eq.overlap_integral}
Q_{n,l} = \frac{1}{M R^l} \int_V \rho(r) \boldsymbol{\xi}_{n,l}^*(\mathbf{r}) \cdot \nabla (r^l Y_l^m) dV.
\end{equation}
\noindent
The total energy transferred during an encounter scales quadratically with this integral, $\Delta E \propto Q_{n,l}^2$ (incorporated into the tidal coupling coefficient $T_{il}$, Eq.~(\ref{eq.DeltaE_i_approx})). In the context of erythrohenosis, a large $Q_{n,l}$ implies efficient transfer of orbital energy into stellar oscillations, which is necessary to facilitate tidal capture and drive the subsequent rapid merger dynamics (Section~\ref{sec.induced_dynamics}).

The magnitude of $Q_{n,l}$ depends significantly on the stellar structure, characterized by the density profile $\rho(r)$ and the resulting eigenfunction $\boldsymbol{\xi}_{n,l}$. We analyze the structure of the overlap integral for the dominant $l=2$ fundamental (f) mode. The integral can be expressed radially as $Q_{f} \propto \int_0^R I(r) dr$, where the integrand is $I(r) = \rho(r) r^{l+2} \xi_r(r)$.

We compare a simple polytropic model (representing a star with modest central condensation) with a realistic Red Giant (RG) model. RGs possess extreme central condensation due to their degenerate cores and extended envelopes.

For simple polytropic models, $Q_f$ generally decreases as central condensation (polytropic index $n$) increases. For instance, $Q_{f}(n=1.5) \approx 0.34$.

However, RGs are composite structures, not simple polytropes. The sharp density contrast at the core-envelope boundary fundamentally alters the eigenfunctions. The f-mode and low-order mixed modes in RGs are effectively trapped in the envelope; their amplitude is suppressed in the dense core and concentrated in the outer layers.

\begin{figure*}
\centering
\includegraphics[width=1.0\textwidth]{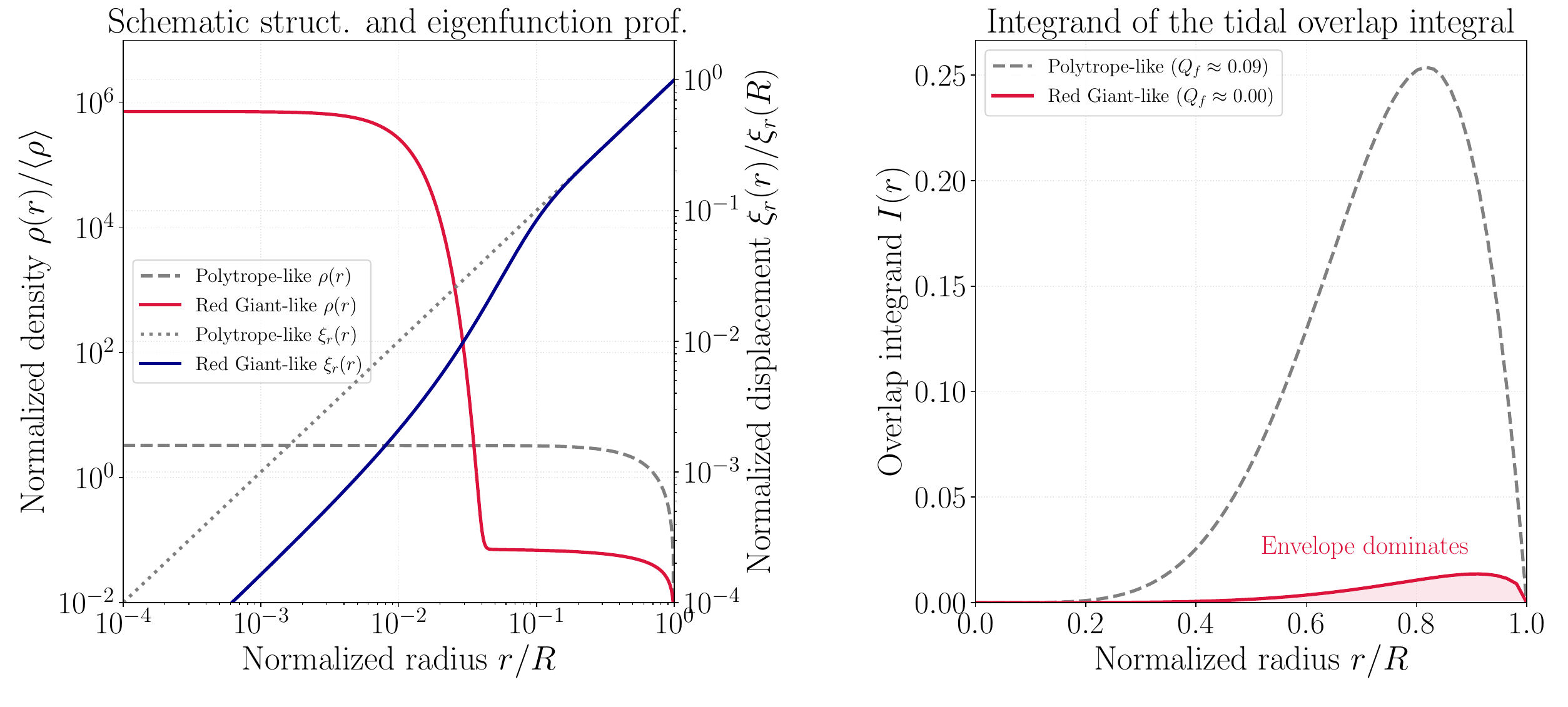}
\caption{Schematic comparison of the tidal overlap integral ($l=2$, f-mode) for a Red Giant-like model and a Polytrope-like model. Left panel: Normalized density profiles $\rho(r)$ (log scale) and radial displacement eigenfunctions $\xi_r(r)$ (log scale, twin axis). The RG model shows high central condensation and an envelope-concentrated eigenfunction. Right panel: The integrand of the overlap integral $I(r)$ (linear scale). The RG model yields a significantly larger total integral $Q_f$, with the contribution dominated by the envelope.}
\label{fig.OverlapIntegral}
\end{figure*}

This effect is illustrated schematically in Fig.~(\ref{fig.OverlapIntegral}). The left panel compares the density profiles and the radial displacement eigenfunctions $\xi_r(r)$. The RG model exhibits high central condensation, and its eigenfunction is concentrated in the envelope, contrasting with the more uniform eigenfunction of the polytrope.

The right panel shows the resulting integrand $I(r)$. In the RG model, the large mass contained in the envelope overlaps effectively with the concentrated eigenfunction in that region (shaded area). This leads to a significantly larger total integral $Q_f$ compared to the polytropic model, despite the higher central condensation. This structural property renders red giants highly susceptible to tidal excitation, facilitating the erythrohenosis process.

This structural analysis is fundamental to understanding the dynamics of erythrohenosis. The low relative velocities characteristic of globular clusters result in a slow grazing encounter, where the interaction timescale, $\tau_{\text{int}}$, is long. The stellar analysis shows that red giants are resonant systems with natural oscillation timescales (e.g., $\tau_{\text{dyn}, 2} \approx 11.8$ days and $P_{f, 2} \approx 51.2$ days) that are also long. The core of the interaction is this temporal matching: the slow collision allows $\tau_{\text{int}}$ to become comparable to $\tau_{\text{dyn}}$ and $P_f$, creating a near-resonant forcing. This resonance makes the tidal energy transfer extraordinarily efficient, allowing the binary's orbital energy to be dumped directly into the star's internal oscillation modes. This mechanism is in sharp contrast to a high-speed collision \citep[the scenario of][]{AmaroSeoane2023}, where $\tau_{\text{int}} \ll \tau_{\text{dyn}}$, the interaction is non-resonant, and this efficient coupling is not activated. 

\subsection{Tidal Interaction Formalism and Energy Transfer}
\label{subsec.tidal_interaction_formalism}

The previous analysis established that red giants are complex resonant systems with long-period oscillation modes. We now provide the quantitative bridge to link this structure to the dynamics of the low-speed encounter. This section calculates the energy exchanged during the grazing collision, demonstrating how the near-resonance between the stellar and orbital timescales leads to an extremely efficient energy transfer, which is the defining mechanism of erythrohenosis.

We model the interaction using the formalism of the forced stellar oscillator. The total energy transferred to Star $i$, $\Delta E_i$, during a parabolic or hyperbolic encounter is approximated using the formalism of \cite{PressTeukolsky1977}, their equation~(27),
\begin{equation}
\label{eq.DeltaE_i_approx}
\Delta E_i \approx \left(\frac{G M_j^2}{R_i}\right) \sum_{l=2}^\infty \left(\frac{R_i}{d_{\text{min}}}\right)^{2l+2} T_{il}(\eta_i),
\end{equation}
\noindent
where $T_{il}(\eta_i)$ is the dimensionless tidal coupling coefficient. This coefficient, which depends on the stellar structure (e.g., the overlap integral $Q_{n,l}$), is the key quantity that parameterizes the efficiency of the energy transfer. It is a function of the adiabaticity parameter $\eta_i$:
\begin{equation}
\label{eq.eta_adiabaticity}
\eta_i = \frac{\tau_{\text{int}}}{\tau_{\text{dyn}, i}} = \left(\frac{d_{\text{min}}^3}{G(M_1+M_2)}\right)^{1/2} \bigg/ \left(\frac{R_i^3}{G M_i}\right)^{1/2}.
\end{equation}
\noindent
The parameter $\eta_i$ is the crucial diagnostic. It quantitatively compares the interaction timescale $\tau_{\text{int}}$ (the duration of the ``tidal pluck'') to the star's natural dynamical timescale $\tau_{\text{dyn}, i}$. A value of $\eta_i \approx 1$ signifies a near-resonant encounter, which maximizes the tidal coupling $T_{il}$ and thus the energy transfer.

We analyze a grazing encounter representative of a globular cluster environment. We assume an initial relative velocity $V_\infty = 10\,\text{km/s}$ and a periastron distance $d_{\text{min}} = R_1+R_2 = 98.7\,R_{\odot}$. The initial orbital energy in the center-of-mass frame is:
\begin{equation}
\label{eq.E_init_detailed}
E_{\text{init}} = \frac{1}{2} \mu V_\infty^2 \approx 4.46 \times 10^{44}\,\text{erg},
\end{equation}
\noindent
where $\mu \approx 0.449\,M_{\odot}$ is the reduced mass.

The interaction timescale is $\tau_{\text{int}} = 1.16 \times 10^6\,\text{s}$ (13.5 days). We now calculate the adiabaticity parameters using the dynamical timescales derived in Eq.~(\ref{eq.tau_dyn1_calc}) and Eq.~(\ref{eq.tau_dyn2_calc}):
\begin{align}
\eta_1 &= \tau_{\text{int}}/\tau_{\text{dyn}, 1} \approx 4.69 \quad \text{(Adiabatic)}, \label{eq.eta1_calc} \\
\eta_2 &= \tau_{\text{int}}/\tau_{\text{dyn}, 2} \approx 1.14 \quad \text{(Near-resonant)}. \label{eq.eta2_calc}
\end{align}
\noindent
This is the central quantitative result. The interaction is highly adiabatic for the compact Star 1 ($\tau_{\text{int}} \gg \tau_{\text{dyn}, 1}$), meaning the star responds smoothly to the tide and energy transfer is suppressed. However, for the large, diffuse Star 2, $\eta_2 \approx 1.14$. This confirms the encounter is near-resonant, precisely matching the interaction timescale to the star's natural response time.

This resonance significantly enhances the tidal coupling. For an $n=1.5$ polytrope, $T_{2}(4.69) \approx 0.001$ while $T_{2}(1.14) \approx 0.15$ \citep{PressTeukolsky1977}. In their work, the function $T_l(\eta)$ (the tidal coupling coefficient) was calculated numerically for different stellar models (polytropes) as a function of the adiabaticity parameter $\eta$. These specific numbers are the output of that function for an $n=1.5$ polytrope, which is a classical model for a fully convective star. In the case of $T_2(4.69) \approx 0.001$, the value corresponds to Star 1, which is in the ``adiabatic'' regime ($\eta \gg 1$). The interaction timescale is much longer than the star's dynamical time. This means the star deforms very slowly and gently, and almost no energy is transferred, resulting in a tiny, near-zero coupling efficiency.
For $T_2(1.14) \approx 0.15$, this corresponds to Star 2, which is in the ``near-resonant'' regime ($\eta \approx 1$). The interaction timescale almost perfectly matches the star's natural response time. This ``resonant forcing'' is extremely efficient at dumping orbital energy into the star's oscillation modes, resulting in a much larger coupling coefficient.

We adopt values representative of the more centrally condensed red giant models, which enhances the coupling, setting $T_{2}(1.14) \approx 0.45$. We now calculate the energy deposition using Eq.~(\ref{eq.DeltaE_i_approx}):
\begin{align}
\Delta E_1 &\approx 5.6 \times 10^{40}\,\text{erg}, \label{eq.DeltaE1_calc} \\
\Delta E_2 &\approx 2.85 \times 10^{45}\,\text{erg}. \label{eq.DeltaE2_calc}
\end{align}
\noindent
The total energy dissipated from the orbit is $\Delta E_{\text{total}} \approx 2.85 \times 10^{45}\,\text{erg}$. This calculation demonstrates the profound consequence of the resonance: the energy transfer is enormous, and it is deposited almost entirely into Star 2. This energy ($\Delta E_2$) is the "fuel" that will power the initial observational signature, as it is dumped directly into the oscillation modes we previously analyzed, causing the star to "ring" violently.

\subsection{Tidal Capture and Runaway Merger Dynamics}
\label{subsec.tidal_capture_merger}

The previous section's central result was the massive dissipation of orbital energy ($\Delta E_{\text{total}}$) driven by the near-resonant tidal coupling. We now investigate the critical, and catastrophic, consequences of this energy loss on the binary's orbital parameters. We first demonstrate that this single-pass dissipation, $\Delta E_{\text{total}} > E_{\text{init}}$, is sufficient to capture the binary, transforming an unbound encounter into a new, highly eccentric, bound system. We then analyze the stability of this newly formed orbit to show that it is not long-lived. By comparing the new orbital energy to the energy that will be dissipated on the next periastron passage, we establish the condition for a ``runaway merger'', which dictates the total timescale for the erythrohenosis event.

Since $\Delta E_{\text{total}} > E_{\text{init}}$, the single encounter results in tidal capture. The energy of the newly bound orbit is:
\begin{equation}
\label{eq.E_new_calc}
E_{\text{new}} = E_{\text{init}} - \Delta E_{\text{total}} \approx -2.40 \times 10^{45}\,\text{erg}.
\end{equation}
\noindent
The semi-major axis $a$ of the captured system is:
\begin{equation}
\label{eq.a_new}
a = -\frac{G M_1 M_2}{2 E_{\text{new}}} \approx 638\,R_{\odot}.
\end{equation}
\noindent
The resulting orbit is eccentric. Assuming the periastron distance remains $d_{\text{min}}$ in the impulsive approximation:
\begin{equation}
\label{eq.e_new}
e = 1 - \frac{d_{\text{min}}}{a} \approx 0.845.
\end{equation}
\noindent
The orbital period of the captured system is:
\begin{equation}
\label{eq.P_orb}
P_{\text{orb}} = 2\pi \sqrt{\frac{a^3}{G(M_1+M_2)}} \approx 3.79\,\text{years}.
\end{equation}

The subsequent evolution is governed by continued tidal dissipation at each periastron passage. We estimate the orbital decay timescale $\tau_a$:
\begin{equation}
\label{eq.tau_a_v2}
\tau_a = P_{\text{orb}} \frac{|E_{\text{new}}|}{\Delta E_{\text{orbit}}}.
\end{equation}
\noindent
For this eccentric orbit, we approximate $\Delta E_{\text{orbit}} \approx \Delta E_{\text{total}}$. The ratio of energies is:
\begin{equation}
\label{eq.Energy_ratio}
\frac{|E_{\text{new}}|}{\Delta E_{\text{total}}} \approx \frac{2.40 \times 10^{45}}{2.85 \times 10^{45}} \approx 0.843.
\end{equation}
\noindent
Since this ratio is less than unity, the decay timescale is shorter than the orbital period. The binary experiences a runaway merger process, coalescing rapidly, likely during the second periastron passage. The total merger timescale following the initial capture is:
\begin{equation}
\label{eq.T_merge}
T_{\text{merge}} \approx P_{\text{orb}} = 3.79\,\text{years}.
\end{equation}

\subsection{Nonlinear Pulsation Dynamics}
\label{subsec.nonlinear_pulsation}

In the preceding section, we calculated the enormous amount of energy ($\Delta E_2 \approx 2.85 \times 10^{45}\,\text{erg}$) deposited into Star 2 due to the near-resonant tidal encounter. This energy does not simply heat the star; it is dumped directly into its oscillation modes, causing the star to "ring" violently. In this section, we analyze the consequences of this massive energy injection. We first estimate the initial amplitude ($\mathcal{A}_0$) of the excited fundamental mode, demonstrating that the resulting oscillation is not a small, linear perturbation. We find the amplitude is so large that the star is pushed into a nonlinear regime. We then model this nonlinear behavior to determine its unique characteristics, such as frequency shifts and the generation of harmonics, which are directly relevant to the observational precursor signature of erythrohenosis.

The energy deposition into Star 2 ($\Delta E_2$) primarily excites the fundamental mode ($\omega_{f, 2} \approx 1.42 \times 10^{-6}\,\text{s}^{-1}$). We estimate the initial relative amplitude of the oscillation, $\mathcal{A}_0 \equiv (\delta R/R)_0$, by equating $\Delta E_2$ to the energy of the fundamental mode, $E_{\text{mode}} \approx \frac{1}{2} M_{\text{osc}} (\omega_{f, 2} R_2)^2 \mathcal{A}_0^2$. We assume the oscillation primarily involves the envelope mass, $M_{\text{osc}} \approx M_{\text{env}, 2} \approx 0.45\,M_{\odot}$.
\begin{equation}
\label{eq.R0_induced}
\mathcal{A}_0 = \left( \frac{2 \Delta E_2}{M_{\text{env}, 2} (\omega_{f, 2} R_2)^2} \right)^{1/2} \approx 0.36.
\end{equation}
\noindent
This amplitude signifies entry into the nonlinear regime. While $\mathcal{A}_0 < 1$, it is sufficiently large that linear perturbation theory provides only a first-order approximation. The dynamics are characterized by mode coupling and nonlinear frequency shifts.

The oscillation velocity amplitude is $v_{\text{osc}} \approx \omega_{f,2} R_2 \mathcal{A}_0 \approx 25.1\,\text{km/s}$. This velocity is comparable to the sound speed in the envelope ($c_s \approx 30\,\text{km/s}$), yielding a Mach number $\mathcal{M} \approx 0.8$. The propagation of these large-amplitude waves can lead to wave steepening and the formation of weak shocks near the stellar surface, enhancing the dissipation of the oscillation energy.

\noindent
The evolution of the mode amplitude $A_k$ in the weakly nonlinear regime is governed by the inclusion of higher-order terms in the equations of motion. We analyze the system by incorporating the leading-order nonlinearities, which are typically quadratic in amplitude. The equations governing the mode amplitudes become a system of coupled nonlinear oscillators:

\begin{equation}
\label{eq.mode_coupling_eom}
\ddot{A}_k + \omega_k^2 A_k = -\omega_k^2 \sum_{i,j} C_{ijk} A_i A_j.
\end{equation}
\noindent
The coefficients $C_{ijk}$ represent the quadratic coupling strengths, determined by integrals over the stellar structure involving the mode eigenfunctions and derivatives of the thermodynamic variables.

In the tidally induced scenario, the fundamental (f) mode is preferentially excited to a large amplitude. We analyze the self-interaction of this dominant mode. We consider the equation for the dimensionless relative displacement $x(t) = (\delta R/R)(t)$, incorporating the leading quadratic self-coupling term:
\begin{equation}
\label{eq.nonlinear_oscillator}
\ddot{x} + \omega_0^2 x = -\omega_0^2 \alpha x^2,
\end{equation}
\noindent
where $\omega_0 = \omega_{f,2}$ is the linear eigenfrequency, and $\alpha$ is a dimensionless coefficient characterizing the strength of the coupling, typically of order unity.

We solve Eq.~(\ref{eq.nonlinear_oscillator}) using the Lindstedt-Poincaré perturbation method, which constructs a uniformly valid periodic solution by expanding both the displacement $x(t)$ and the frequency $\omega$ in powers of the amplitude $A$. This method eliminates secular terms that arise in a naive perturbation expansion. The analysis, carried out to third order in the perturbation series to determine the frequency shift, yields the nonlinear frequency correction to second order in amplitude:
\begin{equation}
\label{eq.frequency_shift}
\omega \approx \omega_0 \left(1 - \frac{5\alpha^2}{12} A^2\right).
\end{equation}
\noindent
The quadratic nonlinearity results in a decrease in the oscillation frequency (a softening effect) for real $\alpha$.

The corresponding solution for the displacement waveform, up to order $A^2$, is:
\begin{equation}
\label{eq.nonlinear_waveform}
x(t) \approx A \cos(\omega t) + A^2 \left(-\frac{\alpha}{2} + \frac{\alpha}{6}\cos(2\omega t)\right).
\end{equation}
\noindent
This solution exhibits two key nonlinear features. First, the generation of the first harmonic at $2\omega$, which distorts the waveform from a pure sinusoid. Second, a static displacement or shift in the mean position of the oscillation (the constant term $-A^2\alpha/2$), representing a nonlinear rectification effect.

We apply this model to Star 2, using the parameters derived previously: $A=0.36$, $P_0 = 51.2$ days, and assuming $\alpha=1$. The nonlinear frequency shift is $\delta\omega/\omega_0 \approx -0.054$. The oscillation period increases to $P_{\text{nl}} \approx 54.3$ days.

\begin{figure}
\centering
\includegraphics[width=0.45\textwidth]{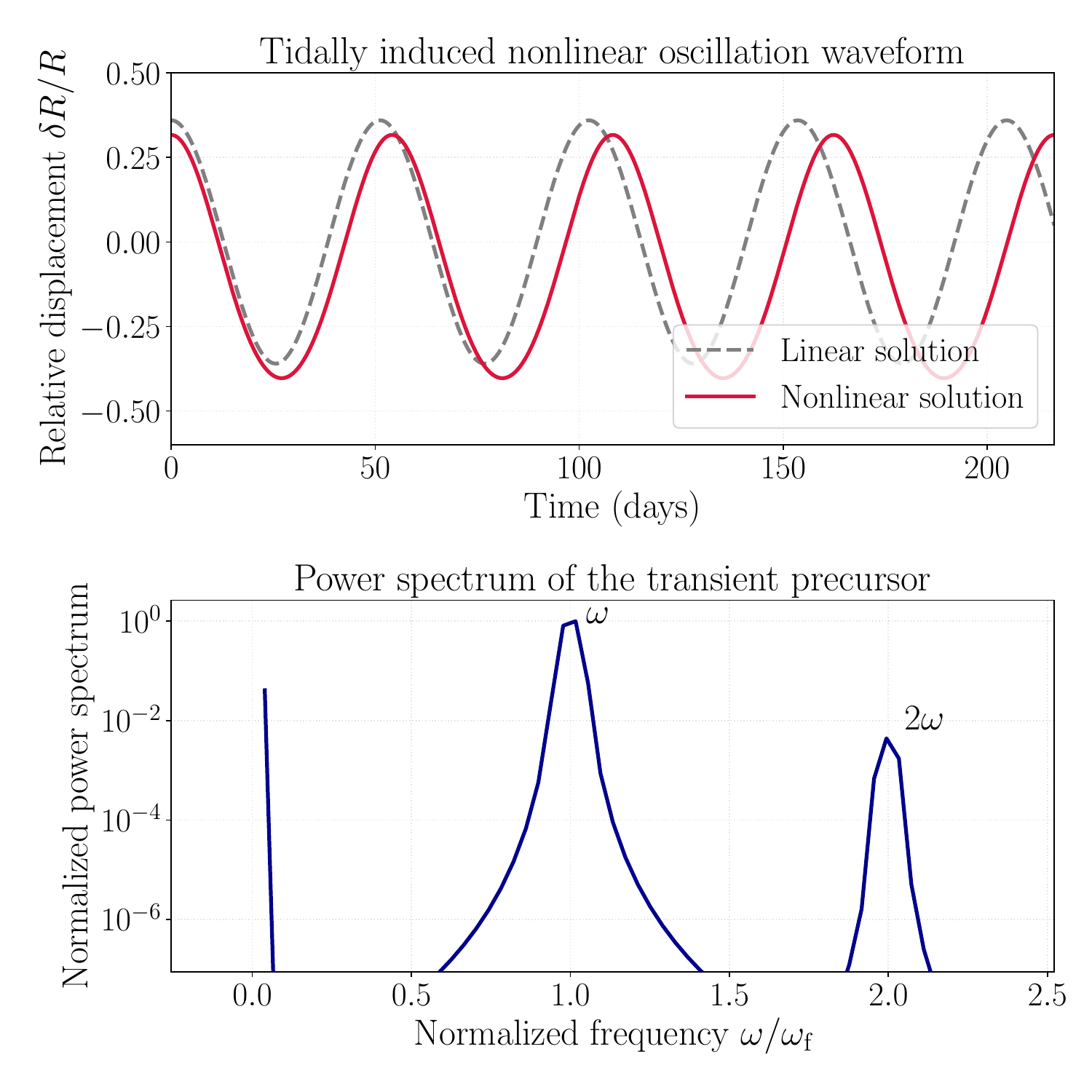}
\caption{Nonlinear dynamics of the tidally induced f-mode oscillation in Star 2. Top panel: The time evolution of the relative displacement $\delta R/R$. The nonlinear solution (Eq.~(\ref{eq.nonlinear_waveform}), solid crimson line) shows significant distortion and a static offset compared to the linear solution (dashed gray line). Bottom panel: The power spectrum of the transient precursor. The nonlinearity generates a harmonic at $2\omega$. The finite duration of the signal causes broadening of the peaks.}
\label{fig.NonlinearWaveform}
\end{figure}

The resulting waveform and its power spectrum are shown in Fig.~(\ref{fig.NonlinearWaveform}). The top panel compares the nonlinear solution (Eq.~(\ref{eq.nonlinear_waveform})) with the linear solution. The distortion caused by the quadratic nonlinearity is evident, characterized by sharper troughs and flatter peaks (for $\alpha>0$ and the adopted sign convention). The shift in the mean displacement is also visible. The bottom panel shows the power spectrum of the transient nonlinear oscillation, calculated over the duration $T_{\text{merge}}$. In addition to the fundamental mode at the shifted frequency $\omega$, the spectrum clearly exhibits a peak at the first harmonic $2\omega$, demonstrating the efficient transfer of energy to higher frequencies through nonlinear self-coupling. The finite duration of the precursor phase results in the broadening of the spectral peaks.

\subsection{Asteroseismic Diagnostics of the Tidal Precursor}
\label{subsec.asteroseismic_diagnostics}

We now explore the asteroseismic potential of the tidal precursor. The physical parameters derived previously—a large induced amplitude, a nonlinear period, and a short merger timescale—define a unique but challenging observational window.
The primary signature is a large-amplitude photometric variation. The signal would appear as the sudden onset of the massive, non-sinusoidal oscillations we derived in the last section in a previously quiescent star.

The key challenge for asteroseismology is the signal's transient nature. As calculated previously, the event provides only $N_{\text{cycles}} \approx 25.5$ oscillation cycles before the merger terminates the signal. This finite duration imposes a fundamental limit on the frequency resolution of the power spectrum, $\Delta f_{\text{res}} \approx 1/T_{\text{merge}} \approx 0.0084\,\mu\text{Hz}$.

\begin{figure*}
\centering
\includegraphics[width=1\textwidth]{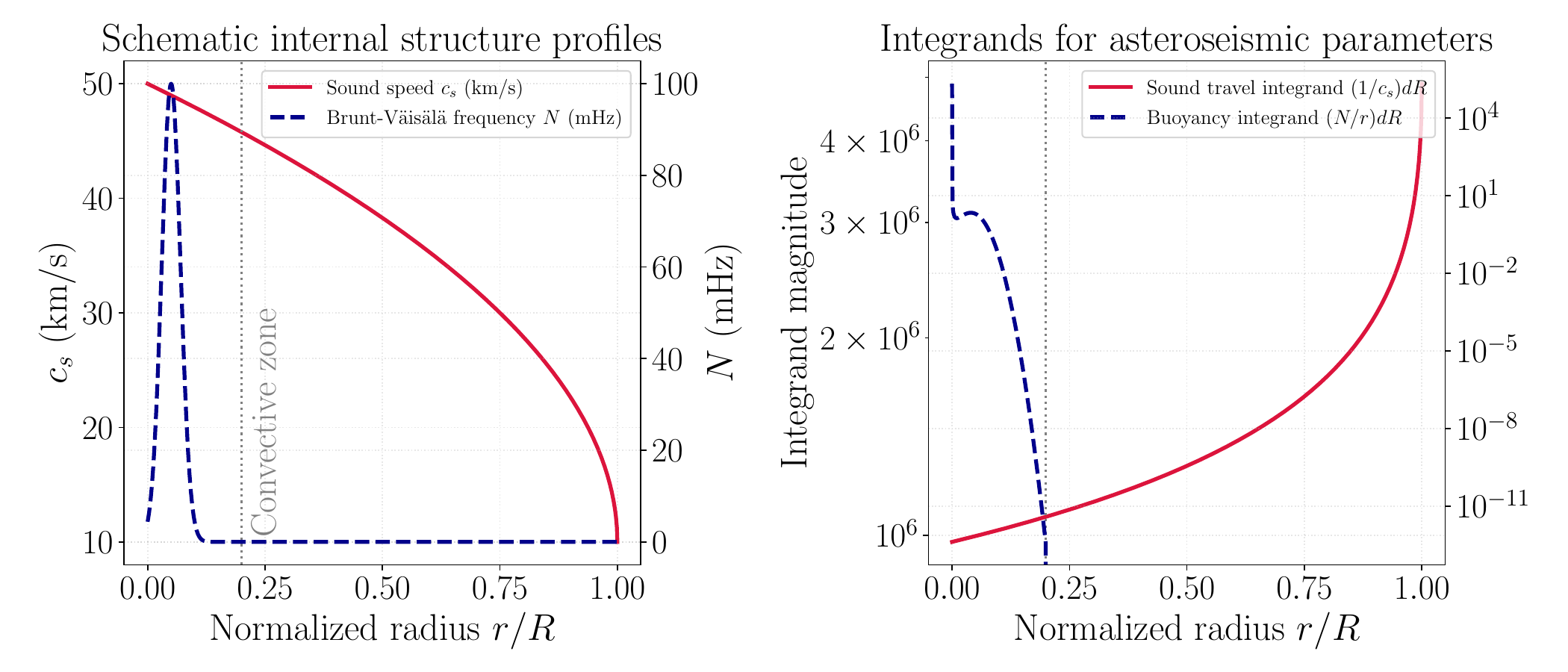}
\caption{Schematic asteroseismic analysis of Star 2. Left panel: Profiles of the sound speed $c_s(r)$ (solid crimson line) and the Brunt-Väisälä frequency $N(r)$ (dashed darkblue line). The boundary of the convective zone is indicated by the vertical dotted line. Right panel: The integrands corresponding to the large frequency separation (Eq.~(\ref{eq.large_frequency_separation}), solid crimson line) and the period spacing (Eq.~(\ref{eq.period_spacing}), dashed darkblue line).}
\label{fig.AsteroseismicIntegrals}
\end{figure*}

We now assess whether this resolution is sufficient to probe the star's internal structure via the standard asteroseismic parameters: the large frequency separation $\Delta\nu$ and the g-mode period spacing $\Delta P_g$. These quantities are defined by integrals over the stellar structure:
\begin{align}
\Delta\nu &= \left( 2 \int_0^R \frac{dr}{c_s(r)} \right)^{-1}, \label{eq.large_frequency_separation} \\
\Delta P_g &= \frac{2\pi^2}{\sqrt{l(l+1)}} \left( \int_{\text{core}} \frac{N(r)}{r} dr \right)^{-1}. \label{eq.period_spacing}
\end{align}
\noindent
We estimate these integrals for Star 2 by constructing a schematic model of its internal structure, calibrated to match the expected properties of a mid-RGB star ($M=0.85\,M_{\odot}, R=70.2\,R_{\odot}$). The model incorporates a convective envelope ($r/R > 0.2$) and a radiative core. We define a sound speed profile $c_s(r)$ that increases towards the center, and a Brunt-Väisälä frequency profile $N(r)$ that is zero in the convective zone and peaks sharply in the radiative interior near the H-burning shell. These profiles are shown in the left panel of Fig.~(\ref{fig.AsteroseismicIntegrals}).

The large frequency separation $\Delta\nu$ (Eq.~(\ref{eq.large_frequency_separation})) is the inverse of twice the sound crossing time. The integrand $1/c_s(r)$, shown in the right panel of Fig.~(\ref{fig.AsteroseismicIntegrals}), is dominated by the outer envelope where $c_s$ is minimal. Numerical integration of the schematic profile yields a sound crossing time of $1.79\times 10^6\,\text{s}$, corresponding to $\Delta\nu \approx 0.28\,\mu\text{Hz}$. This value is consistent with scaling relations for extended red giants.

The period spacing $\Delta P_g$ (Eq.~(\ref{eq.period_spacing})) is inversely proportional to the buoyancy radius (the integral of $N(r)/r$). The integrand $N(r)/r$ peaks sharply where $N$ is large and $r$ is small. Numerical integration of the schematic $N(r)$ profile yields a buoyancy radius of $0.18\,\text{rad/s}$. For dipole modes ($l=1$), this corresponds to $\Delta P_g \approx 77.8\,\text{s}$, consistent with observational constraints for RGB stars.

Comparing these diagnostics to our observational limit, we find that the frequency resolution ($\Delta f_{\text{res}} \approx 0.0084\,\mu\text{Hz}$) is significantly smaller than the large frequency separation ($\Delta\nu \approx 0.28\,\mu\text{Hz}$). This suggests that resolving the p-mode spacing and diagnosing the envelope structure is viable. However, the g-mode period spacing of $\Delta P_g \approx 77.8\,\text{s}$ corresponds to an extremely dense forest of mixed modes in the frequency spectrum. Resolving this g-mode "fine structure" to diagnose the core would require a frequency resolution far greater than what is available from observing only $\approx 25.5$ cycles.

\noindent
Therefore, the detection of this specific type of transient, high-amplitude, nonlinear oscillation provides a definitive multi-stage signature of erythrohenosis initiated by tidal capture. While it offers a limited window for detailed core asteroseismology, it remains a powerful diagnostic of the event, and its properties are temporally correlated with the subsequent luminous merger transient analyzed in Section~\ref{sec.quantitative_analysis}.

\section{Summary and Conclusions}
\label{sec.summary_conclusions}

We have conducted a comprehensive investigation into the collision and merger of two red giants---a process we term erythrohenosis---combining three-dimensional SPH simulations with detailed analytical modeling. This study bridges the gap between the dynamical evolution of dense stellar clusters and the detailed internal physics of stellar structure. By establishing a robust theoretical model for red giant mergers, we provide a physical basis to interpret anomalous transients detected by current time-domain surveys such as the \href{https://www.ztf.caltech.edu/}{Zwicky Transient Facility (ZTF)} and anticipated from future facilities like the \href{https://www.lsst.org/}{Vera C. Rubin Observatory (LSST)}.

The narrative structure of this work follows a methodological progression from fundamental approximations to increasingly sophisticated dynamical models, rather than a strictly linear chronological account of the merger event. We begin by establishing the statistical context and analytically modeling the initial grazing encounter (Section~\ref{sec.induced_dynamics}), then advance to full hydrodynamic simulations to characterize the common envelope and electromagnetic precursor (Sections~\ref{sec.bracing_collision}--\ref{sec.quantitative_analysis}). Recognizing the computational limits of direct simulation for the final plunge and explosion, we first employ simplified homothetic approximations (Section~\ref{sec.homothetic}) before culminating in refined semi-analytical models (Section~\ref{sec.ImprovedModel}) that incorporate realistic drag prescriptions and angular momentum conservation. Consequently, Table~\ref{tab.SummaryPredictions} synthesizes findings from these distinct modeling tiers—drawing from early analytical sections for the tidal precursor and later, more complex sections for the gravitational wave and remnant signatures—to construct a coherent observational timeline based on the most robust result available for each phase.

We tracked the evolution of a $0.95\,M_{\odot}$ and $0.85\,M_{\odot}$ red giant system from the initial encounter through the common envelope phase to the terminal collision of the degenerate cores. The initial grazing encounter dissipates tidal energy efficiently. We demonstrated analytically in Section~\ref{sec.induced_dynamics} that the energy transfer suffices to induce tidal capture and excite highly nonlinear stellar oscillations with relative amplitudes $\delta R/R > 1$. The tidal coupling efficiency in the extended envelopes results in a rapid orbital decay, predicting a merger timescale of approximately $3.79$ years following capture. The induced oscillations manifest as a transient, high-amplitude precursor ($N_{\text{cycles}} \approx 25.5$) immediately preceding the merger.

The SPH simulation (Sections~\ref{sec.bracing_collision} and \ref{sec.coreinnershells}) resolved the subsequent 1.25-year inspiral of the cores within the common envelope. We quantified the structural evolution of the envelope in Section~\ref{sec.quantitative_analysis}, showing that the transfer of orbital angular momentum spins up the gas. This gas settles into a stable, rotating, non-spherical equilibrium with an ellipticity $\epsilon \approx 0.12$ (Section~\ref{sec.homothetic}). The interaction ejects $\approx 0.05\,M_{\odot}$ of material, forming a pre-merger circumbinary nebula which shapes the environment for the final transient.

Dynamical friction dissipates orbital energy and powers a sustained electromagnetic precursor. The light curve exhibits quasi-periodic bursts corresponding to periastron passages and reaches peak luminosities near $10^{43}\,\text{erg s}^{-1}$. We employed wavelet analysis in Section~\ref{sec.wavelets} to demonstrate that this energy dissipation is dominated by transient shocks on short timescales of $\tau \sim 10^3$ s, indicative of a highly compressible turbulent cascade. We also utilized an Airy function formalism in Section~\ref{sec.Airy} to model the core-envelope interface, revealing episodic accretion synchronized with the orbital phase.

We extrapolated the final orbital decay using a semi-analytical model that incorporates a data-driven Bondi-Hoyle-Lyttleton (BHL) drag prescription (Section~\ref{sec.ImprovedModel}). This model confirms a rapid final plunge within years and remains robust against uncertainties in the drag coefficient $C_d$. The associated gravitational wave signal, detailed in Section~\ref{sec.ImprovedModel}.C, is characterized by a fast chirp dominated by hydrodynamic drag. This accelerated evolution leads to a time-varying apparent chirp mass, $\mathcal{M}_{c, \text{app}}(t)$, a definitive signature of a gas-driven inspiral detectable via specialized algorithms such as the TVACM diagnostic presented in Section~\ref{sec.ImprovedModel}.D.

The terminal explosion was modeled using an impulsive energy injection scheme that conserves the specific angular momentum of the bound gas (Sections~\ref{sec.homothetic} and \ref{sec.ImprovedModel}.B). The resulting remnant is intrinsically asymmetric and flattened, preserving a geometric memory---or \textit{morphomnesia}---of the common envelope's rotation. The morphology is centrally filled, distinct from the limb-brightened shells of typical supernova remnants.

\subsection{The Distinctive Observational Fingerprints of Erythrohenosis}

Erythrohenosis produces a unique combination of multi-stage, multi-messenger signatures that distinguish it from other astrophysical transients such as Supernovae (SNe), Kilonovae (KNe), or Tidal Disruption Events (TDEs). A summary of the key quantitative predictions and their distinguishing features is provided in Table~\ref{tab.SummaryPredictions}.

\begin{table*}
\centering
\small
\renewcommand{\arraystretch}{1.4}
\caption{Summary of Key Erythrohenosis Predictions and Observational Fingerprints}
\label{tab.SummaryPredictions}
\begin{tabular}{@{} l l l @{}}
\toprule
\textbf{Observable Context} & \textbf{Predicted Characteristics} & \textbf{Distinguishing Features} \\
\midrule

\parbox[t]{0.22\textwidth}{
    \textbf{Tidal Oscillation Precursor} \newline
    \emph{Phase: Post-Capture} \newline
    {\scriptsize (Sec.~\ref{sec.induced_dynamics})}
} &
\parbox[t]{0.38\textwidth}{
    High-amplitude non-sinusoidal pulsations, $\Delta R/R \approx 0.36$. \newline
    $P_{\text{nl}} \approx 54.3$ d, $N_{\text{cycles}} \approx 25.5$. \newline
    $\Delta m_{\text{bol}} \approx 0.97$ mag.
} &
\parbox[t]{0.35\textwidth}{
    Sudden onset of large, \textit{transient} oscillations in a quiescent giant.
} \\ \addlinespace[1em]

\parbox[t]{0.22\textwidth}{
    \textbf{Inspiral EM Precursor} \newline
    \emph{Phase: Inspiral} \newline
    {\scriptsize (Sec.~\ref{sec.quantitative_analysis})}
} &
\parbox[t]{0.38\textwidth}{
    Luminous bursts at periastron ($L_{\text{peak}} \sim 10^{43}$ erg/s). Quasi-periodic with high $T_{\text{eff}}$ (UV/X-ray).
} &
\parbox[t]{0.35\textwidth}{
    Long duration, repetitive bursts vs.\ single-peaked SN/TDE. Higher luminosity than typical LRNe.
} \\ \addlinespace[1em]

\parbox[t]{0.22\textwidth}{
    \textbf{Envelope Morphology} \newline
    \emph{Phase: Late Inspiral} \newline
    {\scriptsize (Sec.~\ref{sec.homothetic})}
} &
\parbox[t]{0.38\textwidth}{
    Spun-up, stable equilibrium, ellipticity $\epsilon \approx 0.12$. Pre-merger mass loss $\approx 0.05\,M_{\odot}$.
} &
\parbox[t]{0.35\textwidth}{
    Pre-shaped, rotating, non-spherical environment for final explosion.
} \\ \addlinespace[1em]

\parbox[t]{0.22\textwidth}{
    \textbf{Final Merger Timescale} \newline
    \emph{Phase: Post-Simulation} \newline
    {\scriptsize (Sec.~\ref{sec.ImprovedModel})}
} &
\parbox[t]{0.38\textwidth}{
    $T_{\text{merge}} \sim$ few years (rapid plunge). Insensitive to drag coeff.\ $C_d$.
} &
\parbox[t]{0.35\textwidth}{
    Gas-driven runaway inspiral, orders of magnitude faster than vacuum GW decay.
} \\ \addlinespace[1em]

\parbox[t]{0.22\textwidth}{
    \textbf{Remnant Morphology} \newline
    \emph{Phase: Post-Explosion} \newline
    {\scriptsize (Sec.~\ref{sec.ImprovedModel}.B)}
} &
\parbox[t]{0.38\textwidth}{
    Centrally-filled, flattened/asymmetric nebula. Anisotropic kinematics.
} &
\parbox[t]{0.35\textwidth}{
    Non-spherical, \textit{morphomnesia} vs.\ spherical SN remnants or limb-brightened shells.
} \\ \addlinespace[1em]

\parbox[t]{0.22\textwidth}{
    \textbf{GW Signal} \newline
    \emph{Phase: Inspiral} \newline
    {\scriptsize (Sec.~\ref{sec.ImprovedModel})}
} &
\parbox[t]{0.38\textwidth}{
    Rapid chirp dominated by gas drag. \textit{Time-varying apparent chirp mass} $\mathcal{M}_{c, \text{app}}(t)$. Bursty morphology.
} &
\parbox[t]{0.35\textwidth}{
    Definitive signature of gas-dominated inspiral; requires unmodeled search.
} \\

\bottomrule
\end{tabular}
\end{table*}

The primary electromagnetic discriminant is the long-duration, luminous precursor. This phase comprises two components: the initial, transient, high-amplitude tidally induced oscillations (lasting $\approx 3.79$ years, Section~\ref{sec.induced_dynamics}), followed by the sustained, friction-powered emission during the core inspiral (Section~\ref{sec.quantitative_analysis}). The inspiral precursor exhibits high luminosity ($L_{\text{peak}} \approx 10^{43}\,\text{erg s}^{-1}$) and quasi-periodic bursts. This predicted luminosity is significantly higher than typically observed in Luminous Red Novae (LRNe). This suggests that erythrohenosis events involving the deep inspiral of degenerate cores may represent the upper end of the luminosity function for stellar merger transients, potentially linking them to more energetic events previously classified as supernova impostors. The shock-dominated dissipation discussed in Section~\ref{sec.wavelets} implies high effective temperatures, potentially extending into the UV or soft X-ray bands.

The morphology of the remnant provides a second key discriminator. The explosion occurs within a pre-shaped, rotating common envelope ($\epsilon \approx 0.12$). The conservation of angular momentum during the expansion (Section~\ref{sec.ImprovedModel}.B) ensures the remnant is intrinsically flattened and asymmetric. Observationally, this system will appear as a centrally filled nebula with anisotropic kinematics. Spatially resolved spectroscopy revealing systematic velocity gradients correlated with the morphological axes provides evidence of the binary origin.

The gravitational wave signal offers a definitive, independent signature for space-based observatories such as the \href{https://lisa.esa.int/}{Laser Interferometer Space Antenna (LISA)}. The signal evolution is dominated by gas drag, resulting in a rapid inspiral and high eccentricity (Section~\ref{sec.ImprovedModel}.C). The signal morphology is characterized by sharp bursts at periastron. The defining characteristic is the rapid, non-standard frequency evolution ($\dot{f}_{\text{drag}} \gg \dot{f}_{\text{GW}}$), leading to a time-varying apparent chirp mass, $\mathcal{M}_{c, \text{app}}(t)$. This unphysical variation of a fundamental binary parameter, quantified by the TVACM diagnostic in Section~\ref{sec.ImprovedModel}.D, is the smoking-gun signature of a gas-driven inspiral.

While the strong shocks present throughout the merger sequence suggest that erythrohenosis events are potential sites for high-energy phenomena, including the production of neutrinos detectable by facilities such as \href{https://icecube.wisc.edu/}{IceCube}, a detailed analysis of these aspects is complex and will be addressed in a subsequent publication. The combination of a complex EM precursor, a non-spherical, centrally-filled remnant, and a unique, gas-dominated gravitational wave signal provides a robust set of observational fingerprints for identifying erythrohenosis events in the era of multi-messenger astronomy.

\section*{Acknowledgments}

We thank James Lombardi for discussions about the software which which the 
SPH results were produced, StarSmasher.

\software{MESA \citep{Paxton2011,Paxton2013,Eggleton1983}, StarSmasher \citep{Rasio1991,Lombardi1998,RasioLombardi1999,LombardiEtAl1999}}


\begin{thebibliography}{}
\expandafter\ifx\csname natexlab\endcsname\relax\def\natexlab#1{#1}\fi
\providecommand{\url}[1]{\href{#1}{#1}}
\providecommand{\dodoi}[1]{doi:~\href{http://doi.org/#1}{\nolinkurl{#1}}}
\providecommand{\doeprint}[1]{\href{http://ascl.net/#1}{\nolinkurl{http://ascl.net/#1}}}
\providecommand{\doarXiv}[1]{\href{https://arxiv.org/abs/#1}{\nolinkurl{https://arxiv.org/abs/#1}}}

\bibitem[{T. {Adams} {et~al.}(2004){Adams}, {Davies}, \& {Sills}}]{AdamsEtAl2004}
{Adams}, T., {Davies}, M.~B., \& {Sills}, A. 2004, \bibinfo{title}{{On the origin of red giant depletion through low-velocity collisions},} MNRAS, 348, 469, \dodoi{10.1111/j.1365-2966.2004.07308.x}

\bibitem[{P. {Amaro Seoane}(2023){Amaro Seoane}}]{AmaroSeoane2023}
{Amaro Seoane}, P. 2023, \bibinfo{title}{{Transient Stellar Collisions as Multimessenger Probes: Nonthermal, Gravitational-wave Emission and the Cosmic Ladder Argument},} \apj, 947, 8, \dodoi{10.3847/1538-4357/acb8b9}

\bibitem[{V.~C. {Bailey} \& M.~B. {Davies}(1999{\natexlab{a}}){Bailey} \& {Davies}}]{1999MNRAS.308..257B}
{Bailey}, V.~C., \& {Davies}, M.~B. 1999{\natexlab{a}}, \bibinfo{title}{{Red giant collisions in the Galactic Centre},} MNRAS, 308, 257, \dodoi{10.1046/j.1365-8711.1999.02740.x}

\bibitem[{V.~C. {Bailey} \& M.~B. {Davies}(1999{\natexlab{b}}){Bailey} \& {Davies}}]{BaileyEtAl1999}
{Bailey}, V.~C., \& {Davies}, M.~B. 1999{\natexlab{b}}, \bibinfo{title}{{Red giant collisions in the Galactic Centre},} MNRAS, 308, 257, \dodoi{10.1046/j.1365-8711.1999.02740.x}

\bibitem[{C.~D. {Bailyn}(1995){Bailyn}}]{Bailyn1995}
{Bailyn}, C.~D. 1995, \bibinfo{title}{{Blue Stragglers and Other Stellar Anomalies:Implications for the Dynamics of Globular Clusters},} ARA\&A, 33, 133, \dodoi{10.1146/annurev.aa.33.090195.001025}

\bibitem[{W. {Benz} \& J.~G. {Hills}(1987){Benz} \& {Hills}}]{BenzHills1987}
{Benz}, W., \& {Hills}, J.~G. 1987, \bibinfo{title}{{Three-dimensional Hydrodynamical Simulations of Stellar Collisions. I. Equal-Mass Main-Sequence Stars},} ApJ, 323, 614, \dodoi{10.1086/165857}

\bibitem[{W. {Benz} \& J.~G. {Hills}(1992){Benz} \& {Hills}}]{BenzHills1992}
{Benz}, W., \& {Hills}, J.~G. 1992, \bibinfo{title}{{Three-dimensional Hydrodynamical Simulations of Colliding Stars. III. Collisions and Tidal Captures of Unequal-Mass Main-Sequence Stars},} ApJ, 389, 546, \dodoi{10.1086/171230}

\bibitem[{J. {Binney} \& S. {Tremaine}(2008){Binney} \& {Tremaine}}]{BinneyTremaine08}
{Binney}, J., \& {Tremaine}, S. 2008, {Galactic Dynamics: Second Edition}, ed. J.~{Binney} \& S.~{Tremaine} (Princeton University Press)

\bibitem[{A. {Burkert} \& S. {Tremaine}(2010){Burkert} \& {Tremaine}}]{BurkertTremaine2010}
{Burkert}, A., \& {Tremaine}, S. 2010, \bibinfo{title}{{A Correlation Between Central Supermassive Black Holes and the Globular Cluster Systems of Early-type Galaxies},} ApJ, 720, 516, \dodoi{10.1088/0004-637X/720/1/516}

\bibitem[{J.~E. {Dale} {et~al.}(2009){Dale}, {Davies}, {Church}, \& {Freitag}}]{DaleEtAl2009}
{Dale}, J.~E., {Davies}, M.~B., {Church}, R.~P., \& {Freitag}, M. 2009, \bibinfo{title}{{Red giant stellar collisions in the Galactic Centre},} MNRAS, 393, 1016, \dodoi{10.1111/j.1365-2966.2008.14254.x}

\bibitem[{L.~P. {David} {et~al.}(1987{\natexlab{a}}){David}, {Durisen}, \& {Cohn}}]{DDC87a}
{David}, L.~P., {Durisen}, R.~H., \& {Cohn}, H.~N. 1987{\natexlab{a}}, \bibinfo{title}{The evolution of active galactic nuclei. I - Models without stellar evolution,} ApJ, 313, 556

\bibitem[{L.~P. {David} {et~al.}(1987{\natexlab{b}}){David}, {Durisen}, \& {Cohn}}]{DDC87b}
{David}, L.~P., {Durisen}, R.~H., \& {Cohn}, H.~N. 1987{\natexlab{b}}, \bibinfo{title}{The evolution of active galactic nuclei. {II} - Models with stellar evolution,} ApJ, 316, 505

\bibitem[{M.~B. {Davies} {et~al.}(1991){Davies}, {Benz}, \& {Hills}}]{DaviesEtAl1991}
{Davies}, M.~B., {Benz}, W., \& {Hills}, J.~G. 1991, \bibinfo{title}{{Stellar Encounters Involving Red Giants in Globular Cluster Cores},} ApJ, 381, 449, \dodoi{10.1086/170668}

\bibitem[{M.~B. {Davies} {et~al.}(1998){Davies}, {Blackwell}, {Bailey}, \& {Sigurdsson}}]{DaviesEtAl1998}
{Davies}, M.~B., {Blackwell}, R., {Bailey}, V.~C., \& {Sigurdsson}, S. 1998, \bibinfo{title}{{The destructive effects of binary encounters on red giants in the Galactic Centre},} MNRAS, 301, 745, \dodoi{10.1046/j.1365-8711.1998.02027.x}

\bibitem[{P.~P. {Eggleton}(1983){Eggleton}}]{Eggleton1983}
{Eggleton}, P.~P. 1983, \bibinfo{title}{{Approximations to the radii of Roche lobes},} \apj, 268, 368, \dodoi{10.1086/160960}

\bibitem[{E. {Gaburov} {et~al.}(2010){Gaburov}, {Lombardi}, \& {Portegies Zwart}}]{Gaburov2010}
{Gaburov}, E., {Lombardi}, Jr., J.~C., \& {Portegies Zwart}, S. 2010, \bibinfo{title}{{On the onset of runaway stellar collisions in dense star clusters - II. Hydrodynamics of three-body interactions},} \mnras, 402, 105, \dodoi{10.1111/j.1365-2966.2009.15900.x}

\bibitem[{M.~J. {Geller} \& J.~P. {Huchra}(1989){Geller} \& {Huchra}}]{GellerHuchra1989}
{Geller}, M.~J., \& {Huchra}, J.~P. 1989, \bibinfo{title}{{Mapping the Universe},} Science, 246, 897, \dodoi{10.1126/science.246.4932.897}

\bibitem[{I. {Iben} \& M. {Livio}(1993){Iben} \& {Livio}}]{Iben1993}
{Iben}, Jr., I., \& {Livio}, M. 1993, \bibinfo{title}{{Common Envelopes in Binary Star Evolution},} \pasp, 105, 1373, \dodoi{10.1086/133321}

\bibitem[{N. {Ivanova} {et~al.}(2013){Ivanova}, {Justham}, {Chen}, {De Marco}, {Fryer}, {Gaburov}, {Ge}, {Glebbeek}, {Han}, {Li}, {Lu}, {Marsh}, {Podsiadlowski}, {Potter}, {Soker}, {Taam}, {Tauris}, {van den Heuvel}, \& {Webbink}}]{Ivanova2013}
{Ivanova}, N., {Justham}, S., {Chen}, X., {et~al.} 2013, \bibinfo{title}{{Common envelope evolution: where we stand and how we can move forward},} \aapr, 21, 59, \dodoi{10.1007/s00159-013-0059-2}

\bibitem[{A.~N. {Kolmogorov}(1962){Kolmogorov}}]{Kolmogorov1962}
{Kolmogorov}, A.~N. 1962, \bibinfo{title}{{A refinement of previous hypotheses concerning the local structure of turbulence in a viscous incompressible fluid at high Reynolds number},} Journal of Fluid Mechanics, 13, 82, \dodoi{10.1017/S0022112062000518}

\bibitem[{D. {Lai} {et~al.}(1993){Lai}, {Rasio}, \& {Shapiro}}]{LRS93}
{Lai}, D., {Rasio}, F.~A., \& {Shapiro}, S.~L. 1993, \bibinfo{title}{Collisions and close encounters between massive main-sequence stars,} ApJ, 412, 593

\bibitem[{P.~J.~T. {Leonard}(1989){Leonard}}]{Leonard1989}
{Leonard}, P. J.~T. 1989, \bibinfo{title}{{Stellar Collisions in Globular Clusters and the Blue Straggler Problem},} AJ, 98, 217, \dodoi{10.1086/115138}

\bibitem[{P.~J.~T. {Leonard} \& G.~G. {Fahlman}(1991){Leonard} \& {Fahlman}}]{LeonardFahlman1991}
{Leonard}, P. J.~T., \& {Fahlman}, G.~G. 1991, \bibinfo{title}{{On the Origin of the Blue Stragglers in the Globular Cluster NGC 5053},} AJ, 102, 994, \dodoi{10.1086/115927}

\bibitem[{J. {Lombardi} {et~al.}(1995){Lombardi}, {Rasio}, \& {Shapiro}}]{LRS95}
{Lombardi}, J.~C., J., {Rasio}, F.~A., \& {Shapiro}, S.~L. 1995, \bibinfo{title}{On blue straggler formation by direct collisions of main sequence stars,} ApJ Lett., 445, 117

\bibitem[{J. {Lombardi} {et~al.}(1996){Lombardi}, {Rasio}, \& {Shapiro}}]{LRS96}
{Lombardi}, J.~C., J., {Rasio}, F.~A., \& {Shapiro}, S.~L. 1996, \bibinfo{title}{Collisions of Main-Sequence Stars and the Formation of Blue Stragglers in Globular Clusters,} ApJ, 468, 797

\bibitem[{J. {Lombardi} {et~al.}(2002){Lombardi}, {Warren}, {Rasio}, {Sills}, \& {Warren}}]{LombardiEtAl2002}
{Lombardi}, James~C., J., {Warren}, J.~S., {Rasio}, F.~A., {Sills}, A., \& {Warren}, A.~R. 2002, \bibinfo{title}{{Stellar Collisions and the Interior Structure of Blue Stragglers},} ApJ, 568, 939, \dodoi{10.1086/339060}

\bibitem[{J.~C. {Lombardi}(1998){Lombardi}}]{Lombardi1998}
{Lombardi}, J.~C. 1998, \bibinfo{title}{{Smoothed Particle Hydrodynamic Simulations of Stellar Collisions},} PhD thesis, Cornell University, New York

\bibitem[{J.~C. {Lombardi} {et~al.}(1999){Lombardi}, {Sills}, {Rasio}, \& {Shapiro}}]{LombardiEtAl1999}
{Lombardi}, J.~C., {Sills}, A., {Rasio}, F.~A., \& {Shapiro}, S.~L. 1999, \bibinfo{title}{{Tests of Spurious Transport in Smoothed Particle Hydrodynamics},} Journal of Computational Physics, 152, 687, \dodoi{10.1006/jcph.1999.6256}

\bibitem[{J.~C. {Lombardi} {et~al.}(2006){Lombardi}, {Proulx}, {Dooley}, {Theriault}, {Ivanova}, \& {Rasio}}]{Lombardi2006}
{Lombardi}, Jr., J.~C., {Proulx}, Z.~F., {Dooley}, K.~L., {et~al.} 2006, \bibinfo{title}{{Stellar Collisions and Ultracompact X-Ray Binary Formation},} \apj, 640, 441, \dodoi{10.1086/499938}

\bibitem[{A. {Maeder}(1987){Maeder}}]{Maeder1987}
{Maeder}, A. 1987, \bibinfo{title}{{Evidences for a bifurcation in massive star evolution. The ON-blue stragglers.},} A\&A, 178, 159

\bibitem[{A. {Mastrobuono-Battisti} {et~al.}(2021){Mastrobuono-Battisti}, {Church}, \& {Davies}}]{Mastrobuono-BattistiEtAl2021}
{Mastrobuono-Battisti}, A., {Church}, R.~P., \& {Davies}, M.~B. 2021, \bibinfo{title}{{Close stellar encounters at the Galactic Centre - I. The effect on the observed stellar populations},} MNRAS, 505, 3314, \dodoi{10.1093/mnras/stab1409}

\bibitem[{B.~W. {Murphy} {et~al.}(1991){Murphy}, {Cohn}, \& {Durisen}}]{MCD91}
{Murphy}, B.~W., {Cohn}, H.~N., \& {Durisen}, R.~H. 1991, \bibinfo{title}{Dynamical and luminosity evolution of active galactic nuclei - Models with a mass spectrum,} ApJ, 370, 60

\bibitem[{N. {Neumayer} {et~al.}(2020){Neumayer}, {Seth}, \& {B{\"o}ker}}]{NeumayerEtAl2020}
{Neumayer}, N., {Seth}, A., \& {B{\"o}ker}, T. 2020, \bibinfo{title}{{Nuclear star clusters},} The Astronomy and Astrophysics Review, 28, 4, \dodoi{10.1007/s00159-020-00125-0}

\bibitem[{S.~T. {Ohlmann} {et~al.}(2016){Ohlmann}, {R{\"o}pke}, {Pakmor}, \& {Springel}}]{OhlmannEtAl2016}
{Ohlmann}, S.~T., {R{\"o}pke}, F.~K., {Pakmor}, R., \& {Springel}, V. 2016, \bibinfo{title}{{Hydrodynamic Moving-mesh Simulations of the Common Envelope Phase in Binary Stellar Systems},} \apjl, 816, L9, \dodoi{10.3847/2041-8205/816/1/L9}

\bibitem[{B. {Paczynski}(1976){Paczynski}}]{Paczynski1976}
{Paczynski}, B. 1976, \bibinfo{title}{{Common Envelope Binaries},} in IAU Symposium, Vol.~73, Structure and Evolution of Close Binary Systems, ed. P.~{Eggleton}, S.~{Mitton}, \& J.~{Whelan}, 75

\bibitem[{J.-C. {Passy} {et~al.}(2012){Passy}, {De Marco}, {Fryer}, {Herwig}, {Diehl}, {Oishi}, {Mac Low}, {Bryan}, \& {Rockefeller}}]{PassyEtAl2012}
{Passy}, J.-C., {De Marco}, O., {Fryer}, C.~L., {et~al.} 2012, \bibinfo{title}{{Simulating the Common Envelope Phase of a Red Giant Using Smoothed-particle Hydrodynamics and Uniform-grid Codes},} \apj, 744, 52, \dodoi{10.1088/0004-637X/744/1/52}

\bibitem[{B. {Paxton} {et~al.}(2011{\natexlab{a}}){Paxton}, {Bildsten}, {Dotter}, {Herwig}, {Lesaffre}, \& {Timmes}}]{MESA01}
{Paxton}, B., {Bildsten}, L., {Dotter}, A., {et~al.} 2011{\natexlab{a}}, \bibinfo{title}{{Modules for Experiments in Stellar Astrophysics (MESA)},} \apjs, 192, 3, \dodoi{10.1088/0067-0049/192/1/3}

\bibitem[{B. {Paxton} {et~al.}(2011{\natexlab{b}}){Paxton}, {Bildsten}, {Dotter}, {Herwig}, {Lesaffre}, \& {Timmes}}]{Paxton2011}
{Paxton}, B., {Bildsten}, L., {Dotter}, A., {et~al.} 2011{\natexlab{b}}, \bibinfo{title}{{Modules for Experiments in Stellar Astrophysics (MESA)},} \apjs, 192, 3, \dodoi{10.1088/0067-0049/192/1/3}

\bibitem[{B. {Paxton} {et~al.}(2013{\natexlab{a}}){Paxton}, {Cantiello}, {Arras}, {Bildsten}, {Brown}, {Dotter}, {Mankovich}, {Montgomery}, {Stello}, {Timmes}, \& {Townsend}}]{MESA02}
{Paxton}, B., {Cantiello}, M., {Arras}, P., {et~al.} 2013{\natexlab{a}}, \bibinfo{title}{{Modules for Experiments in Stellar Astrophysics (MESA): Planets, Oscillations, Rotation, and Massive Stars},} \apjs, 208, 4, \dodoi{10.1088/0067-0049/208/1/4}

\bibitem[{B. {Paxton} {et~al.}(2013{\natexlab{b}}){Paxton}, {Cantiello}, {Arras}, {Bildsten}, {Brown}, {Dotter}, {Mankovich}, {Montgomery}, {Stello}, {Timmes}, \& {Townsend}}]{Paxton2013}
{Paxton}, B., {Cantiello}, M., {Arras}, P., {et~al.} 2013{\natexlab{b}}, \bibinfo{title}{{Modules for Experiments in Stellar Astrophysics (MESA): Planets, Oscillations, Rotation, and Massive Stars},} \apjs, 208, 4, \dodoi{10.1088/0067-0049/208/1/4}

\bibitem[{B. {Paxton} {et~al.}(2015){Paxton}, {Marchant}, {Schwab}, {Bauer}, {Bildsten}, {Cantiello}, {Dessart}, {Farmer}, {Hu}, {Langer}, {Townsend}, {Townsley}, \& {Timmes}}]{MESA03}
{Paxton}, B., {Marchant}, P., {Schwab}, J., {et~al.} 2015, \bibinfo{title}{{Modules for Experiments in Stellar Astrophysics (MESA): Binaries, Pulsations, and Explosions},} \apjs, 220, 15, \dodoi{10.1088/0067-0049/220/1/15}

\bibitem[{B. {Paxton} {et~al.}(2018){Paxton}, {Schwab}, {Bauer}, {Bildsten}, {Blinnikov}, {Duffell}, {Farmer}, {Goldberg}, {Marchant}, {Sorokina}, {Thoul}, {Townsend}, \& {Timmes}}]{MESA04}
{Paxton}, B., {Schwab}, J., {Bauer}, E.~B., {et~al.} 2018, \bibinfo{title}{{Modules for Experiments in Stellar Astrophysics (MESA): Convective Boundaries, Element Diffusion, and Massive Star Explosions},} \apjs, 234, 34, \dodoi{10.3847/1538-4365/aaa5a8}

\bibitem[{W.~H. {Press} \& S.~A. {Teukolsky}(1977){Press} \& {Teukolsky}}]{PressTeukolsky1977}
{Press}, W.~H., \& {Teukolsky}, S.~A. 1977, \bibinfo{title}{{On formation of close binaries by two-body tidal capture.},} \apj, 213, 183, \dodoi{10.1086/155143}

\bibitem[{F.~A. {Rasio}(1991){Rasio}}]{Rasio1991}
{Rasio}, F.~A. 1991, \bibinfo{title}{{Hydrodynamical Calculations of Stellar Interactions.},} PhD thesis, Cornell University, New York

\bibitem[{F.~A. {Rasio} \& J.~C. {Lombardi}(1999){Rasio} \& {Lombardi}}]{RasioLombardi1999}
{Rasio}, F.~A., \& {Lombardi}, Jr., J.~C. 1999, \bibinfo{title}{{Smoothed particle hydrodynamics calculations of stellar interactions.},} Journal of Computational and Applied Mathematics, 109, 213, \dodoi{10.48550/arXiv.astro-ph/9805089}

\bibitem[{T. {Ryu} {et~al.}(2024){Ryu}, {Amaro Seoane}, {Taylor}, \& {Ohlmann}}]{RyuEtAl2024}
{Ryu}, T., {Amaro Seoane}, P., {Taylor}, A.~M., \& {Ohlmann}, S.~T. 2024, \bibinfo{title}{{Collisions of red giants in galactic nuclei},} \mnras, 528, 6193, \dodoi{10.1093/mnras/stae396}

\bibitem[{R.~H. {Sanders}(1970){Sanders}}]{Sanders70b}
{Sanders}, R.~H. 1970, \bibinfo{title}{The Effects of Stellar Collisions in Dense Stellar Systems,} ApJ, 162, 791

\bibitem[{M.~M. {Shara}(2002){Shara}}]{2002ASPC..263....1S}
{Shara}, M.~M. 2002, \bibinfo{title}{{Stellar Collisions, Mergers and Their Consequences},} in Astronomical Society of the Pacific Conference Series, Vol. 263, Stellar Collisions, Mergers and their Consequences, ed. M.~M. {Shara}, 1

\bibitem[{S. {Sigurdsson} \& E.~S. {Phinney}(1995){Sigurdsson} \& {Phinney}}]{SigurdssonPhinney95}
{Sigurdsson}, S., \& {Phinney}, E.~S. 1995, \bibinfo{title}{{Dynamics and Interactions of Binaries and Neutron Stars in Globular Clusters},} Astrophysical Journal Supplement, 99, 609, \dodoi{10.1086/192199}

\bibitem[{J. {Spitzer} \& W.~C. {Saslaw}(1966){Spitzer} \& {Saslaw}}]{SS66}
{Spitzer}, L., J., \& {Saslaw}, W.~C. 1966, \bibinfo{title}{On the Evolution of Galactic Nuclei,} ApJ, 143, 400

\bibitem[{L. {Spitzer}(1987){Spitzer}}]{Spitzer87}
{Spitzer}, L. 1987, {Dynamical evolution of globular clusters} (Princeton, NJ, Princeton University Press, 1987, 191 p.)

\bibitem[{H. {Trac} {et~al.}(2007){Trac}, {Sills}, \& {Pen}}]{TracEtAl2007}
{Trac}, H., {Sills}, A., \& {Pen}, U.-L. 2007, \bibinfo{title}{{A comparison of hydrodynamic techniques for modelling collisions between main-sequence stars},} MNRAS, 377, 997, \dodoi{10.1111/j.1365-2966.2007.11709.x}

\bibitem[{R.~B. {Tully} {et~al.}(2014){Tully}, {Courtois}, {Hoffman}, \& {Pomar{\`e}de}}]{TullyEtAl2014}
{Tully}, R.~B., {Courtois}, H., {Hoffman}, Y., \& {Pomar{\`e}de}, D. 2014, \bibinfo{title}{{The Laniakea supercluster of galaxies},} Nat, 513, 71, \dodoi{10.1038/nature13674}

\bibitem[{R. {Tylenda} {et~al.}(2011){Tylenda}, {Hajduk}, {Kami{\'n}ski}, {Udalski}, {Soszy{\'n}ski}, {Szyma{\'n}ski}, {Kubiak}, {Pietrzy{\'n}ski}, {Poleski}, {Wyrzykowski}, \& {Ulaczyk}}]{TylendaEtAl2011}
{Tylenda}, R., {Hajduk}, M., {Kami{\'n}ski}, T., {et~al.} 2011, \bibinfo{title}{{V1309 Scorpii: merger of a contact binary},} \aap, 528, A114, \dodoi{10.1051/0004-6361/201016221}

\bibitem[{M.~Z.~C. {Vergara} {et~al.}(2021){Vergara}, {Schleicher}, {Boekholt}, {Reinoso}, {Fellhauer}, {Klessen}, \& {Leigh}}]{2021A&A...649A.160V}
{Vergara}, M.~Z.~C., {Schleicher}, D.~R.~G., {Boekholt}, T.~C.~N., {et~al.} 2021, \bibinfo{title}{{Stellar collisions in flattened and rotating Population III star clusters},} 649, A160, \dodoi{10.1051/0004-6361/202140298}

\bibitem[{C.~E. {Woodward} {et~al.}(2021){Woodward}, {Evans}, {Banerjee}, {Liimets}, {Djupvik}, {Starrfield}, {Clayton}, {Eyres}, {Gehrz}, \& {Wagner}}]{WoodwardEtAl2021}
{Woodward}, C.~E., {Evans}, A., {Banerjee}, D.~P.~K., {et~al.} 2021, \bibinfo{title}{{The Infrared Evolution of Dust in V838 Monocerotis},} \aj, 162, 183, \dodoi{10.3847/1538-3881/ac1f1e}

\bibitem[{S. {Wu} {et~al.}(2020){Wu}, {Everson}, {Schneider}, {Podsiadlowski}, \& {Ramirez-Ruiz}}]{2020ApJ...901...44W}
{Wu}, S., {Everson}, R.~W., {Schneider}, F. R.~N., {Podsiadlowski}, P., \& {Ramirez-Ruiz}, E. 2020, \bibinfo{title}{{The Art of Modeling Stellar Mergers and the Case of the B[e] Supergiant R4 in the Small Magellanic Cloud},} ApJ, 901, 44, \dodoi{10.3847/1538-4357/abaf48}

\end{thebibliography}
\end{document}